\documentclass{mn2e} 
\usepackage{color}
\usepackage{dcolumn}
\usepackage{bm}
\usepackage{epsfig}
\usepackage{ulem}

\def\reference{\par\noindent\hangindent\parindent}

 at 13 truept

\begin{document}

\twocolumn

\title[GAMA, G10-COSMOS and 3D-HST]{GAMA/G10-COSMOS/3D-HST: The $0<z<5$ cosmic star-formation history, stellar- and dust-mass densities}
\author[Driver et al.] 
{Simon P.~Driver$^{1,2}$\thanks{SUPA, Scottish Universities Physics Alliance}, 
Stephen K.~Andrews$^{1}$,
Elisabete da Cunha$^{3}$,
Luke J.~Davies$^{1}$,
\newauthor
Claudia~Lagos$^{1}$,
Aaron S.G.~Robotham$^{1,2}$,
Kevin Vinsen$^{1}$,
Angus H.~Wright$^{1}$,
\newauthor
Mehmet Alpaslan$^{4}$,
Joss Bland-Hawthorn$^{5}$,
Nathan Bourne$^{6}$,
Sarah Brough$^{7}$,
\newauthor
Malcolm N.~Bremer$^{8}$,
Michelle Cluver$^{9}$,
Matthew Colless$^{3}$, 
Christopher J.~Conselice$^{9}$,
\newauthor
Loretta Dunne$^{6,10}$,
Steve A.~Eales$^{10}$,
Haley Gomez$^{10}$
Benne Holwerda$^{11}$,
\newauthor
Andrew M.~Hopkins$^{12}$,
Prajwal R.~Kafle$^{1}$,
Lee S.~Kelvin$^{13}$,
Jon Loveday$^{14}$,
\newauthor
Jochen Liske$^{15},$
Steve J.~Maddox$^{6,10}$,
Steven Phillipps$^{8}$,
Kevin Pimbblet$^{16}$,
\newauthor
Kate Rowlands$^{17}$,
Anne E.~Sansom$^{18}$,
Edward Taylor$^{19}$,
Lingyu Wang$^{20}$,
\newauthor
Stephen M. Wilkins$^{14}$ \\
$^1$ International Centre for Radio Astronomy Research (ICRAR), University of Western Australia, Crawley, WA 6009, Australia \\
$^2$ School of Physics \& Astronomy, University of St Andrews, North Haugh, St Andrews, KY16 9SS, UK; SUPA \\
$^3$ Research School of Astronomy and Astrophysics, Australian National University, Canberra, ACT 2611, Australia \\
$^4$ NASA Ames Research Centre, N244-30, Moffett Field, Mountain View, CA 94035, USA \\
$^{5}$ Sydney Institute for Astronomy, School of Physics A28, University of Sydney, NSW 2006, Australia \\ 
$^6$ Institute for Astronomy, University of Edinburgh, Royal Observatory, Edinburgh, EH9 3HJ, UK \\
$^7$ School of Physics, University of New South Wales, NSW 2052, Australia \\
$^8$ H H Wills Physics Laboratory, University of Bristol, Tyndall Avenue, Bristol, BS8 1TL, UK \\
$^{9}$ Astrophysics Group, The University of Western Cape, Robert Sobukwe Road, Bellville 7530, South Africa, \\
$^9$ School of Physics and Astronomy, University of Nottingham, University Park, Nottingham, NG7 2RD, UK \\
$^{10}$ School of Physics and Astrophysics, Cardiff University, Queens buildings, The Parade, Cardiff, CF24 3AA, UK \\
$^{11}$ Department of Physics and Astronomy, University of Louisville, 102 Natural Science Building, Louisville, KY 40292, USA \\
$^{12}$ Australian Astronomical Observatory, PO Box 915, North Ryde, NSW 1670, Australia \\
$^{13}$ Astrophysics Research Institute, Liverpool, IC2, Liverpool Science Park, 146 Brownlow Hill, Liverpool, L3 5RF, UK \\
$^{14}$ Astronomy Centre, Department of Physics and Astronomy, University of Sussex, Falmer, Brighton BN1 9QH, UK\\
$^{15}$ Hamburg Sternwarte, Universit{\"a}t Hamburg, Gojenbergsweg 112, 21029 Hamburg, Germany\\
$^{16}$ Department of Physics and Mathematics, University of Hull, Cottingham Road, Hull, HU6 7RX, UK \\
$^{17}$ Johns Hopkins University, Department of Physics \& Astronomy, 3400 N. Charles St, Baltimore, MD 21218, USA \\
$^{18}$ Jeremiah Horrocks Institute, University of Central Lancashire, Preston, Lancashire, PR1 2HE, UK \\
$^{19}$ Centre for Astrophysics and Supercomputing, Swinburne University of Technology, PO Box 218, Hawthorn, Victoria 3122, Australia \\
$^{20}$ SRON Netherlands Institute for Space Research, Landleven 12, 9747 AD, Groningen, The Netherlands}
\pubyear{2012} \volume{000}
\pagerange{\pageref{firstpage}--\pageref{lastpage}}

\maketitle
\label{firstpage}

\vspace{-2.0cm}

\begin{abstract}
We use the energy-balance code MAGPHYS to determine stellar and dust
masses, and dust corrected star-formation rates for over 200,000 GAMA
galaxies, 170,000 G10-COSMOS galaxies and 200,000 3D-HST galaxies. Our
values agree well with previously reported measurements and constitute
a representative and homogeneous dataset spanning a broad range in
stellar mass ($10^{8}$---$10^{12}$M$_{\odot}$), dust mass
($10^{6}$---$10^{9}$M$_{\odot}$), and star-formation rates
($0.01$---$100$M$_{\odot}$yr$^{-1}$), and over a broad redshift range
($0.0 < z < 5.0$). We combine these data to measure the cosmic
star-formation history (CSFH), the stellar-mass density (SMD), and the
dust-mass density (DMD) over a 12 Gyr timeline. The data mostly agree
with previous estimates, where they exist, and provide a
quasi-homogeneous dataset using consistent mass and star-formation
estimators with consistent underlying assumptions over the full time
range. As a consequence our formal errors are significantly reduced
when compared to the historic literature. Integrating our cosmic
star-formation history we precisely reproduce the stellar-mass density
with an ISM replenishment factor of $0.50 \pm 0.07$, consistent with
our choice of Chabrier IMF plus some modest amount of stripped stellar
mass.  Exploring the cosmic dust density evolution, we find a gradual
increase in dust density with lookback time. We build a simple
phenomenological model from the CSFH to account for the dust mass
evolution, and infer two key conclusions: (1) For every unit of
stellar mass which is formed $0.0065$---$0.004$ units of dust mass is
also formed; (2) Over the history of the Universe approximately 90 to
95 per cent of all dust formed has been destroyed and/or ejected.
\end{abstract}


\section{Introduction}
Since recombination the baryonic mass in the Universe has transformed
from a smooth atomic distribution of neutral gas, to ionised gas (i.e.,
reionisation), and thereafter into a number of distinct forms. Most
notably residual ionised gas, neutral gas (HI), molecular gas, stars,
dust, and super-massive black holes (SMBHs). The redistribution of the
primordial re-ionised plasma over time is of pertinent scientific
interest. Most of the action, in terms of transformational processes,
occur in the context of galaxy formation and evolution. This is
moderated by the dominating gravitational field of the underlying dark
matter halo, galaxy-galaxy interactions, and gas accretion, all of
which drive a multitude of astrophysical processes which give rise to
changes in the cosmic gas, stellar, dust, and SMBH densities over
time.

The current baryon inventory (see Shull, Smith \& Daniforth 2012),
suggests that today's baryonic mass can be roughly broken down into
the following forms:

~

\noindent
UNBOUND:

\noindent
--- hot ionised plasma (28 per cent; Fukugita, Hogan \& Peebles 1998; Shull, Smith \& Danforth 2012)

\noindent
--- the Warm Hot Intergalactic Medium (29 per cent; Shull, Smith \& Danforth 2012)

~

\noindent
BOUND TO CLUSTER AND GROUP HALOSs:

\noindent
--- the intra-cluster light (4 per cent; Shull, Smith \& Danforth 2012)

\noindent
--- the intra-group light ($<1$ per cent; Driver et al.~2016)

~

\noindent
BOUND TO GALAXY HALOS:

\noindent
--- stars (6 per cent Baldry, Glazebrook \& Driver 2008, 2012; Peng et al.~2010; Moffett et al.~2016; Wright et al.~2017a)

\noindent
--- neutral gas (2 per cent; Zwaan et al.~2005; Martin et al.~2010; Delhaize et al.~2013; Martindale et al.~2017)

\noindent
--- circum-galactic medium (5 per cent Shull, Smith \& Danforth 2012; Stocke et al.~2013)

\noindent
--- molecular gas (0.2 per cent; Keres et al., 2003; Walter et al.~2014)

\noindent
--- dust (0.1 per cent Vlahakis, Dunne \& Eales 2005; Driver et al. 2007; Dunne et al.~2011; Clemens et al.~2013; Beeston et al.~2017)

\noindent
--- SMBHs (0.01 per cent Shankar et al.~2004; Graham et al.~2007 ; Vika et al.~2009 ; Mutlu Pakdil, Seigar \& Davis 2016)

~

\noindent
UNACCOUNTED FOR: 

\noindent
--- missing baryons (25 per cent; see also Shull, Smith \& Danforth 2012)

~

\begin{figure}
\begin{center}
\includegraphics[width=\columnwidth]{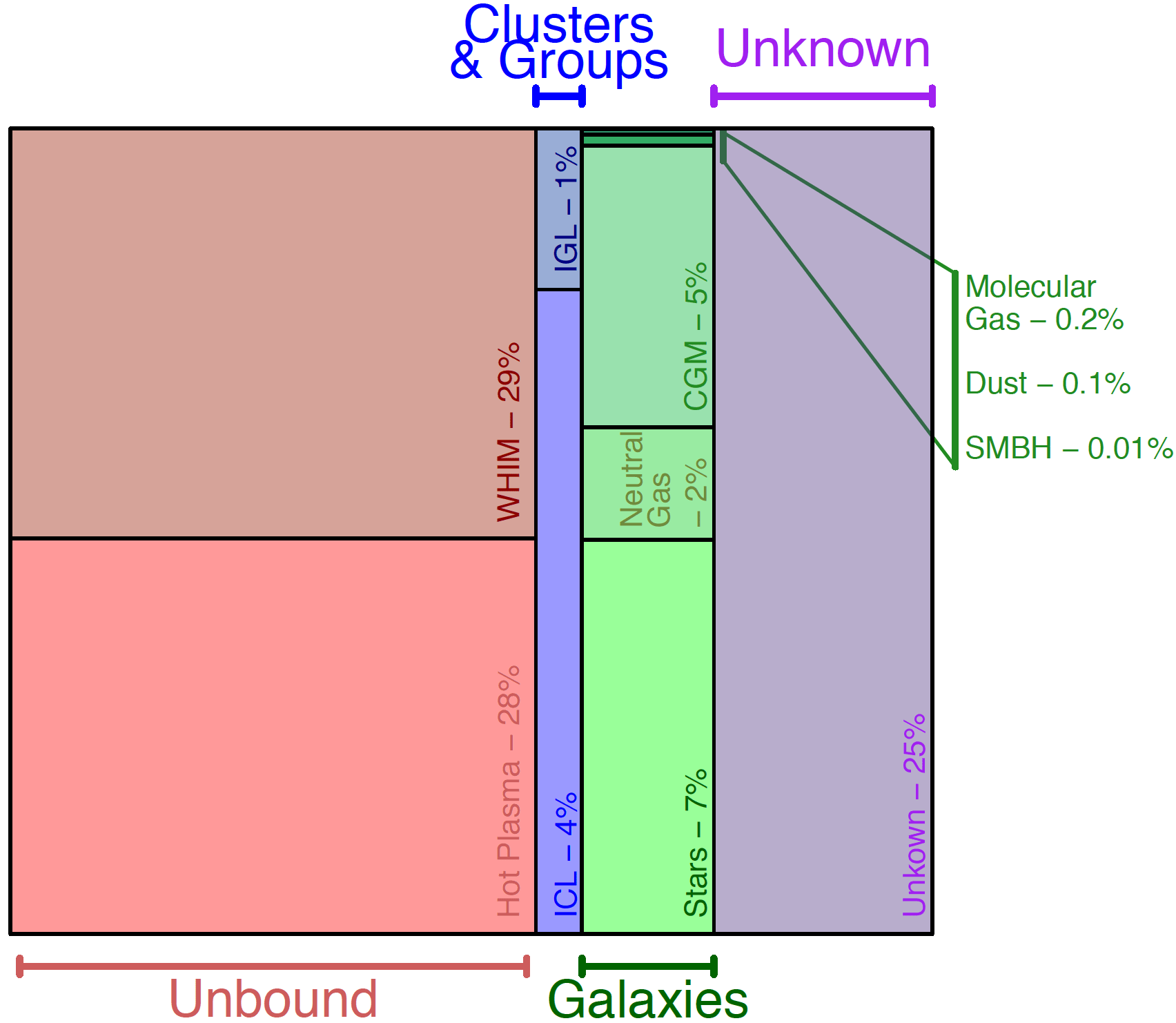}
\caption{ {{{ The baryon budget divided into bound and unbound
        repositories as well as gas, dust, and stellar
        sub-components. Data mostly derived from Shull, Smith \&
        Danforth (2012) with updates as described in the text.}}}
\label{fig:baryons}}
\end{center}
\end{figure}

These components (see Fig.~\ref{fig:baryons}) sum to form the baryon
budget (Fukugita, Hogan \& Peebles 1998), which can be compared to the
baryon density implied from cosmological experiments, e.g., {\it
  Wilkinson Microwave Anisotropy Probe} (WMAP; Hinshaw et al.~2013),
Planck (Ade et al.~2016), and various constraints on Big Bang
Nucleosynthesis (BBN; Cyburt et al.~2016). At the moment some tension
exists between cosmological versus local inventories (Shull, Smith \&
Danforth 2012). However significant leeway (i.e., $\pm 50$ per cent)
is available in almost all of the mass repositories listed above. The
dominant ionised component, in particular, is extremely hard to
robustly constrain and can be crudely divided into: unbound
free-floating and very hot ionised gas ($T \sim 10^{6-8}$ K); the
loosely bound Warm Hot Inter-galactic Medium (WHIM; $T \sim
10^{4-6}$K); the bound hot intra-cluster/group light (ICL/IGL; $T \sim
10^{6-7}$K); and the bound circum-galactic plasma ($T \sim
10^6$K).Cooler components also cannot be ruled out (i.e., $10^2 -
10^4$K). These components, illustrated in Fig.~\ref{fig:baryons}, and
their associated errors, are discussed in Shull, Smith \& Danforth
(2012) who first articulated concerns over the missing $\sim 30$\% of
baryons as compared to WMAP and BBN analyses. A more statistical
approach based on the kinematic Sunyaev-Zeldovich effect in the Planck
Cosmic Microwave Background dataset (Hern\'andez-Monteagudo et
al.~2015) does suggest that the bulk of the baryons closely follow the
dark-matter distribution and, based on opacity arguments, argue for an
additional ionised component beyond that seen via traditional X-ray
absorption lines. Recently an additional hot-WHIM component has been
reported by Bonamente et al.~(2016), as well as an overdensity of the
WHIM along the cosmic web (Eckert et al.~2015). However, conversely,
Danforth et al.~(2015) revisited the estimates of Shull et al., and
reported a lower value for the directly detected gas. Essentially
sufficient uncertainty exists which suggests that the bulk of the
missing baryons is most likely in an ionised component, that closely
follows the underlying dark-matter distribution.

Comparable uncertainty at a similar $\pm 50$ per cent level
potentially exists in the more minor components, i.e., the neutral
gas, molecular gas, stellar, dust and SMBH components associated with
galaxies (with all other repositories considered insignificant
compared to these, e.g., planets and planetesimals). To some extent
these are linked, i.e., if one identifies more stellar mass in the
form of an additional galaxy population the other components would
likely increase too (i.e., the associated CGM, HI etc). Also of
interest is the change in these bound components with time as gas is
converted into stars, metals, and dust.

Measurements of the galaxy population suggests that over the past few
Gyrs the stellar mass and HI cosmic co-moving density has plateaued
(see Wilkins, Trentham \& Hopkins.~2008; Delhaize et al.~2013), while
the molecular mass density has declined with time (Walter et al.~2014;
Decarli et al.~2016), and the cosmic dust density declined rapidly over
late epochs (Dunne et al.~2011). The latter molecular gas and dust
density declines are arguably driven by the decline in the cosmic
star-formation history (Lilly et al.~1996; Hopkins \& Beacom 2006;
Madau \& Dickinson 2014). One of the key goals of the Galaxy And Mass
Assembly (GAMA) project (Driver et al.~2009; 2011) is to quantify the
baryon components contained within galaxies, and to empirically
recover their recent evolution. In this study we focus in particular
on the cosmic star-formation history (CSFH), the stellar mass density
(SMD), and dust mass density (DMD).

Central to a robust estimate of the bound mass components, is the
determination of consistent stellar and dust mass estimates over a
sufficiently large area to overcome cosmic variance (Driver \&
Robotham 2010), and over a sufficiently large redshift baseline to
probe time evolution. This inevitably requires extensive observations
on multiple ground and space-based facilities. Over the past 7 years
we have assembled an extensive database of panchromatic photometry
(Driver et al.~2016) extending from the UV to the far-IR over a
combined 230sq deg region of sky and which builds upon a deep
Australian/European spectroscopic campaign of 300,000 galaxies (with
$r<19.8$ mag and $z \leq 0.5$; Driver et al.~2011; Liske et
al.~2015). This dataset has been constructed from a number of
independent survey programs including GALEX (Martin et al.~2005), SDSS
(York et al.~2000), VIKING (Sutherland et al.~2015), WISE (Wright et
al.~2010), and Herschel-ATLAS (Eales et al.~2010; Valiante et
al.~2016; Bourne et al.~2017).  In parallel, a similar
US/European/Japanese effort has obtained extremely deep panchromatic
imaging over the {\it Hubble Space Telescope} (HST) Cosmology
Evolution Survey (COSMOS) region (Scoville et al.~2007a,b), while a
USA-led group have built extensive multiwavelength and GRISM
observations of notable HST deep fields as part of the 3D-HST study
(see van Dokkum et al.~2013; Brammer et al.~2012 and Momcheva et
al.~2016).

We have recently completed the task of assimilating and homogenising
the first two of these datasets (GAMA and G10-COSMOS) into the GAMA
database, using an identical software analysis pathway for object
detection (SExtractor; Bertin et al.~2017), redshift estimation
(Baldry et al.~2014; Liske et al.~2015; Davies et al.~2015), and
panchromatic flux measurement (Wright et al.~2016; Andrews et
al.~2017a).  {{{ For the 3D-HST dataset we use the online database
      of panchromatic photometry and redshifts as provided by the
      3D-HST team (see Momcheva et al.~2016 and references
      therein).}}}

A crucial step in homogenising these three surveys, is to obtain
consistent star-formation rate, stellar mass, and dust mass
estimates. For this purpose we look to the MAGPHYS energy balance code
provided by da Cunha, Charlot \& Elbaz (2008). MAGPHYS takes as input
a redshift and a series of flux measurements (and errors) spanning the
UV to far-IR wavelength range. It then compares the observed flux
measurements to an extensive stellar spectral and dust emission
library to obtain an optimal spectral energy distribution (SED); and
where the energy attenuated by dust in the UV/optical/near-IR, balances
with the energy radiated in the far-IR. Parameters constrained by this
process include the unattenuated star-formation rate, stellar mass,
and total dust mass along with information on the opacity, temperature
of the ISM and birth clouds, age, metallicity and the unattenuated and
attenuated best fit spectral energy distributions.

In Section~2 we provide summary information on our three adopted
datasets: GAMA, G10-COSMOS and 3D-HST. In Section~3 we describe the
process of MAGPHYS analysis of almost 600,000 galaxy SEDs using the
Australian Research Council Pawsey Supercomputing Facility. We explore
and validate the datasets in the latter part of Section~3 before
finally presenting the cosmic star-formation history and the evolution
of the stellar and dust mass densities since $z=5$ in Section~4. {{{ In
Section~5 we discuss the implications of our results compared to
numerical simulations, attempt to build a phenomenological model to
explain the stellar mass and dust density from the CSFH and finish by
placing our measurements into the context of the evolution of the
bound baryon budget.}}}

This empirical paper therefore
forms the basis for a series of further papers which explore: the very
faint-end of the stellar mass function and the prospect of missing
diffuse low-surface brightness galaxies (Wright et al.~2017a); the HI
and baryonic mass function (Wright et al.~2017b); the faint-end of the
low-redshift dust mass function (Beeston et al.~2017); the evolution
of the cosmic spectral energy distribution (Andrews et al.~2017b); and
detailed modelling of the evolution of the cosmic spectral energy
distribution with time (Andrews et al.~2017c).

In general these studies extend our existing knowledge by providing
consistent {\it homogeneous} measurements over very large volumes,
across a very broad range of stellar mass and lookback times, thereby
minimising the impact of cosmic (sample) variance.

Throughout we use a concordance cosmological model of $\Omega_M=0.3$,
$\Omega_{\Lambda}=0.7$ and $H_{0} = 70 h_{\rm 70}$ km/s/Mpc and work with a
time-invariant Chabrier (2003) IMF.

\section{Data}
We bring together three complementary datasets: GAMA (Driver et
al.~2011; Liske et al.~2015), G10-COSMOS (Davies et al.~2015; Andrews
et al.~2017a) and 3D-HST (Momcheva et al.~2016). All three studies
contain extensive panchromatic photometry extending from the
ultra-violet to mid-infrared wavelengths allowing for robust stellar
mass estimates. The GAMA and G10-COSMOS data also contain far-IR
measurements or constraints from the {\it Herschel Space Observatory's} SPIRE and
PACS instruments (Herschel) by the Herschel-ATLAS (Eales et al.~2010),
and HerMES (Oliver et al.~2012) teams respectively. This allows for
measurement of dust masses and dust-corrected star formation
rates. Collectively all three datasets extend from nearby ($z \leq
0.5$; GAMA), to the intermediate ($z < 1.75$; G10-COSMOS), and high-z
Universe ($z<5.0$; 3D-HST). Each dataset contains approximately 200k
galaxies and collectively sample a broad range in stellar mass,
morphological types and lookback time.  

{{{ In this section we first introduce each of the contributing
      datasets.}}}

\begin{figure*}
\begin{center}
\includegraphics[width=\textwidth]{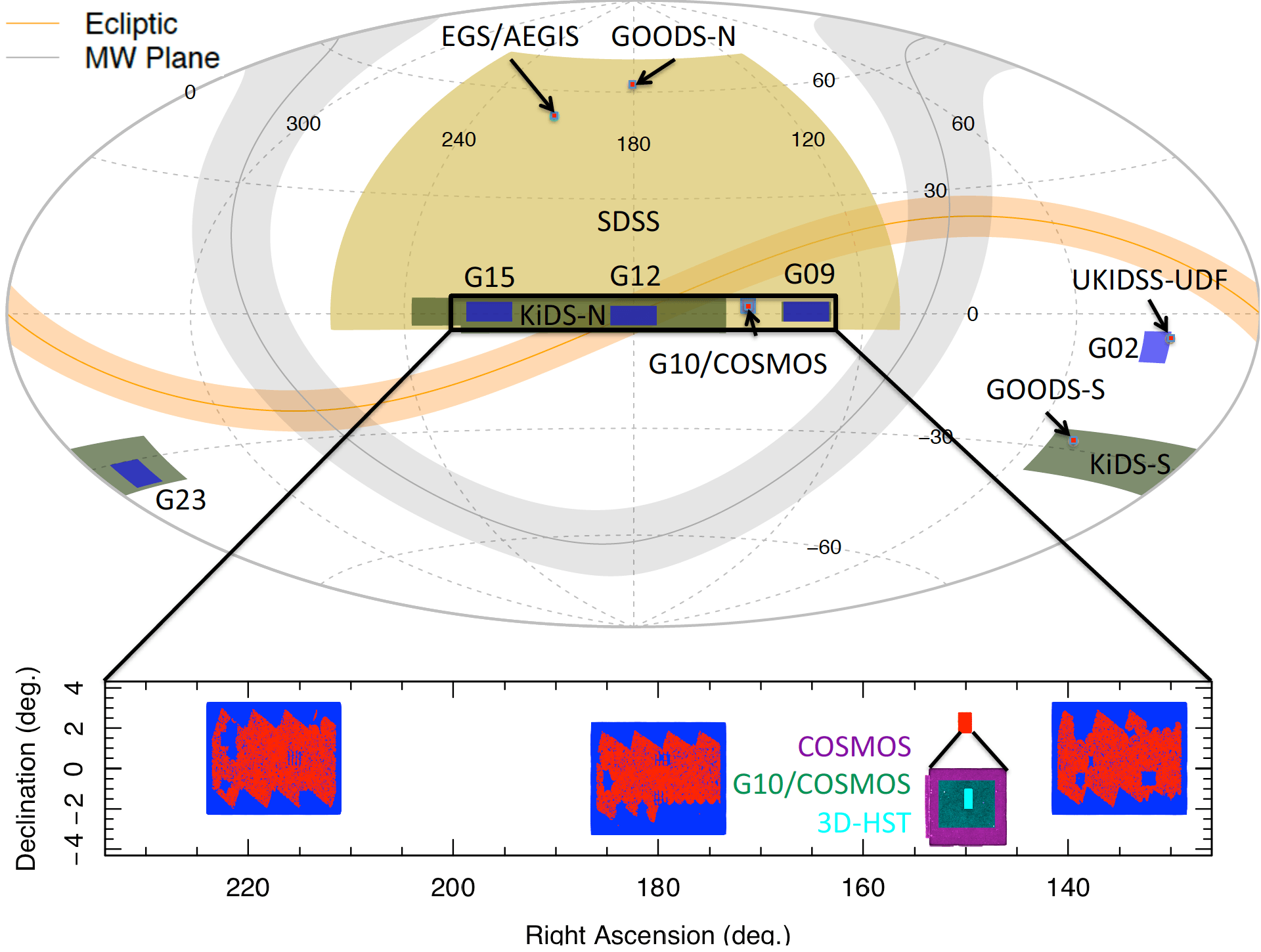}
\caption{The GAMA, G10-COSMOS and 3D-HST (GOODS-N \& -S, UKIDSS-UDS, EGS/AGEIS,
  COSMOS) survey regions shown on an Aitoff projection of the sky (as
  indicated). The lower zoom panel highlights the equatorial GAMA
  regions where blue denotes the full survey regions, and red the
  distribution of galaxies with complete panchromatic coverage. Also
  shown in the lower zoom is the various definitions of the COSMOS
  region, where COSMOS (purple) denotes $\sim 2$ square degree region
  covered by HST, G10-COSMOS the central square degree with consistent
  spectroscopic coverage (emerald), and 3D-HST the sub-region with HST
  GRISM (G141) coverage (cyan). The Aitoff projection is generated
  via AstroMap: http://astromap.icrar.org/}
\label{fig:coverage}
\end{center}
\end{figure*}

\subsection{Galaxy And Mass Assembly (GAMA)}
The Galaxy And Mass Assembly (GAMA) Survey (Driver et al.~2009; 2011)
consists primarily of a dedicated spectroscopic campaign to
$r<19.8$mag (Driver et al.~2011; Hopkins et al.~2013; Liske et al
2015). It builds upon the two-degree field galaxy redshift survey
(2dFGRS; Colless et al.~2001) and the Sloan Digital Sky Survey (SDSS;
York et al.~2000). With the latter providing the basis of the GAMA
input catalogue for the three equatorial fields using colour and size
selection criteria (see Baldry et al.~2010 for details). GAMA overall
covers five distinct survey regions (see Fig.~\ref{fig:coverage}),
including the three equatorial fields at $9^h$ (G09), $12^h$ (G12) and
$14.5^h$ (G15). Each of the equatorial survey fields (see
Fig.~\ref{fig:coverage}, zoom panel) covers a region of $5 \times 12$
sq degrees and, to the spectroscopic survey limits, contains
approximately 70,000 galaxies within each region. Redshifts have been
obtained for $>98$ per cent (see Liske et al.~2015 for the final
spectroscopic survey report), with the majority measured by the GAMA
team using the AAOmega facility at the Anglo Australian Telescope. In
addition to the spectroscopic component, GAMA contains imaging
observations from a broad range of ground and space based facilities
including: UV (GALEX), optical (SDSS, VST),
near-IR (UKIRT, VISTA), mid-IR (WISE), and far-IR (Herschel)
imaging. These data have been aggregated and made publicly available
through the GAMA Panchromatic Data Release (Driver et al.~2016; see
http://gama-psi.icrar.org). Photometric flux measurements in 21
bandpasses (FUV, NUV, $ugriz$, $ZYJHK$, W1234, PACS100/160, SPIRE
250/350/500) have been completed using in-house software (LAMBDAR;
Wright et al.~2016). LAMBDAR uses the elliptical apertures obtained
via SExtractor and convolves them with the appropriate facility PSF,
and manages flux-sharing for blended objects including a contamination
target list if provided (e.g., stars in the UV to mid-IR bands and high-z
systems in the far-IR bands). For more details on LAMBDAR and access
to the photometric catalogue and source code see
http://gama-psi.icrar.org/LAMBDAR.php and Wright et al.~(2016).

Here we use LAMBDARCatv01 which contains 200,246 objects and extract
those with redshifts with quality $nQ \geq 3$ using a name match with
TilingCatv43.  We remove systems with $z < 0.001$, replace
measured negative fluxes with zeros (i.e., where MAGPHYS will ignore
the flux and use the flux error as an upper limit), and replace fluxes
where there is no imaging coverage in that band, with a flux value of
$-999$ (i.e., ignored by MAGPHYS). The catalogue is then parsed to the
MAGPHYS input format which consists of: ID, redshift, $21 \times
$[flux, flux-error] (in Jy). Fig.~\ref{fig:coverage} shows the on-sky
area, with the GAMA regions shown in blue and the area with complete
wavelength coverage in all 21 bands shown in red. Restricting our
dataset to this latter area reduces the galaxies with complete SED
coverage and valid redshifts from 197,494 to 128,568 and our effective
survey area from 180.0 sq deg to 117.2 sq deg. {{{ Within this
      region our final sample is 98 per cent spectroscopically
      complete to $r<19.8$ mag, with no obvious surface brightness or
      colour bias (see Liske et al.~2015).}}}

\subsection{G10-COSMOS}
G10-COSMOS (Davies et al.~2015; Andrews et al.~2017a) is a 1 sq deg
sub-region of the HST COSMOS survey (Scoville et al.~2007a,b). It
enjoys contiguous coverage from ultra-violet to far-IR wavelengths
(Andrews et al.~2017a), including: UV (GALEX), optical (CFHT, Subaru,
HST), near-IR (VISTA), mid-IR (Spitzer), and far-IR (Herschel)
imaging. These deep data have been obtained from a variety of public
websites, and processed in a similar manner to the GAMA data using
LAMBDAR (Andrews et al.~2017a). For G10-COSMOS we adopt an $i<25$ mag
defined catalogue based on a Source Extractor analysis of the $i$-band
Subaru observations (Capak et al.~2007; Taniguchi et al.~2007). This
has been followed by extensive efforts to refine the aperture
definitions and reject spurious detections (see Andrews et al.~2017a
for details). For redshift information we use the updated Davies et
al.~(2015) catalogue. This includes our independent redshift
extraction of the zCOSMOS-Bright sample, combined with spectroscopic
redshifts from PRIMUS, VVDS, SDSS (Cool et al.~2013; Le Fevre et
al.~2013; Ahn et al.~2014), and photometric redshift estimates from
COSMOS2015 (Laigle et al.~2016). The Andrews et al.  photometric and
updated Davies et al. spectroscopic catalogues are publicly available
from http://gama-psi.icrar.org/G10/dataRelease.php

To generate our MAGPHYS input file we adopt
G10CosmosLAMBDARCatv06\footnote{Note that this catalogue has an
  extended far-IR sampling which we briefly describe in Appendix A}
and extract all objects classified as galaxies and produce an input
catalogue with : ID, redshift, $22 \times$[flux, fluxerr] containing
142,260 objects (with $z<1.75$). Explicitly the G10-COSMOS dataset has
the following filters:
FUV,NUV,ugrizYJHK,IRAC1234,MIPS24/70,PACS100/160,
SPIRE250/350/500. Note that we do not include the $B$ and $V$ bands
because their zeropoints remain somewhat uncertain, and their
inclusion would also have the potential to over-resolve the SED fits,
particularly given that the majority of these data have photometric
rather than spectroscopic redshifts. Because the data arise from
multiple facilities, where zero-point errors cannot be entirely ruled
out, we implement an error-floor where we set the flux error in each
band for each galaxy to be the largest of either the quoted error, or
10 per cent of the flux. Fig.~\ref{fig:coverage} shows the on-sky area
with the G10-COSMOS region indicated in both the main panel and the
blow-up region.

{{{ The G10/COSMO sample is therefore 100 per cent redshift
      complete to the specified flux limit, with a combination of both
      spectroscopic and photometric redshifts. For the photometric
      redshifts the accuracy has been shown to be $\pm 0.0007$ with a
      catastrophic redshift failure rate of below 0.5 per cent.  (see
      Laigle et al.~2015).}}}

\subsection{3D-HST}
To extend our stellar mass coverage to the very distant Universe we
also include the 3D-HST dataset (Momcheva et al.~2016; Brammar et
al.~2012). This was downloaded from the 3D-HST website (version
4.1.5): http://3dhst.research.yale.edu and constitutes a sample of
207,967 galaxies, stars and AGN from five notable deep HST
studies. The 3D-HST fields are themselves subregions of the AEGIS,
COSMOS, GOODS-S, GOODS-N, and UKIDSS-UDS HST CANDELS fields, for which
there is GRISM coverage (WFC3/G141 and/or WFC3/G800L), providing
coarse photometric or spectroscopic redshifts over a total of 0.274 sq
arcmins (of which 0.174 sq deg is covered by the GRISM data, see
Momcheva et al.~2016). Overall the sample has been shown to have a
redshift accuracy of $\Delta z/(1+z) = \pm 0.003$, with some
expectation that this accuracy will decrease somewhat below $z=0.7$
and towards fainter magnitudes where the bulk of the redshift
estimates are purely photometric (see Momcheva et al.~2016 their
figures 13 and 14 in particular). In total the 3D-HST catalogue
contains 204,294 galaxies and AGN with either a spectroscopic (3839),
GRISM (15518), or photometric (185843) redshift estimate. In addition
the 3D-HST catalogues also include stellar mass estimates based on SED
fitting under the assumption of a Kroupa (2001) IMF (see Skelton et
al.~2014). Unfortunately far-IR photometry and hence dust mass
estimates do not currently exist for 3D-HST but are in progress as
part of the {\it Herschel Extra-galactic Legacy Project} (HELP; Hurley
et al.~2016).  Star-formation rates are estimated via the FAST code of
Kriek et al.~(2009) as described in Whitaker et al.~(2014). These are
based on a Chabrier IMF and include consideration of both the UV and
mid-IR flux (24 $\mu$m). To be fully consistent with the GAMA and
G10-COSMOS datasets we download the panchromatic photometry provided
by the 3D-HST team for each field, reformatted and once again apply
MAGPHYS to re-determine stellar masses and star-formation rates {{{
      in a manner consistent with our GAMA and G10-COSMOS derived
      values. }}}

\begin{figure}
\begin{center}
\includegraphics[width=\columnwidth]{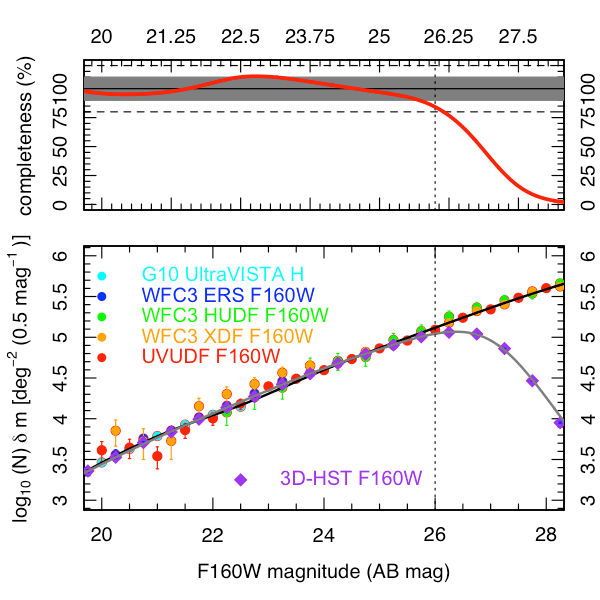}
\caption{(main panel) Literature galaxy number-counts from HST in the
  F160W band (Driver e tal.~2016) compared to those from the 3D-HST
  dataset. (top panel) the deviation as a percentage between a spline
  fit to the literature values and the 3D-HST data. We adopt a flux
  limit of F160W=26.0mag which equates to an 20 per cent
  incompleteness level (i.e., comparable to cosmic variance
  uncertainties).
\label{fig:counts}}
\end{center}
\end{figure}

Fig.~\ref{fig:coverage} shows the location of the five 3D-HST fields
on the sky (see also Table.~1 of Momcheva et al.~2016 and their
figures 1 and 2). Note that the 3D-HST data is of variable depth with
some sub-regions deeper than others. To explore the impact of this
``ragged edge'' we compare the 3D-HST galaxy number-counts to
literature values assembled by Driver et al.~(2016) in the F160W
band. Fig.~\ref{fig:counts} shows this comparison which agree well
with the 3D-HST data deviating only at very faint magnitudes. The
upper panel shows the deviation as a percentage.  We see that 3D-HST
appears to be 90 percent complete at F160W=25.0 mag (in line with the
conclusions of Skelton et al.~2014 and Bourne et al.~2016), reducing to
85 per cent completeness at F160W=26.0 mag. Here we do not adopt a
specific flux limit, but note that our sample is effectively limited
to $F160W \approx 26.0$ mag.

\subsection{AGN contamination}
AGN contamination of all three samples could result in erroneously
high stellar masses, and star-formation rates for a small number of
interlopers. The impact on dust masses is less obvious as the dust
is fairly impervious to the heating mechanism. For our star-formation
and stellar mass census it is therefore important to clean our catalogues of
AGN. First we remove significant outliers in stellar mass, i.e., all
systems with masses greater than $10^{12}$ M$_{\odot}$, this equates
to 32, 2, and 66 objects in GAMA, G10-COSMOS and 3D-HST
respectively. For each catalogue we then adopt the following strategy
to remove AGN contaminants:

~

\noindent
{\bf GAMA:} No AGN removal is attempted, beyond the mass cut mentioned
above, as the density at $z \leq 0.5$ is extremely low and any AGN
component likely to be sub-dominant.

~

\noindent
{\bf G10-COSMOS:} We implement an AGN selection using the criteria described
in Donley et al. (2012, see eqns 1 \& 2) using near and mid-IR
selection. In addition we reject radio-loud sources as identified using
the criteria from Seymour et al. (2008, see fig. 1) using cuts of
$\log_{10}(S_{1.4GHz}/S_{K_s}) > 1.5$ and $\log_{10}(S_{24\mu
  m}/S_{1.4GHz}) < 0.0$. Finally we reject any object with recorded
flux in any of the 3 XMM bands provided in the Laigle et al.~(2016)
catalogue.  The 1.4GHz fluxes were obtained from the VLA-COSMOS survey
(Schinnerer et al.~2007; Bondi et al.~2008). Together these three cuts
should identify naked, obscured, and radio-loud AGN yielding a
superset of 849 AGN which we now remove from our catalogue.

~

\noindent
{\bf 3D-HST:} We downloaded the on-line 3D-HST panchromatic photometry
and once again applied the Donley et al. cut resulting in the removal
of 7,403 AGN. This we recognise is a conservative cut so we also
expand the Donley criteria adding either a 0.5 mag boundary or a 1.0
mag boundary around the Donley criteria resulting in a selection of
13,896, or 33,730 AGN. Later we will use the three AGN cuts (lenient,
fair, extreme) to include an error estimate due to the uncertainty in
AGN removal.

\subsection{N(z) distributions}
\begin{figure}
\begin{center}
\includegraphics[width=\columnwidth]{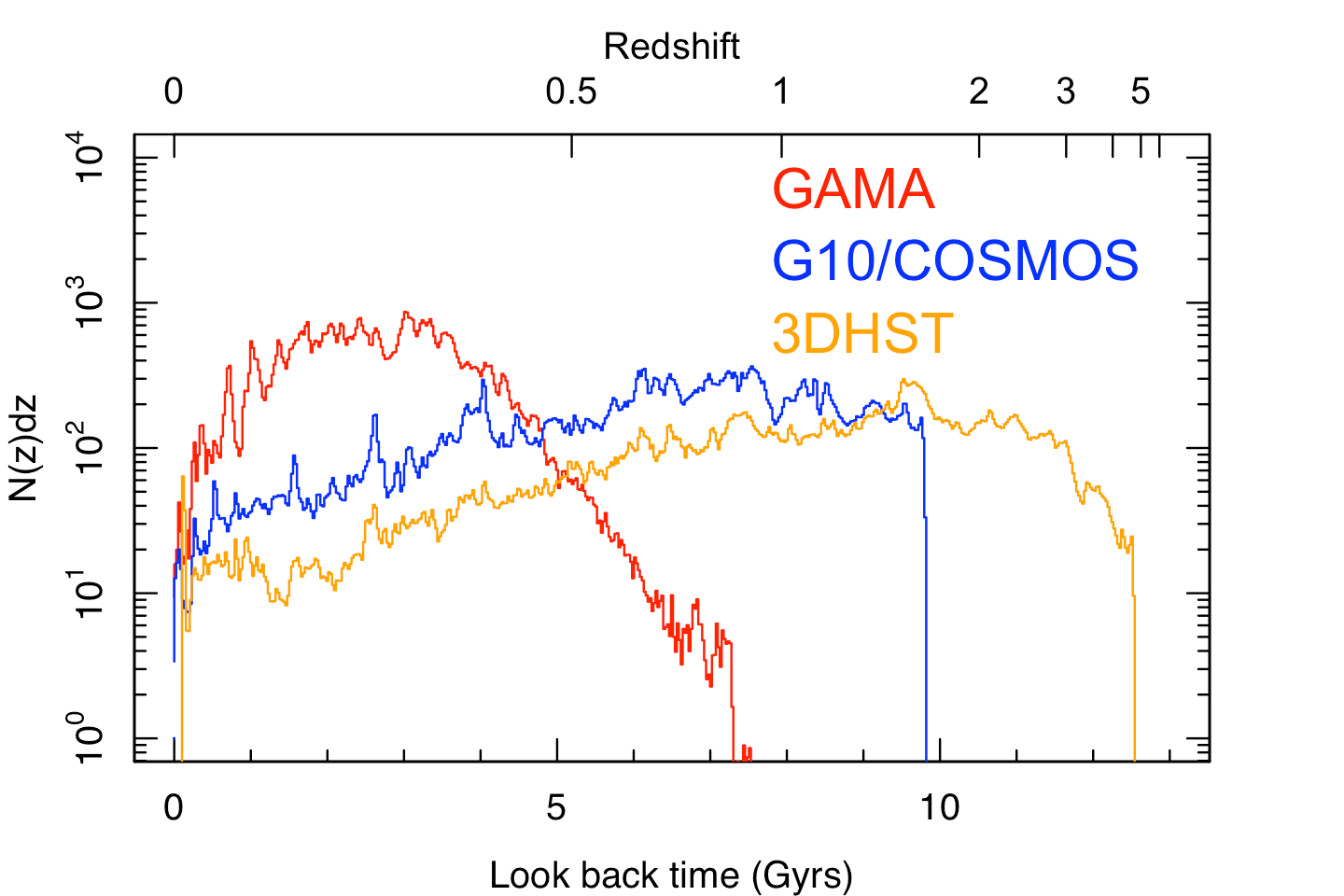}
\caption{The observed redshift distributions of the final selected
  GAMA, G10-COSMOS and 3D-HST datasets (as indicated).
\label{fig:nz}}
\end{center}
\end{figure}

Fig.~\ref{fig:nz} shows the final galaxy number-density for each of
our three catalogues versus lookback time, and includes a combined total of
582,314 galaxies extending over the range 0 to 12 billion years in
lookback-time ($0 < z < 5$). Each dataset, after trimming and AGN
removal, contains approximately 125k galaxies, with GAMA dominating at
very low redshifts, G10-COSMOS at intermediate redshifts, and 3D-HST
at high redshift. The GAMA data dominates out to $z=0.5$, G10-COSMOS
to $z=1.75$, and 3D-HST to $z=5$. It is worth bearing in mind that the
three samples are selected in distinct bands, $r$, $i$ and F160W for
GAMA, G10-COSMOS and 3D-HST respectively and that the distinctive
4000\AA-break passes through these bands at $z \approx 0.5, 1.0,$ and
$3.0$ and one should expect more severe selection biases to start to
occur beyond these limits (see Table~\ref{tab:info}).

\begin{table*}
\caption{Summary information for our three catalogues \label{tab:info}}
\begin{tabular}{cccccc} \hline
Dataset & Selection & Number & Area & Ref\\ \hline 
GAMA & $r < 19.8$ \& $nQ > 2 $ & 128,568 & 117.2 sq deg & Liske et al.~(2015) \\
G10-COSMOS & $i \leq 25.0$ \& $z_{\rm spec~or~photo} < 1.75$ & 142,260 & 1.022 sq deg & Andrews et al.~(2017a) \\
3D-HST & $F160W \leq 26.0$ \& $z_{\rm spec~or~photo} < 5.0$ & 194,728(l); 188,235(f), 168,401(e) & 0.274 sq deg & Momcheva et al.(~2016) \\ \hline
\end{tabular}
\end{table*}

\section{MAGPHYS Analysis}
Here we describe the MAGPHYS fitting process from which we obtain
stellar and dust mass estimates and dust corrected star-formation
rates. MAGPHYS is an SED fitting code (da Cunha, Charlot \& Elbaz
2008) which uses an extensive stellar library based on Bruzual \&
Charlot (2003) synthetic spectra. Here we elect to use the BC03
libraries rather than the more recent CB07 (Charlot \& Bruzual 2007)
which arguably over-predicts the Thermally Pulsing-Asymptotic Giant
Branch (TP-AGB) phase. The library samples spectra with single
(exponentially decaying) star-formation histories over a range of
e-folding timescales ($10^5 - 2 \times 10^{10}$ yr), and over a broad
range of metallicities. Starlight is assumed to be attenuated by both
spherically symmetric birth clouds, as well as the ISM using the
Charlot \& Fall (2000) prescription. The energy lost to dust
attenuation is then projected into the mid- and far-infrared assuming
four key dust components: PAH and associated continuum, and hot, warm
and cold dust components. Sets of optical and far-IR spectra, where the
energy lost in the optical equates to the energy radiated in the
far-IR, are then regressed against the flux measurements and errors to
determine a best fit SED and to determine optimal parameters and
probability density functions for the parameters in question. While
MAGPHYS produces a wide range of measurements here we focus only on
the stellar mass, star-formation rate and dust masses which are
considered robust (Hayward \& Smith 2015).

The explicit version of the MAGPHYS code that we implement here, has
also been adapted by us as follows:

\noindent
(i) the code has been modified to derive fluxes based on photon energy
rather than photon number in the far-IR,

\noindent
(ii) the latest PACS and SPIRE filter curves are used (in particular
the PACS filter curves have changed significantly as the instrument
characteristics have become better defined),

\noindent
(iii) the code has been modified to use upper-limits by identifying
zero flux as a limit and using the error as the upper-bound,

\noindent
(iv) we have extended the upper limit for the output dust mass
probability density distribution, from $10^{9}$ to
$10^{12}$M$_{\odot}$ as a small number of systems were hitting the
$10^9$M$_{\odot}$ upper buffer.

\begin{figure*}
\begin{center}

\vspace{-1.5cm}

\includegraphics[width=\textwidth]{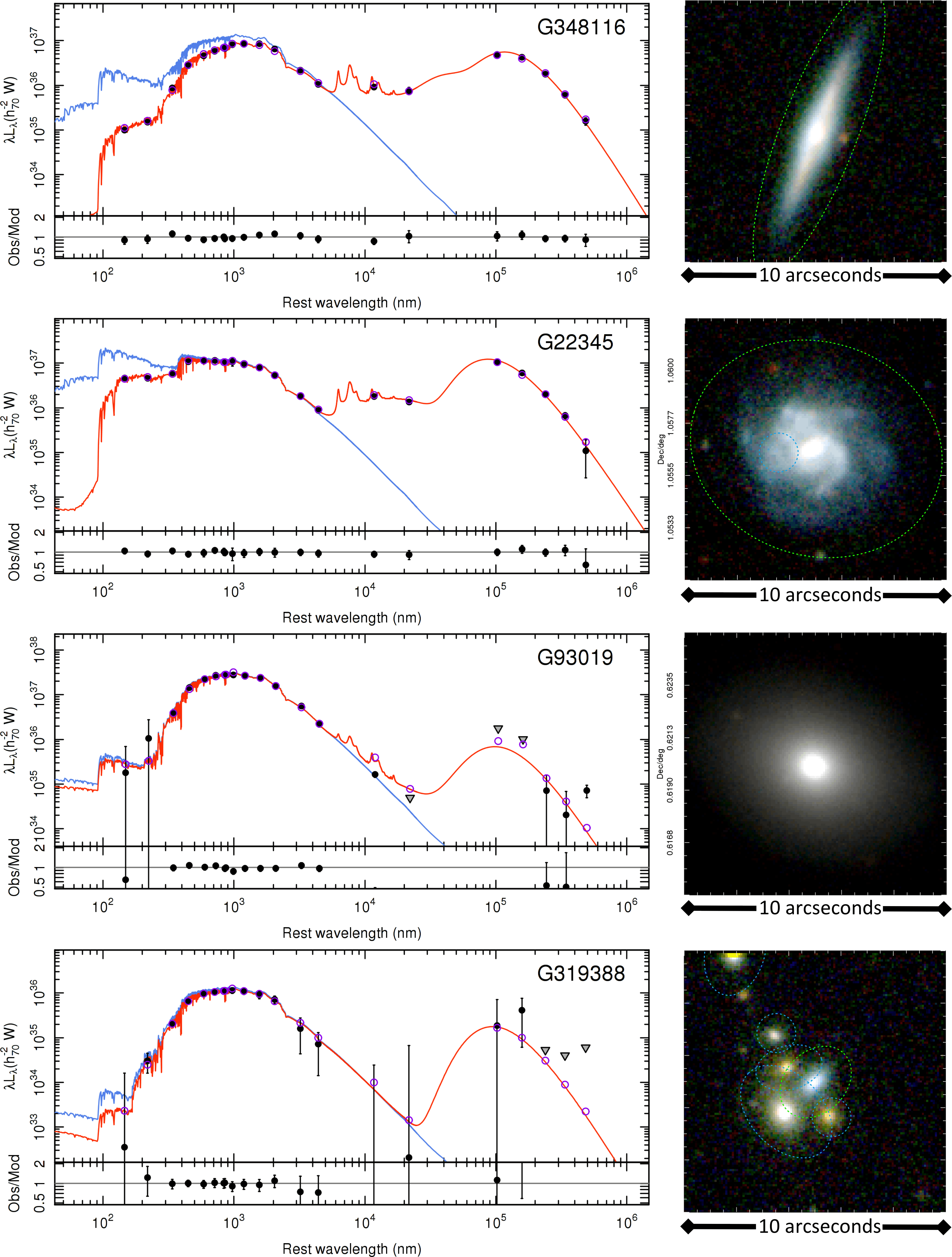}

\caption{Four examples of GAMA galaxies at $z \approx 0.1$ processed
  with MAGPHYS. The left panels show the spectral energy distributions
  showing the data points (black circles), limits (triangles) and the
  dust attenuated (red curve) and dust unattenuated (blue) MAGPHYS
  fits, with residual values shown at the bottom. The right side
  panels show a $KZr$ images from VISTA/VIKING and SDSS, the green
  dotted ellipses denote the apertures used by
  LAMBDAR. \label{fig:examples}}
\end{center}
\end{figure*}

\begin{figure*}
\begin{center}

\includegraphics[width=\textwidth]{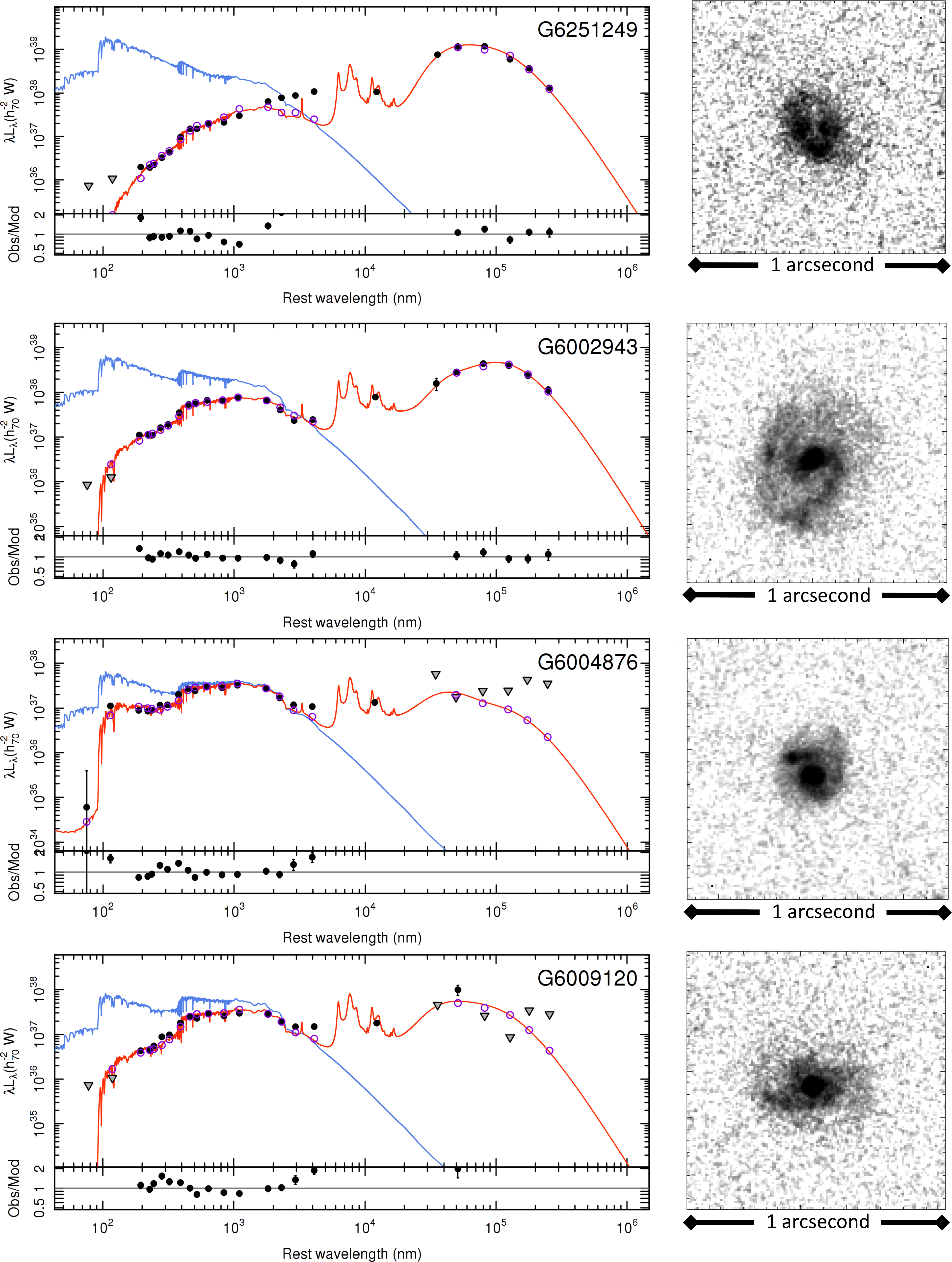}

\caption{Four examples of G10-COSMOS galaxies at $z \approx 1$
  processed with MAGPHYS. The left panels show the spectral energy
  distributions showing the data points (black circles), limits
  (triangles) and the dust attenuated (red curve) and dust
  unattenuated (blue) MAGPHYS fits, with residual values shown at the
  bottom. The right side panels show the HST F814W image ($2' \times
  2'$). Note the top panel shows a system with strong AGN
  contamination. \label{fig:examples2}}
\end{center}
\end{figure*}

\subsection{Data preparation}
The GAMA, G10-COSMOS and 3D-HST datasets are as described in Sections
2.1, 2.2 and 2.3 respectively. Critically the GAMA and G10-COSMOS
panchromatic catalogues are based on LAMBDAR analysis (Wright et
al.~2016) which produces either flux measurements with errors,
upper-limits, or provides a flag (-999) for objects where there is no
imaging coverage in that filter. The 3D-HST data is obtained from the
public download site (See
({\sc http://3dhst.research.yale/edu/Data.php} and associated
documentation). The MAGPHYS code is capable of managing three types of
data: MEASUREMENTS: positive flux and positive flux error; LIMITS:
zero flux and positive flux-error; NODATA: negative flux values.

{{{ For GAMA: Our earlier LAMBDAR analysis provides appropriate
      values by default for all 128,568 galaxies in the common
      coverage region, i.e., every galaxy contains flux measurements
      in all far-IR bands using the r-band optically defined aperture
      convolved with the appropriate instrument PSF. See Wright et
      al.~(2016) for full details of these measurements.}}}

{{{ For G10-COSMOS: In the Andrews et al. LAMBDAR analysis of
  G10-COSMOS data we adopted a cascading selection in the far-IR to
  manage the extreme mismatch in depth between the optical selection
  band and the far-IR Spitzer and Herschel data. In the case of
  objects with non-measurable fluxes in Spitzer 24 $\mu$m, these were
  not propagated for measurement at longer-wavelengths. This process
  was replicated as the analysis progressed to longer wavelengths. For
  objects excluded via this process, and for which LAMBDAR
  measurements were therefore not made, we set the flux limits to zero
  and adopt a flux-error equivalent to the quoted $1\sigma$
  point-source detection limit appropriate for each band (see Andrews
  et al.~2017a, figure~2).

Following the initial analysis a number of systems were identified
with predicted fluxes above the detection threshold.  This then led to
a modified selection and additional flux measurements which expanded
our far-IR measurements from $\sim$12k systems to $\sim$24k systems
and is outlined in Appendix~A. This revised LAMBDAR catalogue was then
prepared and passed through MAGPHYS to generate our final G10-COSMOS
MAGPHYS catalogue (again see Appendix~A for further details).}}}

For 3D-HST: We extracted the following filter combinations from the
panchromatic catalogues provided online by the 3D-HST team:

~

\noindent
AEGIS: $u,g,F606W,r,i,F814W,z,F125W,j1,j2,j3,j,$
$F140W,h1,h2,h,F160W,k,ks,irac1,irac2,irac3,irac4$

~

\noindent
COSMOS: $u,b,g,v,F606W,r,rp,i,ip,F814W,z,zp,Y_{VISTA}$
$F125W,j1,j2,j3,j,J_{VISTA},F140W,h1,h2,H,H_{VISTA}$
$F160W,k,ks,Ks_{VISTA},irac1,irac2,irac3,irac4$

~

\noindent
GOODS-N: $u,F435W,B,V,F606W,r,F775W,z,F850LP$
$F125W,j,F140W,h,F160W,ks,irac1,irac2,irac3,irac4$

~

\noindent
GOODS-S: $u38,u,F435W,b,v,F606WC,F606W,r,rc,F775W$
$i,F814W,F850LP,F850LPc,F125W,j,j_{tenis},F140W,h$
$F160W,k_{tenis},ks,irac1,irac2,irac3,irac4$

~

\noindent
UDS: $u,B,V,F606W,r,i,F814W,z,F125W,j,F140W,h$
$F160W,ks,irac1,irac2,irac3,irac4$

~

{{{ For the 3D-HST data no limits are used, i.e., all measurements
      either have an appropriate measurements or no data recorded. and
      grism or photometric redshifts for all objects. }}}

\subsection{Processing 600,000 files using the MAGNUS Supercomputer}
To optimise the processing of multiple runs of $\sim 600,000$ independent
galaxies we developed a Python script which sorted the galaxies by
redshift (rounded to four decimal places) and batch ran galaxies with
redshifts within $\pm 0.0001$ intervals using the pre-prepared MAGPHYS
libraries. This essentially provided a speed-up factor of $\times 10$
over regenerating redshifted SED libraries for each individual galaxy.

The MAGPHYS SED fitting was run on the {\sc Magnus} machine at the
Pawsey Supercomputering Centre.  {\sc Magnus} is a {\sc Cray XC40}
Series Supercomputer made up of 1,488 compute nodes.  Each node
contains twin Intel Xeon E5-2690V3 Haswell processors (12-core, 2.6
GHz), and has 64GB of DDR4 RAM.  Jobs are then submitted using the
SLURM (Yoo et al.~2003) job scheduler. In effect {\sc Magnus} is
therefore running MAGPHYS independently across 35,712 processors. With
this capacity we are able to process all 600,000 systems within a 24
hr period. In total the MAGPHYS runs were performed approximately six
times for each dataset, as improvements were made in the photometry
during LAMBDAR development and updates to the
MAGPHYS code (as described) or the FILTERBIN.RES file.

For 3D-HST we run both the {{{ standard MAGPHYS template library,
      and the high-z MAGPHYS template library on all data and use the
      $\chi^2$ values returned by MAGPHYS to select whether to adopt
      the standard or high-z results.  In 97 per cent of cases the
      optimal fit is selected from the standard-MAGPHYS output rather
      than the high-z template set.}}}

{{{ The MAGPHYS process, as described above, provides
      star-formation, stellar mass, and dust mass estimates for every
      galaxy within our optically flux selected samples. For GAMA,
      G10-COSMOS and 3D-HST these selection limits are: $r<19.8$ mag,
      $i \leq 25$ mag and $F814W \leq 26.0$ mag respectively (see
      Section~2), and these are the only relevant selection limits. In
      all other bands measurements have been made, using the optically
      defined apertures, except for G10-COSMOS-only where far-IR
      measurements are made for the 24k objects, with the brightest
      predicted 250$\mu$m flux, and upper limits assigned to the
      remainder. For those G10-COSMOS systems with assigned far-IR
      upper-limits the dust mass estimates essentially revert to an
      estimated dust mass based on the Charlot \& Fall (2000)
      prescription.}}}

\subsection{Diagnostics and verification}
Fig.~\ref{fig:examples} shows four examples of the GAMA MAGPHYS
outputs. These examples are relatively bright galaxies extracted from
the GAMA sample and have been selected to illustrate an edge-on
spiral, a face-on spiral, an elliptical, and a crowded field
system. In all cases the MAGPHYS fits, indicated by the unattenuated
(blue) and attenuated (red) lines, are reasonable, and the residuals
are small, indicating plausible fits. As expected the two spirals have
far-IR peaks which are as prominent as their optical peaks, whereas
the elliptical galaxy shows a more suppressed far-IR peak ---
presumably due to a paucity of dust. As a consequence the attenuated
and unattenuated curves are very similar for the elliptical galaxy ---
what you see is what you get --- whereas for the spirals the actual
energy production is significantly higher than the optical light would
indicate, i.e., spiral galaxies are heavily obscured. The edge-on
spiral is significantly more attenuated than the face-on spiral, again
as one would expect, and highlights the inclination dependence of
dust attenuation (Driver et al.~2007). The lower panel, shows an
object in a crowded region, and indicates how the LAMBDAR photometric
errors inflate where flux has been divided, particularly for those
datasets with poorer spatial resolution (i.e. GALEX, WISE and in
particular Herschel). The grey inverted triangles indicate bands where
the flux measurement is found to be less than the flux error, and
hence the flux error becomes the upper limit. Equivalent panels for
all GAMA galaxies are available from the GAMA database.

Fig.~\ref{fig:examples2} shows an equivalent set of four galaxies
drawn from the G10-COSMOS sample at $z \approx 1$. Again we have
selected four galaxies. The first illustrates likely AGN contamination
and excessive far-IR emission, the second a face-on spiral but one
which is clearly dustier than the low-z counterpart, and two examples
of galaxies where upper limits are in play.

In addition to the systems shown in Figs.~\ref{fig:examples} \&
\ref{fig:examples2} individual inspections were made of several
hundred objects drawn randomly from each dataset. In the vast majority
of cases ($>99$ per cent) the MAGPHYS outputs appear appropriate and
the attenuated data accurately describe the measured flux values.

\begin{figure*}
\begin{center}

\includegraphics[width=\textwidth]{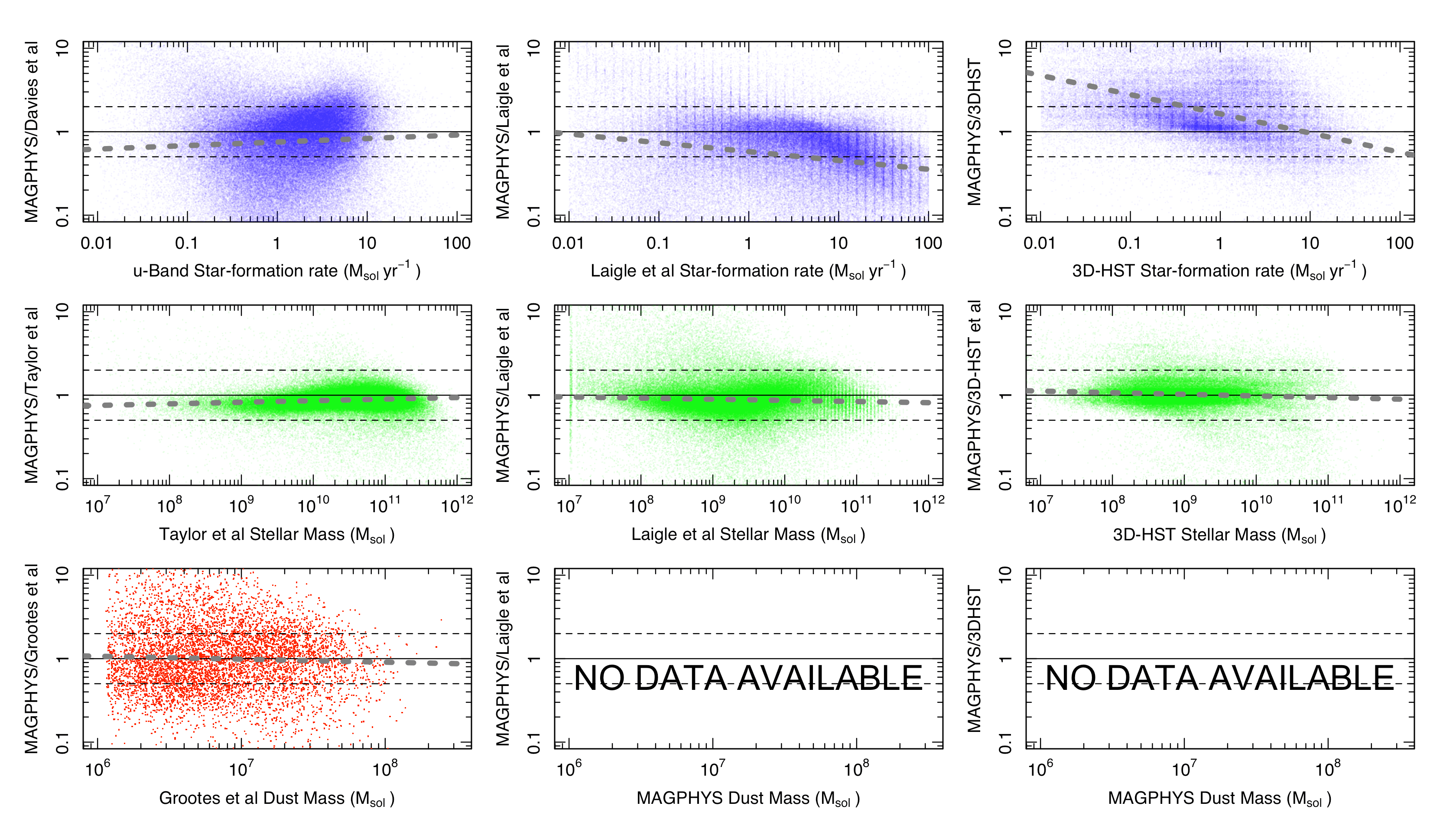}

\caption{ ({\it upper panels}) Comparisons between the star-formation
  rates estimated from the literature to our MAGPHYS measurements for
  GAMA {\it left}), G10-COSMOS ({\it middle panels}), and 3D-HST ({\it
    right}).  ({\it middle panels}) Comparisons between stellar mass
  estimates from the literature to our MAGPHYS measurements for GAMA
  and ({\it lower panels}) comparisons between dust mass estimates from Grootes et al.~(2013) to our MAGPHYS GAMA measurements. In each panel we show a $times 2$ deviation as dotted lines and the fitted linear trend. 
\label{fig:compare}}
\end{center}
\end{figure*}

\subsection{Cross-checking measurements}
Demonstrating the veracity of the full sample is non-trivial, however,
we can compare the MAGPHYS derived stellar and dust masses, and 
dust corrected star-formation rates, to those derived via other
methods/groups. In particular all three samples, GAMA, G10-COSMOS and
3D-HST, have published stellar mass estimates and star-formation
estimates (see Table.~\ref{tab:reference}). Fig.~\ref{fig:compare}
shows a comparison of {{{ the MAGPHYS measurements derived as described in section 3, }}} to the
available star-formation rates (upper), stellar-masses (middle), and
dust masses (lower), for the GAMA (left), G10-COSMOS (centre), and
3D-HST (right) samples.

\begin{table*}
\caption{Literature references against which we compare our stellar masses and cosmic star-formation rates \label{tab:reference}}
\begin{tabular}{cccc} \hline
Dataset & Stellar Masses & Dust Masses & Star-formation rates \\ \hline
GAMA & Taylor et al.~(2011) & Grootes et al.~(2013) & Davies et al.~(2016) \\ 
G10-COSMOS & Laigle et al.~(2016) & N/A & Laigle et al.~(2016) \\
3D-HST & Skeleton et al.~(2013) & N/A & Momcheva et al.~(2016) \\ \hline
\end{tabular}
\end{table*}

On Fig.~\ref{fig:compare} the parity line is indicated in solid black
and variations of $\times 2$ by the dotted tram-lines. The thicker
dashed line shows a robust linear fit to the data.  In the upper
panels of Fig.~\ref{fig:compare} we compare the star-formation
estimates. Note that these are derived in distinct ways. The
star-formation rates for our GAMA and G10-COSMOS samples are based on
UV-far-IR SED template fitting (da Cunha, Charlot \& Elbaz 2008) and
hence include an individual dust correction for each galaxy. The
estimate of Davies et al.~(2016) for GAMA relies on the calibration of
$u$-band fluxes to the late-type disk sample of Grootes et al.~(2013)
which used full radiative transfer modelling. The G10-COSMOS values
are taken from the catalogue provided by Laigle et al.(2016). The
3D-HST values are from Whitaker et al.~(2014) via FAST (Kriek et al.,
2009) fitting.

A fairly significant trend is seen between our MAGPHYS star-formation
measurements for 3D-HST compared to their published values. The trend
is in the sense that MAGPHYS star-formation rates are higher than
3D-HST at lower star-formation rates. There is also a cloud of
outliers which may arise from some inconsistency in the photometry
across the bands. For example, and particularly in the COSMOS region,
we see some inconsistencies between the CFHT and Subaru
photometry. This argues for the need at some point to revisit the
3D-HST photometry using a LAMBDAR-like method to homogenise aperture
measurements across all the bands. At this point we elect to move
forward with the MAGPHYS star-formation measurements for all three
datasets to ensure that our measurements are based on a consistent
methodology, IMF, and dust assumptions.

In the middle panels of Fig.~\ref{fig:compare} the stellar mass
estimates show reasonably good agreement across all three datasets,
mild trends are seen but clearly the bulk of the population have
stellar mass estimates well within the dotted lines. This suggests a
high level of consistency across the three datasets. 

On Fig.~\ref{fig:compare} (lower left) we compare the MAGPHYS derived
dust masses to those derived by Grootes et al.~(2016) using a full
radiative transfer treatment. This is a sample of 6356 late-type
spiral field galaxies, introduced in Grootes et al.~(2013), where the
$\tau$ opacity values were derived. Note that in order to make a valid
comparison we correct the Grootes et al. data from an emissivity based
on Weingartner \& Draine (2001) to that adopted by MAGPHYS, requiring
an upward modification of the Grootes et al. dust masses by
40\%. While the comparison shows scatter the fitted robust linear fit
(thick grey dashed line) shows extremely good agreement of the mean
behaviour with no obvious bias with mass. We note that the Grootes et
al. data has an associated error of $\pm 0.2$~dex suggesting that the
majority of the error seen is coming from the MAGPHYS data with a
1$\sigma$ error of $\pm 0.3$~dex in the MAGPHYS dust mass estimates.
This is in good agreement with the findings of Beeston et al.~(2017)
who see a MAGPHYS error ranging from 0.09 dex to 0.5 dex depending on
whether the GAMA data contain measurements or upper-limits in the
far-IR bands.  Unfortunately no literature data currently exists for
the G10-COSMOS region but is work in progress by a number of teams.

\begin{figure}
\begin{center}

\includegraphics[width=\columnwidth]{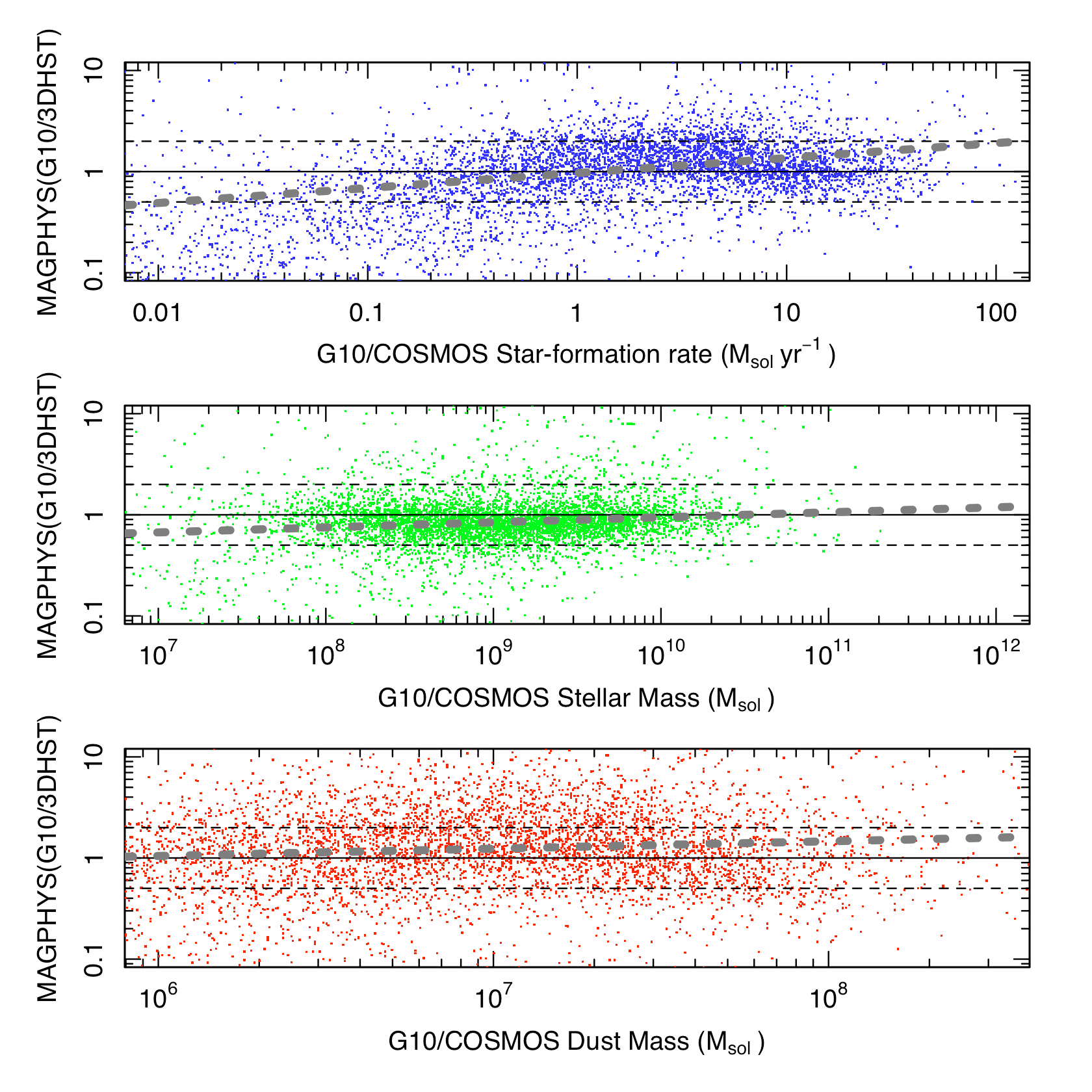}

\caption{ {{{ A comparison of star-formation rates, stellar mass
        measurements and dust mass measurements for $\sim 6000$ galaxies in
        common between our G10-COSMOS and 3D-HST samples.
\label{fig:compareinternal}}}} }
\end{center}
\end{figure}

\begin{figure}
\begin{center}

\includegraphics[width=\columnwidth]{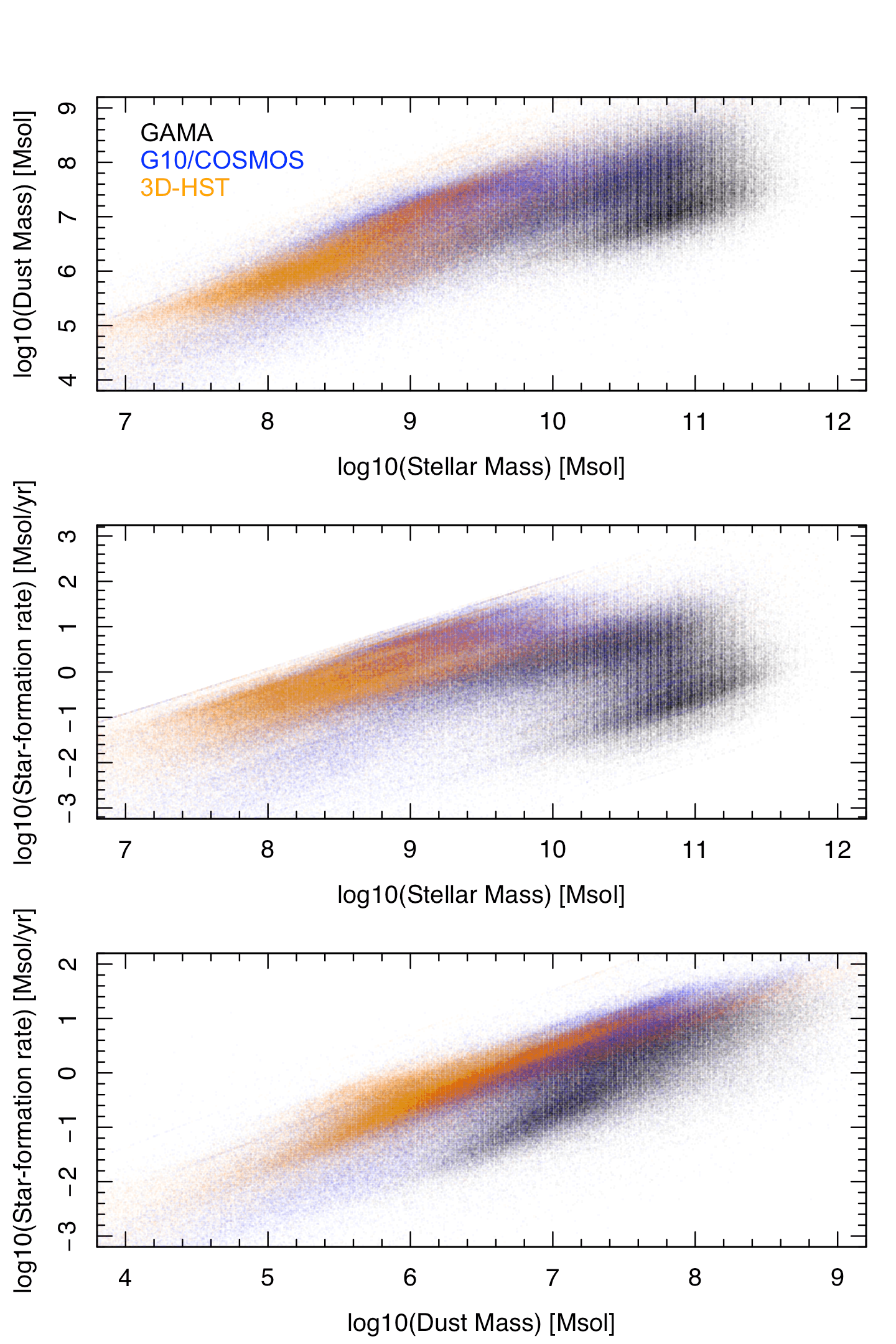}

\caption{ {{{ Panels showing the three 2D projections of the 3D
        cube defined by stellar mass, dust mass, and star-formation
        rate (our key derived quantities), for each of our three
        datasets (as indicated). Some striations, binning and
        boundaries are evident (but not considered problematic), as is
        a separable high stellar mass, inert, and low dust population
        at low redshift which we take to be the Elliptical,
        lenticular and early-type systems. Note that these panels
        shows the samples in their entirety which span the full
        redshift range from nearby to $z=5$.
\label{fig:scatter} }}} }
\end{center}
\end{figure}

{{{ We can also undertake an internal consistency check as one of
      the five 3D-HST fields lies within the G10-COSMOS region. Using
      a 0.5$''$ radial match we find 6198 objects from the 3D-HST
      sample which match our independent G10-COSMOS
      data. Fig.~\ref{fig:compareinternal} compares the derived
      star-formation rates, stellar mass measurements, and dust mass
      measurements. We find close agreement across all three values
      with the majority of data points well within the dashed buffer
      lines. Note that for the 3D-HST derived dust masses no actual
      far-IR information is included at all.

      Finally Fig.~\ref{fig:scatter} shows 2D projections of the
      3D-cube defined by our key derived quantities: stellar mass,
      dust mass, and star-formation rate. The entirety of the three
      datasets are shown which span the full redshift range. In
      general the three populations interleave and this is despite the
      lack of far-IR data constraining the dust masses for
      3D-HST. Most obvious is the separable high stellar mass
      intermediate to low dust mass and inert population in the GAMA
      sample only (i.e., at low redshift only). We take this
      population to correspond to elliptical systems known to be
      mostly devoid of dust with low star-formation rates.  In future
      papers we will explore various trends and scaling relations for
      the combined dataset as a function of redshift.

      Following the above we conclude that we now have consistent and
      reasonable stellar mass and star-formation rate estimators
      across the three catalogues extending from $z=0$ to $z=5$. }}}

\section{The cosmic star-formation history and the build up of stellar mass and dust mass}
\begin{figure*}
\begin{center}
\includegraphics[width=\textwidth]{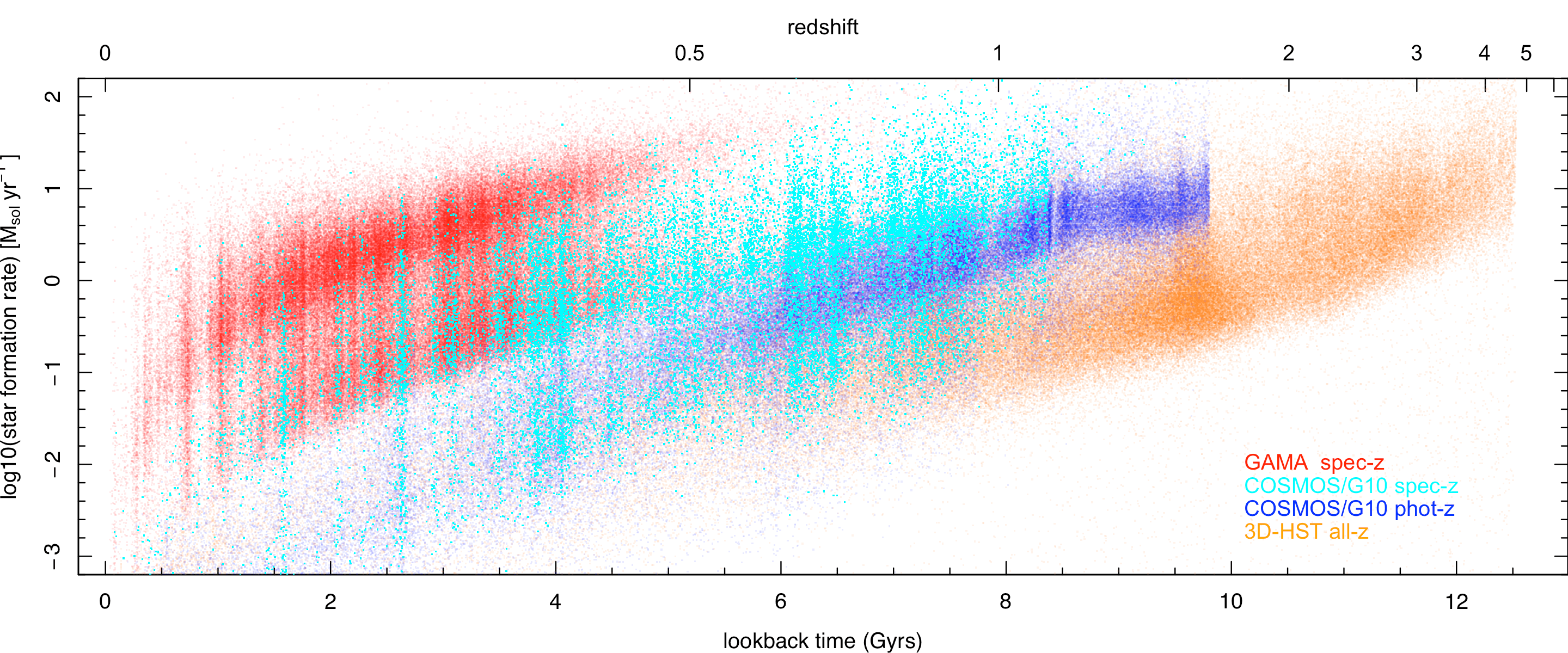}

\includegraphics[width=\textwidth]{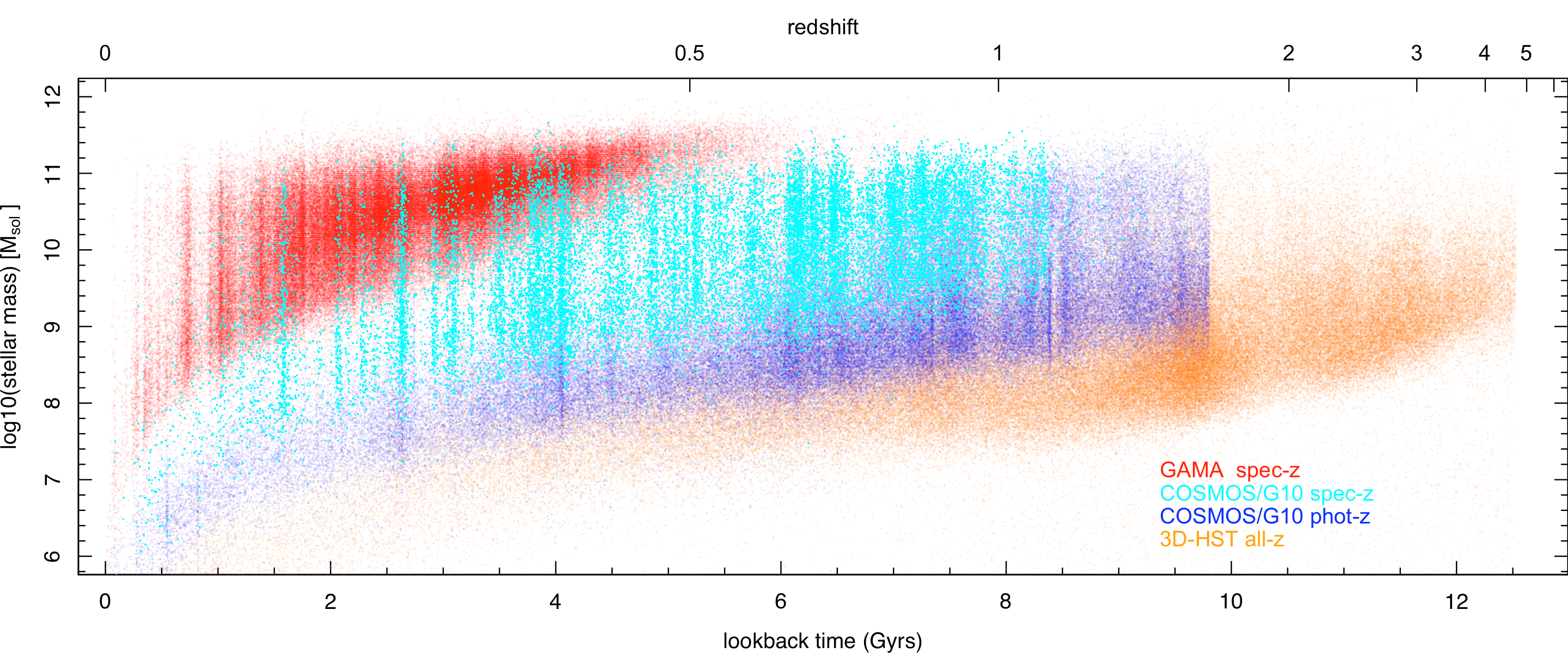}

\includegraphics[width=\textwidth]{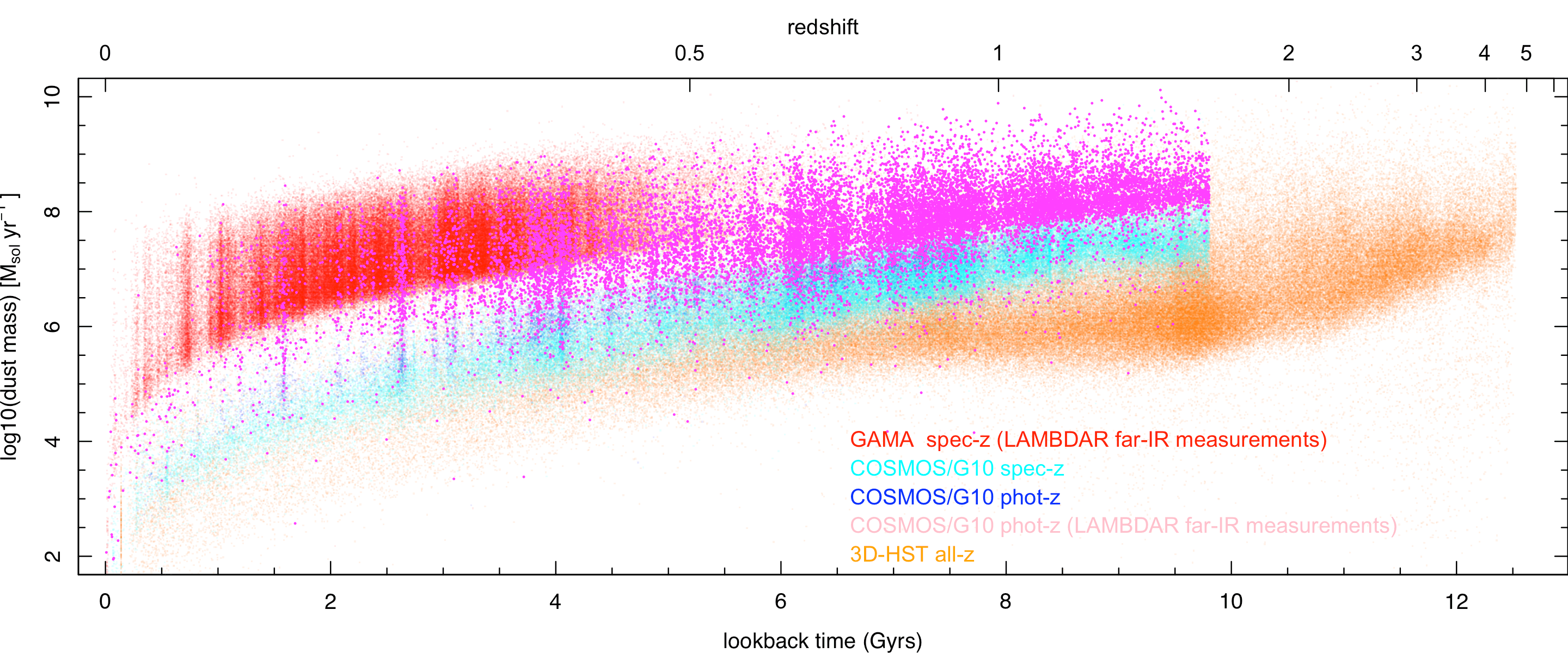}

\caption{({\it upper}) The star-formation v redshift distributions for
  the three samples.  ({\it middle}) The stellar mass redshift
  distribution of our three complementary samples. ({\it lower}) The
  dust mass redshift distribution for the GAMA and G10-COSMOS datasets.
\label{fig:zmass}}
\end{center}
\end{figure*}

Fig.~\ref{fig:zmass} shows the resulting distribution of
star-formation (upper), stellar-mass (middle), and dust-mass (lower)
measurements. Combined these data cover a significant portion of the
star-formation-redshift, stellar-mass-redshift, and dust-mass-redshift
planes with each dataset essentially dominating a distinct portion of
the parameter space. For GAMA all data have secure redshifts. For
G10-COSMOS we highlight those systems with redshifts in light blue and
those with photometric-redshifts in mauve (as indicated). The 3D-HST
data are shown as photometric redshifts throughout however it is worth
noting that both the G10-COSMOS and 3D-HST have quoted photometric
errors of $\Delta z << \pm 0.01$. 

\begin{figure}
\begin{center}

\includegraphics[width=\columnwidth]{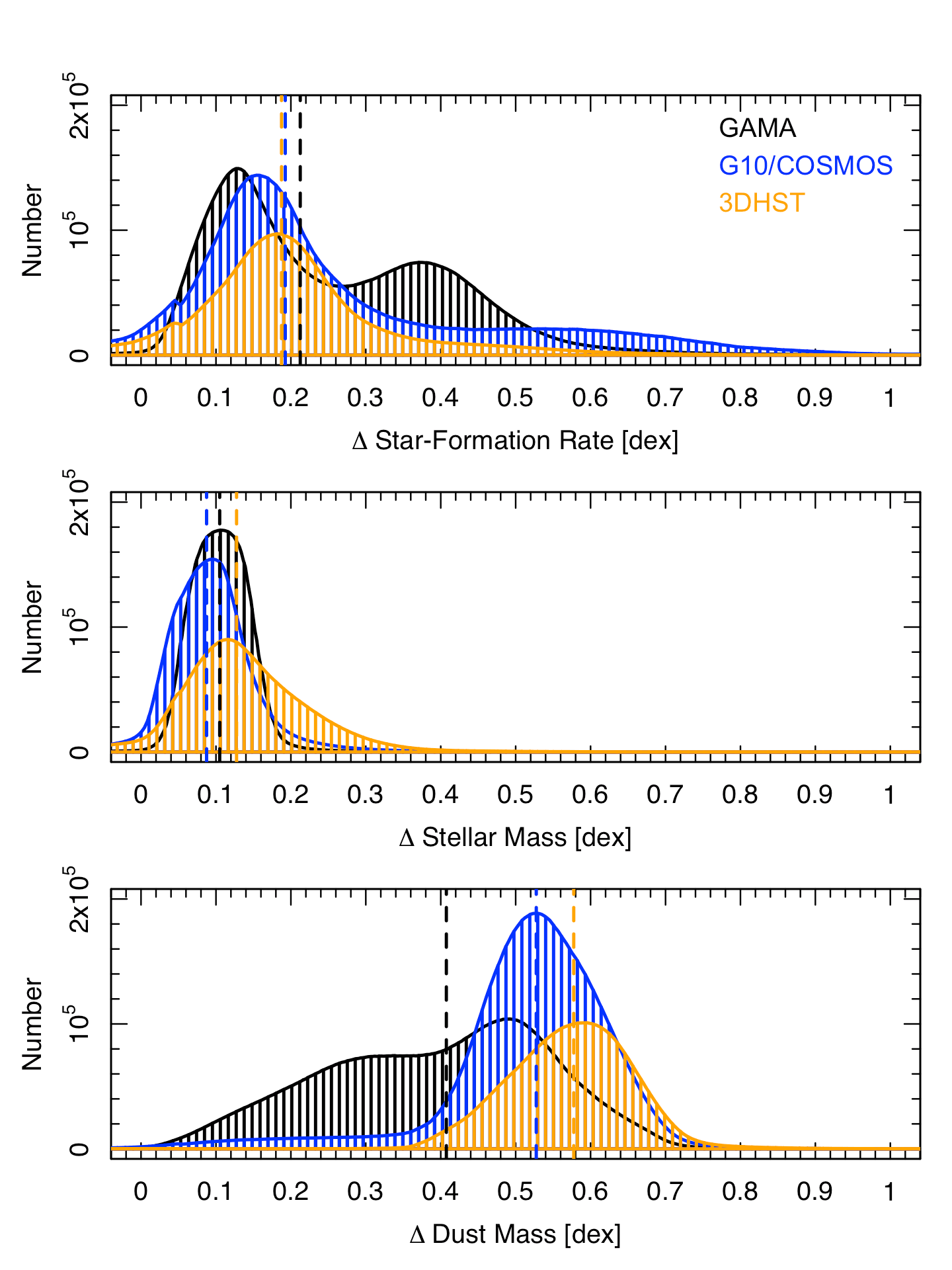}

\caption{The MAGPHYS errors on each of our three key measurements for
  our three samples. The MAGPHYS errors are determined from the 84 to
  16 percentile ranges. \label{fig:hists}}
\end{center}
\end{figure}

Of equal interest to the measurements themselves are the quoted error
values. Fig.~\ref{fig:hists} shows the histogram of errors for each
dataset and for each of our three key parameters. These errors are
directly extracted from the MAGPHYS output and show half the 84 to 16
percentile ranges. For the star-formation rate (top panel) we see a
fairly uniform median error of approximately 0.1---0.2 dex for all
three datasets. However the GAMA distribution is clearly bimodal which
is reflecting the inherent bimodality seen, and well known in the
low-z galaxy population, i.e., star-formation rates for early-types
have a much broader error range. This bimodality is not apparent in
either the G10-COSMOS and GAMA datasets where the incidence of truly
inert systems is significantly less, although both the 3D-HST and
G10-COSMOS data show long tails towards large error values. For the
stellar mass measurements we also see very consistent error
distributions of about 0.1 dex, this error range is narrow in keeping
with the trends seen in Fig.~\ref{fig:compare} (middle
panels). Finally the lower panel shows the dust mass measurements
errors which are significantly broader with median errors of 0.4 and
0.55 for GAMA, and G10-COSMOS. Note because of the lack of far-IR
measurements we do not attempt to use the 3D-HST dust mass estimates.
The GAMA distribution is again broad and bimodal reflecting those
systems for which we have good high signal-to-noise measurements in
the far-IR and those for which we have upper-limits only (i.e.,
$F_{\rm IR} < \Delta F_{\rm IR}$). Recently Beeston et al.~(2017)
explicitly explored the impact on the dust mass estimate as one
reduces from three to zero far-IR filters (see Beeston et al.~2017,
their figures 1 \& 2), finding that while the dust mass error
increases (from 0.09 dex to 0.5 dex), there is no obvious systematic
bias. The range of errors found by Beeston matches well the range of
recovered measurement errors shown in Fig.~\ref{fig:hists}). Similarly
the G10-COSMOS data, for which far-IR measurements exist for a small
fraction (10-20 percentile), also has an error that is consistent with the
findings of Beeston et al.

The range and spread of these errors shown in Fig.~\ref{fig:hists},
have two important implications. One is that the data will be prone to
Eddington bias, particularly for the dust measurements, because the
errors are comparable or larger to our adopted bin sizes in the
upcoming analysis presented in section 4.1 (0.5 dex for stellar and
dust masses). Secondly, a full Monte-Carlo analysis will be necessary
because of the spread in errors, i.e., adopting a single error for
each parameter, for each datasets, would not be appropriate.

\begin{figure*}
\begin{center}
\includegraphics[width=0.9\textwidth]{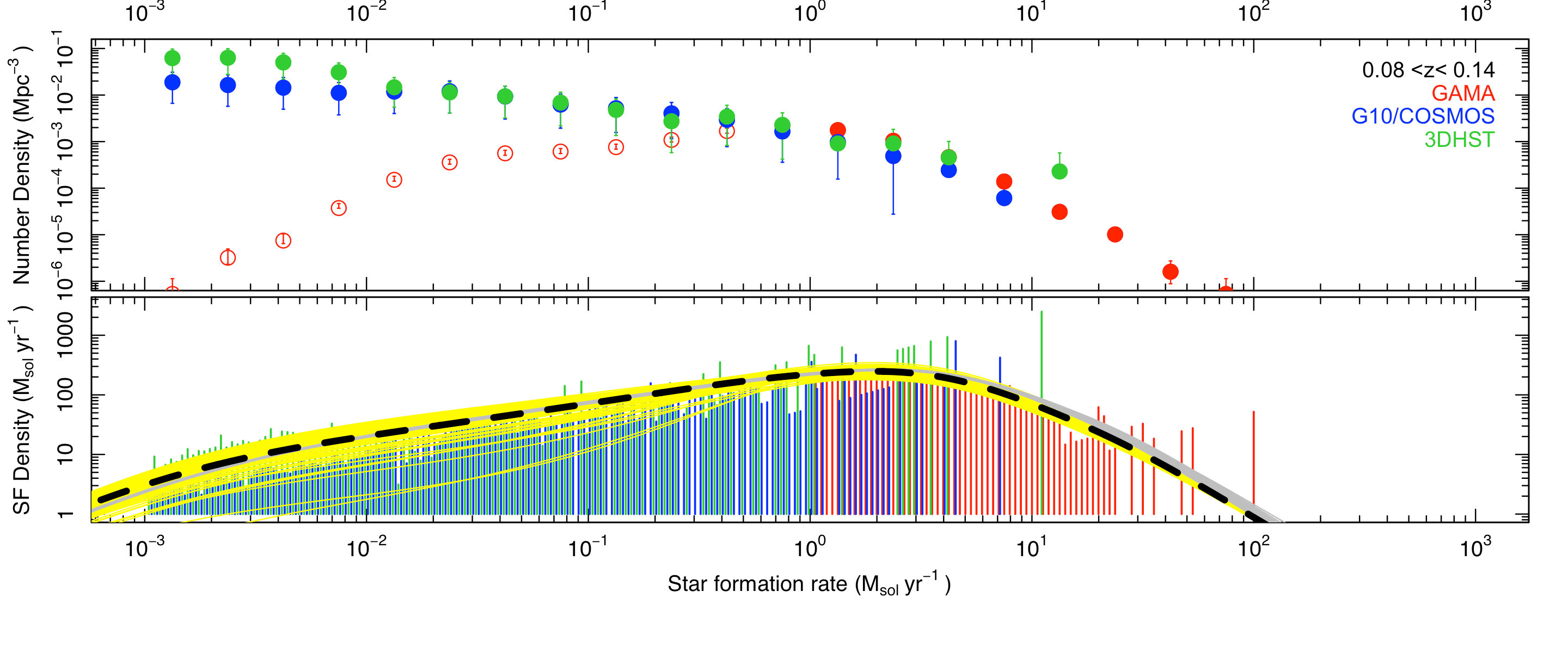}

\vspace{-0.2cm}

\includegraphics[width=0.9\textwidth]{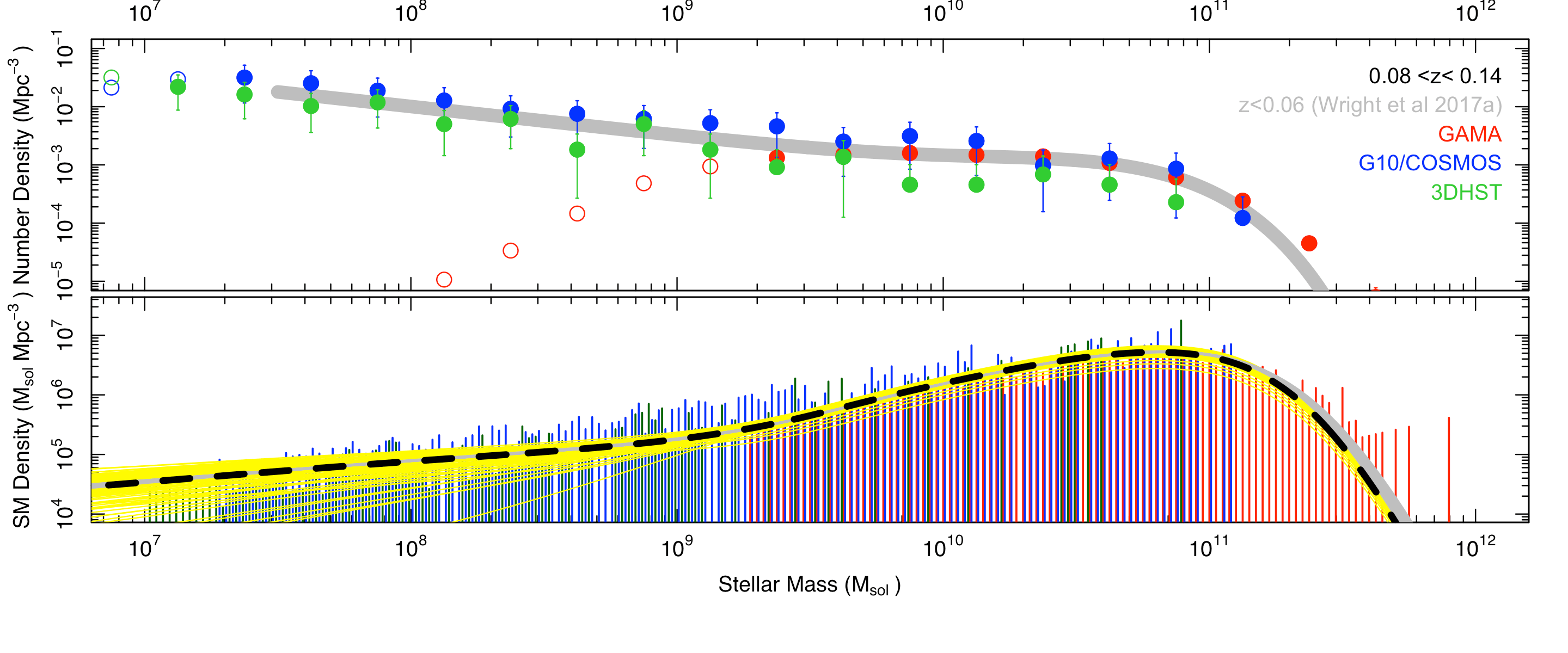}

\vspace{-0.2cm}

\includegraphics[width=0.9\textwidth]{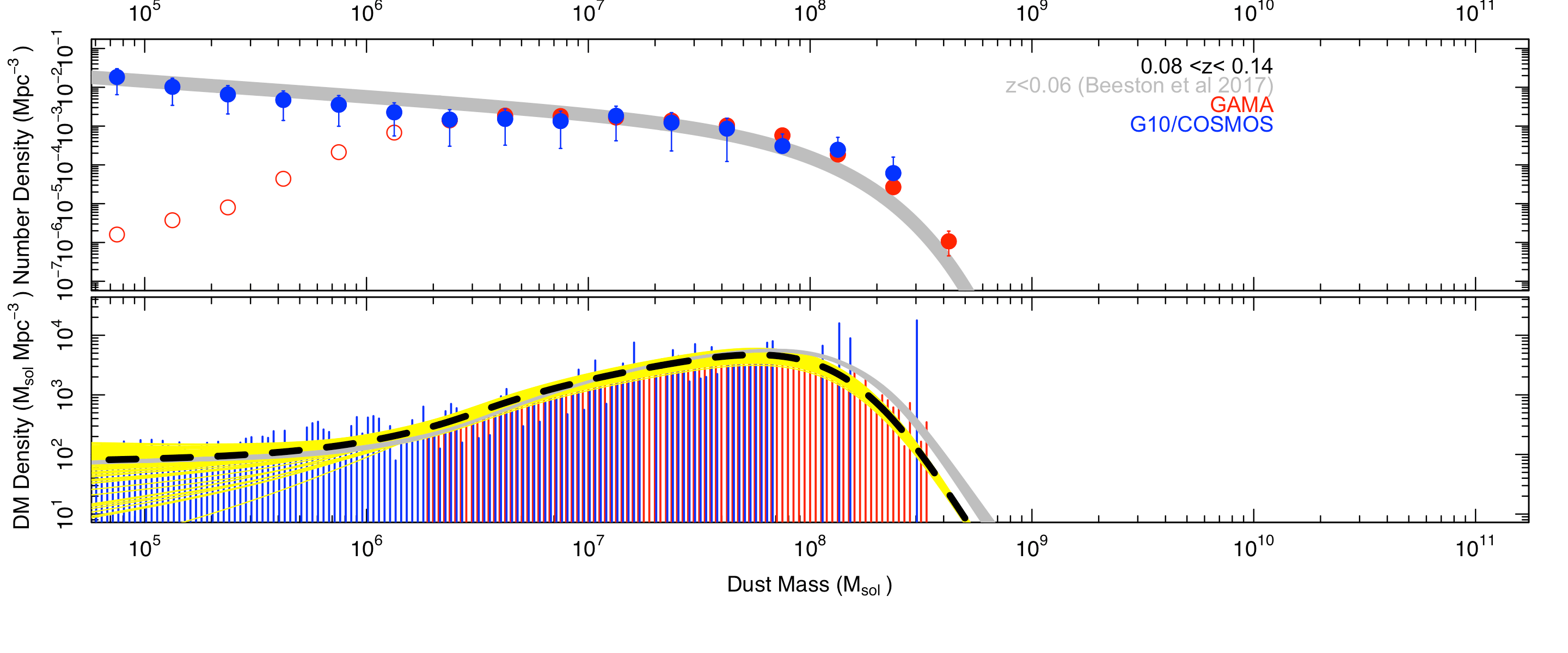}

\vspace{-0.2cm}

\caption{({\it upper}) The star-formation number and density
  distributions for the GAMA, G10-COSMOS and 3D-HST datasets. ({\it
    middle}) The stellar mass number and density distributions for the
  GAMA, G10 and 3D-HST samples, ({\it lower}) the dust mass number and
  density distributions for the GAMA and G10-COSMOS samples. In each
  panel the upper portion shows the number-density without any volume
  corrections. Data points are plotted in solid if the data are deemed
  to sample the full volume limited region, and as open symbols if
  deemed to sample only a fraction of the volume. The thick grey
  shaded line shows the zero redshift fits determined by Wright et al
  (2017b) for the galaxy stellar mass function, or by Beeston et al
  (2017) for the zero redshift dust mass function. In the lower
  portion of each panel we show the contribution of each interval to
  the overall SFH, SMF, or DMF. It is this distribution which we fit
  with a 9-point spline (dashed black line) and integrate to recover
  the total CSFH, SMD or DMD for that redshift interval. Note that the
  grey lines represent our Monte Carlo reruns where we modify each
  galaxy by its individual error (highlighting the Eddington Bias),
  and the yellow lines represent our Monte Carlo reruns where we
  perturb each dataset by the estimates cosmic variance error (see
  Section~4.2 for our error analysis discussion).
\label{fig:methods}}
\end{center}
\end{figure*}

\subsection{Methodology for deriving star-formation and mass densities}
Fig.~\ref{fig:methods} illustrates our methodology for deriving
star-formation (upper), stellar mass (middle), and dust mass (lower)
densities in the redshift interval $0.08 < z < 0.14$.  For each
redshift interval, we start by constructing the star-formation,
stellar mass, or dust mass space-density histograms; having first
divided by the appropriate survey volume (where we take the survey
areas to be: 117.2 sq deg for GAMA, 1.022 sq deg for G10-COSMOS, and
0.274 sq deg for 3D-HST, see Table.~\ref{tab:info}). These
space-density distributions are shown in the upper half of each panel
and constitute the star-formation, stellar mass, or dust mass
distributions respectively for a particular redshift slice. No volume
corrections are applied and so for each sample the measurements are
volume-limited at the right-hand side, but as we move downwards in
star-formation rate (or mass), the contributing systems are no longer
sampled over the full volume range (redshift slice). At this point the
distributions will turn-down due to traditional Malmquist bias.

For each dataset in each redshift interval we can identify this
turn-down by noting where the shallower dataset (e.g., GAMA) deviates
below the deeper dataset (e.g., G10-COSMOS or 3D-HST). For 3D-HST we
simply assume that any sharp downward deviation at low star-formation rate,
or low masses is unphysical, and therefore caused by the diminishing
volume over which these lower star-forming or lower mass systems are
seen. This is a purely empirical constraint and has the distinct
advantage of folding in most hidden biases, but the disadvantage of
being somewhat subjective. The check comes from the overlap regions
between the shallow and deep data.

For example, in the upper-middle panel of Fig.~\ref{fig:methods} the
GAMA stellar mass distribution (red solid points) traces the
stellar-mass function down to $\sim 10^{9.5}$M$_{\odot}$, at which
point the GAMA distribution starts to deviate from the deeper datasets
(blue and green points), indicating the onset of incompleteness. We
highlight the turn-downs by plotting data which we believe is
incomplete using open symbols, and datapoints we consider complete as
solid symbols.

To reiterate, in Fig.~\ref{fig:methods} (upper-middle panel) we see
the three stellar mass distributions, where the high-mass end is well
defined by the GAMA sample (red symbols), and the intermediate-mass
range and low-mass end are well defined by the G10-COSMOS (blue) and
3D-HST (green) samples. As a comparison yardstick the grey curve shows
the GAMA Galaxy Stellar Mass Function recently derived by Wright et
al.~(2017a) for $z < 0.1$ which includes a volume correction which
incorporates density sampling of the underlying large scale
structure. The agreement between the Wright et al. curve and our
composite data from the three distinct datasets provides a good
demonstration that the G10-COSMOS, 3D-HST data are consistent with the
fully volume-corrected low-mass GAMA data.

The lower half of each panel now shows the differential contribution
to the star-formation rate (upper), stellar-mass (middle), and
dust-mass (lower) densities, i.e., $M \phi(M)dM$ in the stellar mass
case. Here the histograms are more finely sampled for plotting clarity
and we can see, in the case of stellar mass densities for example,
that the peak contribution within this redshift interval occurs at
$\sim 10^{10.8}$M$_{\odot}$. Again we see how the density distribution
is defined by the three distinct datasets with GAMA dominating at high
mass (red bars), G10-COSMOS at intermediate mass (blue bars), and
3D-HST at the lowest mass range (green bars). The errors associated
with each data point and adopted in the spline fitting (dashed black
line) are a combination of Poisson error added in quadrature to the
cosmic variance error. The cosmic variance error is derived from
Driver \& Robotham (2011) and shown in Table~\ref{tab:limits}. We fit
a 7-point spline to the full distribution of density spikes weighted
by the inverse fractional error squared. Hence the spline most closely
follows the GAMA data at high masses, then the G10-COSMOS data and
finally the 3D-HST data, i.e., it uses all the data simultaneously but
most closely traces the sample with the lowest errors. Finally to
determine the overall density we integrate the spline over a fixed
star-formation/mass range to get the total density at that redshift.

The use of a spline-fit is necessary, as opposed to just summing the
data, because in the higher redshift bins the distribution is only
partially sampled. Hence the spline allows us to extrapolate over a
fixed star-formation/mass range and to recover star-formation/mass
below the detection limits.  This introduces the scope for
extrapolation error which is managed by the Monte-Carlo error analysis
described in the next section. A spline-fit is also preferable to the
more standard single or double Schechter function fitting, because it
most closely follows the shape of the distributions, however, care
must be taken that the spline is well behaved, for this reason we show
all our fits in Appendix~B (Figs~\ref{fig:methods1} to
Fig.~B18). Across all bins we can see that the distributions from the
three datasets are extremely consistent and well defined. We also note
that our technique allows GAMA to remain useful in constraining the
highest mass/star-formation end up to $z \approx 0.68$ and G10-COSMOS
up to $z \approx 1.75$, with only the 3D-HST data extending to the
very highest redshifts $z < 5$. The adopted mass and star-formation
limits for each dataset, along with the estimated cosmic variance
values from Driver \& Robotham (2011) are shown in
Table~\ref{tab:limits}.

\begin{table*}
\caption{A summary of the cosmic variance, stellar mass, dust mass and star-formation limits used in the analysis. \label{tab:limits}}
\noindent
\small
\begin{tabular} {ccccccccccccc}\hline
Redshift & \multicolumn{4}{c}{GAMA limits} & \multicolumn{4}{c}{G10-COSMOS limits} & \multicolumn{4}{c}{3D-HST limits} \\ \cline{2-12}
interval & CV & $M_{\rm *, lim}$ & $M_{\rm D, lim}$ & $\log_{10} SF_{\rm lim}$ & CV & $M_{\rm *, lim}$ & $M_{\rm D, lim}$ & $\log_{10} SF_{\rm lim}$ & CV & $M_{\rm *, lim}$ & $M_{\rm D, lim}$ & $\log_{10} SF_{\rm lim}$ \\ 
         &    & $(M_{\odot})$ & $(M_{\odot})$ & $(M_{\odot} yr^{-1})$ &   & $(M_{\odot})$ & $(M_{\odot})$ & $(M_{\odot} yr^{-1})$ & & $(M_{\odot})$ & $(M_{\odot})$ & $(M_{\odot} yr^{-1})$ \\ \hline
$0.02$---$0.08$&0.19  &  8.75 & 5.75 &-1.00& 0.77 &6.75 &4.00    &-3.0 & 0.60 & 6.50 &4.00 &-3.00  \\
$0.06$---$0.14$&0.13   & 9.25 & 6.25  & 0.00& 0.59 &7.25 &4.25    &-3.0 & 0.50 & 7.00 &4.00 &-3.00  \\
$0.14$---$0.20$&0.10   & 10.00 & 6.50 & 0.00& 0.51 &7.50 &4.50    &-3.0 & 0.45 & 7.00 &4.50  &-3.00  \\
$0.20$---$0.28$&0.072  & 10.50 & 7.00 & 0.25& 0.39 &7.50 &4.75    &-2.50 & 0.40 & 7.25 &5.25  &-2.50  \\
$0.28$---$0.36$&0.062  & 10.75 & 7.50 & 0.75& 0.35 &7.75 &5.25    &-1.75 & 0.40 & 7.50 &5.25  &-1.75  \\
$0.36$---$0.45$&0.052  & 11.0 & 8.50  & 1.00& 0.30 &8.00 &5.25    &-1.50 & 0.35 & 7.75 &5.25  &-1.50  \\
$0.45$---$0.56$&0.043  & 11.25 & 9.00 & 1.50& 0.26 &8.25 &6.00    &-1.00 & 0.30 & 7.75 &5.50  &-1.00  \\
$0.56$---$0.68$&0.039  & 11.50 & 9.25 & 1.75& 0.23 &8.50 &6.00    &-0.50 & 0.25 & 8.00 &5.75  &-1.00  \\
$0.68$---$0.82$&0.035  & 11.75 & -& -& 0.21 &8.50 &6.25    &-0.50 & 0.20 &8.25&5.75  &-0.75  \\
$0.82$---$1.00$&0.03   & - & -& -& 0.18 &9.00 &6.75    &0.00 & 0.20 &8.25&6.00  &-0.50  \\
$1.00$---$1.20$&0.03   & - & -& -& 0.18 &9.00 &7.00    &0.25 & 0.18 &8.50&6.00  &-0.50  \\
$1.20$---$1.45$&0.03   & - & -& -& 0.18 &9.25 &7.25    &0.50 & 0.18 &8.75&6.25  &-0.25  \\
$1.45$---$1.75$&0.03   & - & -& -& 0.18 &9.75 &7.25    &0.75 & 0.15 &8.50&6.25  &-0.25  \\
$1.75$---$2.20$&0.03   & - & -& -& 0.18 &- &-  & - & 0.15 &9.25&6.75  &0.00  \\
$2.20$---$2.60$&0.03   & - & -& -& 0.18 &- &-  & - & 0.10 &9.25&7.00  &0.50  \\
$2.60$---$3.25$&0.03   & - & -& -& 0.18 &- & -& - & 0.10 &9.50&7.25  &0.75  \\
$3.25$---$3.75$&0.03   & - & -& -& 0.18 &- & -& - & 0.10 &9.50&7.50  &0.75  \\
$3.75$---$4.25$&0.03   & - & -& -& 0.18 &- & -& - & 0.10 &9.50&7.50  &1.00 \\
$4.25$---$5.00$&0.03   & - & -& -& 0.18 &- & -& - & 0.10 &9.50&7.50  &1.25  \\ \hline
\end{tabular}
\end{table*}

\subsection{Measurement and error analysis}
We consider three forms of error: that arising from Poisson
statistics; that arising from the cosmic variance (see
Table~\ref{tab:limits}); and Eddington bias.

To assess the statistical error we jostle all data points individually
by drawing randomly from a Normal distribution of width equal to the
quoted MAGPHYS measurement error for each galaxy, and then rerun the
full analysis and repeat 101 times (sufficiently large to sample the
error range but not so large to be computationally challenging). We
then assess the spread of each of our derived density values (these
alternative fits are shown as the grey lines on the lower panels of
Fig.~\ref{fig:methods}).  These lines highlight both the resilience to
Poisson error, but also the potential impact of any Eddington bias
(based on how well the grey lines cluster around the base measurement
(given by the dashed black line). In each case the star-formation
density, stellar mass density, or dust mass density, is derived from
the integrations of the splines. With our base measurement coming from
the integration of the dashed black line. We can similarly integrate
each of the grey splines to get both the dispersion in measured values
(from the 84-16 percentile range), and an offset from the base
measurement due to the error perturbation process (the Eddington
bias). We correct our derived data for this bias by subtracting the
offset of our base measurement from the median of our Poisson re-fits.

To assess the cosmic variance error we again repeat the analysis but
this time jostle the amplitude of each of the entire three datasets
independently, by drawing from a Normal distribution of width equal to
the estimated CV error, as listed in Table~\ref{tab:limits}. Again we
repeat the analysis 101 times (yellow lines on Fig.\ref{fig:methods})
and once again assess the dispersion for each of our derived values.

Finally, we rerun our base analysis for our three AGN selections for
the 3D-HST dataset (see Section~2.4)  representing the lenient, fair,
and extreme selections and determine an error estimate based on half
the range across the three
measurements. Tables~\ref{tab:sfr},~\ref{tab:smd}~\&~\ref{tab:dmd} show
our final measurements, including the Eddington bias correction, along
with each of the individual errors (Eddington bias, Poisson error, CV
error, and AGN classification uncertainty).

Note when fitting splines we utilise the information of null data in a
high stellar-mass, dust-mass or star-formation bin, where an absence
of data represents significant information, by setting the first
unoccupied bin to have a low value with high significance, this
ensures that the spline fits bound the data at the high mass or
star-formation end and do not diverge or include excess extrapolated
flux.

\begin{figure*}
\includegraphics[width=0.99\textwidth]{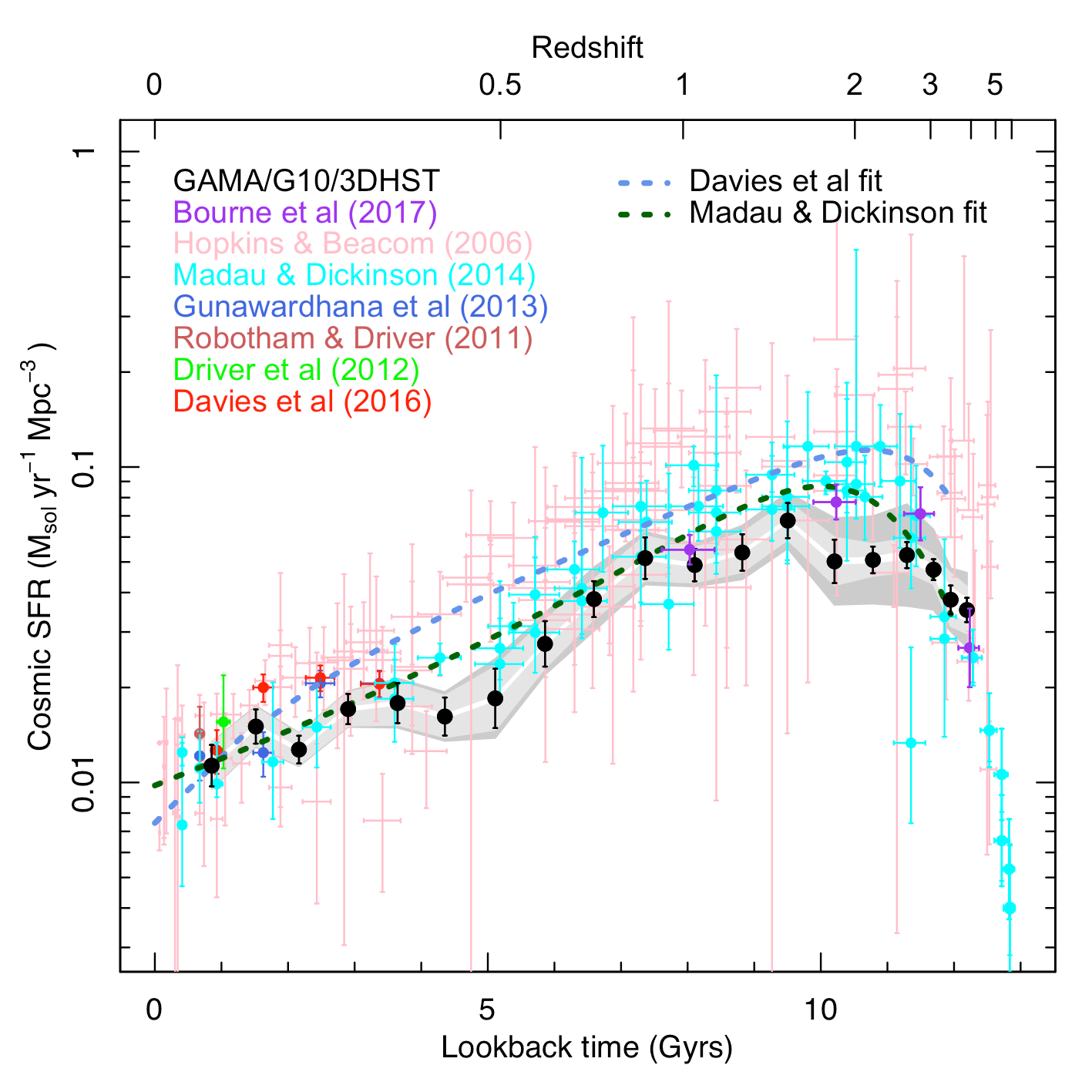}

\caption{ {{{ The cosmic star-formation history as a function of lookback
  time. Shown are literature compendia along with the recently
  trends reported by Davies et al.~(2016) and Madau \&
  Dickinson (2014). Our new measurements from our combined sample are
  shown in black with the three error components indicated by the
  distinct grey shading. The errors are shown as additive with the
  light grey indicating statistical error, the grey as the cosmic
  variance, and the dark grey the AGN classification uncertainty. Note
  that a figure showing the same data versus redshift is given in
  Appendix~C.  \ref{fig:plotall_z}.
\label{fig:data_sfr} }}} }
\end{figure*}

\subsection{The Cosmic Star-formation History}
The cosmic star-formation history is one of the most well studied
cosmic-planes since the original work from the Canada France Redshift
Survey (CFRS; Lilly et al.~ 1996), and the Hubble Space Telescope
Hubble Deep Field (HDF; Madau et al.~1998).  Fig.~\ref{fig:data_sfr}
shows two compendia of recent measurements drawing in disparate
star-formation tracers from diverse surveys. These compendia are taken
from Hopkins \& Beacom (2006; light pink), and Madau \& Dickinson
(2014; cyan). For the genesis of the individual data points please see
the Tables included in these works. Note also that the Madau \&
Dickinson compendium includes many of the Hopkins \& Beacom data but
corrected for various issues which later came to light (i.e.,
recalibrations, treatment of $h$ etc). For this reason we show the
more recent Madau \& Dickinson data in bright cyan and the earlier
older Hopkins \& Beacom compendium in light pink, an offset between
these two compendia is clearly visible. Note we convert from Salpeter
(or Salpeter-A) IMFs to Chabrier IMFs by multiplying by a factor of
0.63 (0.85) (see Madau \& Dickinson 2014 and Driver et al.~2013). Also
obvious is the significant vertical scatter which arises from the use
of distinct tracers, and in some cases the relatively small volumes
probed giving rise to significant cosmic variance fluctuations.

Also shown are recent measurements at very low redshift (Robotham \&
Driver 2011; Driver et al.~2012; Gunawardhana et al.~2013; Davies et
al.~2015, with the latter four of these coming from various distinct
analysis of the GAMA data). Recent fits to the data are shown as
either the dashed green line (Madau \& Dickinson 2014), or the dashed mauve line
(Davies et al.~2016), with Davies et al.~finding a very similar but
marginally higher SFR due to the inclusion of the Hopkins \& Beacom
data in the fitting. Finally we show the most recent data from Bourne
et al.~(2017; purple points) based on SCUBA observations of selected
3D-HST galaxies.

Overlain as solid black discs with error-bars, are data derived from
the combined GAMA/G10-COSMOS/3D-HST dataset, produced in the manner
described in the previous section. Because of the very large size of
the GAMA, G10-COSMOS and 3D-HST datasets the Poisson errors become
vanishingly small (light grey shading) and the dominant error comes
from cosmic variance (grey shading; particularly at the lower redshift
end) and AGN-uncertainty (dark grey shading band; at the high
redshift-end). Note, these errors are shown added linearly. We now
have a complete record of the star-forming history over a 12 Gyr
timeline drawn from the combination of three large datasets with
cross-calibrated star-formation rates. Particularly noticeable is that
the error spread is now almost a factor of $\times 2$ better than the
Madau \& Dickinson compendium and $\times 5$ better than the Hopkins
\& Beacom compendium. Our values agree well with the Madau \&
Dickinson data, although we do see a modest tendency for our data to
lie slightly below the Madau \& Dickinson fit, and in particular a
slump at ~5 Gyrs lookback time ($z \approx 0.5$). Although the Madau
\& Dickinson compendium is fairly thin on the ground we ascribe this
slump to cosmic variance because a similar slump is also seen in the
stellar mass density shown in the middle panel. AT high-z we see our
data falls below the most recent study of Bourne et al.~(2017). There
are two obvious possibilities: {{{ Either our study is incomplete
      for highly obscured star-formation, or the Bourne et al.~study
      is contaminated by AGN. Distinguishing between these two
      possibilities is not trivial and will take significant effort. We
      therefore elect to carry both our data and the Bourne data
      forward. }}}

In later analysis where we fit these data we will also fold in the
highest seven redshift data points taken from Table 1 of Madau \&
Dickinson (but originally reported in Bouwens et al.~2012a,b and
Schenker et al.~2013), and which extend to 12.8 Gyrs in lookback time
(i.e., see Fig.~\ref{fig:sims} upper panel), in addition to the Bourne
et al.~(2017) data (purple dots).

\begin{figure*}
\includegraphics[width=0.99\textwidth]{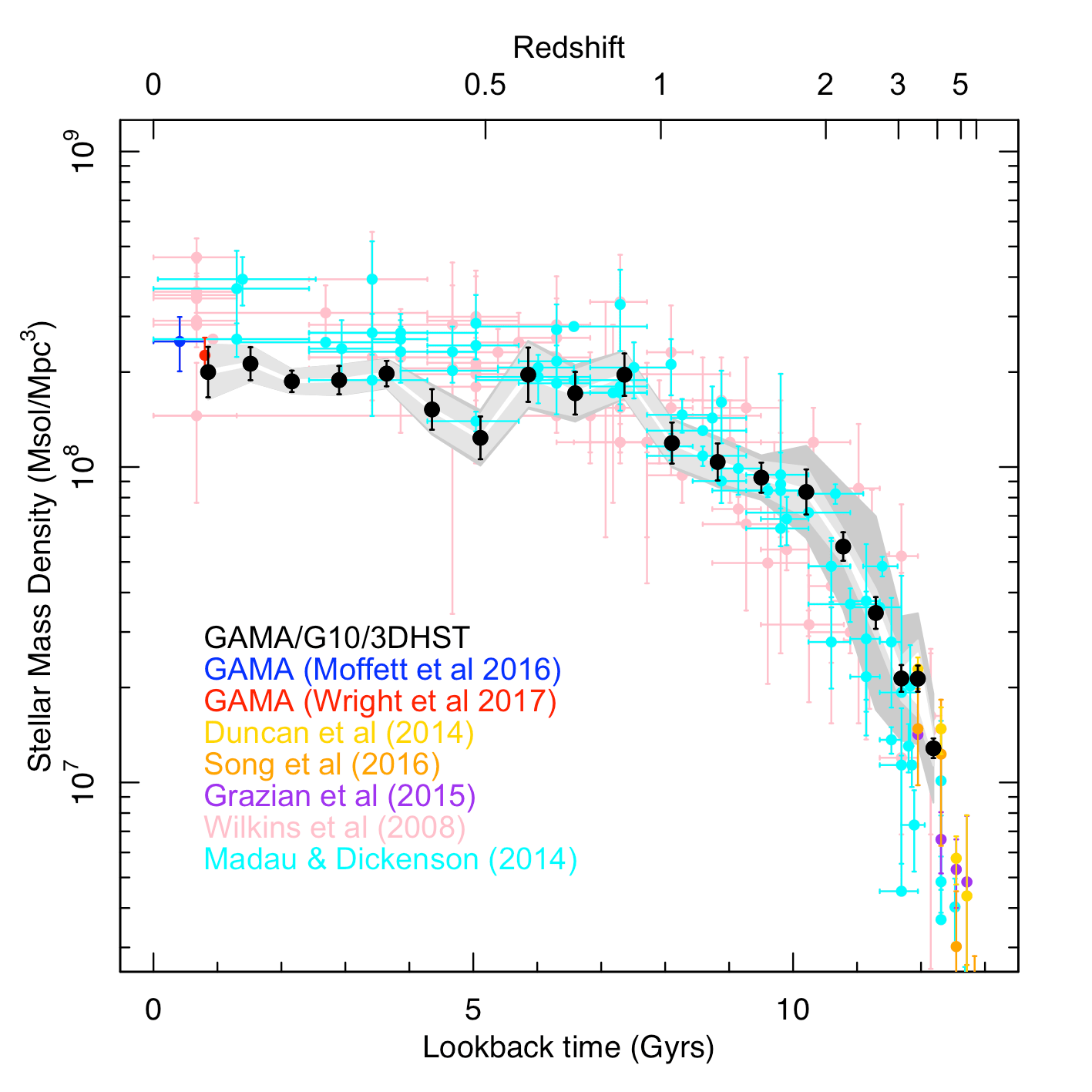}
\caption{ {{{ The stellar mass density versus lookback time. Shown are
  literature compendium and recently measurements along with our
  measurements form our combined sample (black dots). The grey bands
  show the error budget, plotted in additive fashion with the light
  grey representing the statistical uncertainty, the grey line the
  cosmic variance and the dark grey the uncertainty from AGN
  classification. }}}
\label{fig:data_smd}}
\end{figure*}

\subsection{The build-up of the stellar mass density}
Fig.~\ref{fig:data_smd} shows the stellar mass density as a
function of cosmic time, compared to the Wilkins et al.~(2008; light pink
data points), and the Madau \& Dickinson (2014; cyan) compilations of
literature estimates. As before we show the more recent compendium in
bright cyan and the older Wilkins et al. compendium in light pink and
have scaled down the stellar masses by 0.63 to convert from a Salpeter
IMF to a Chabrier IMF. The blue point represents a recent stellar mass
estimate based on GAMA for $z < 0.06$ by Moffett et al.~(2016), and
the red point shows the recent estimate by Wright et al.~(2017)
from an analysis of GAMA to $z<0.1$. 

The astute will wonder that given the leftmost black data point and
the blue and red are essentially three estimates of the same dataset,
why do we see any scatter at all? The answer is in the extrapolation
and the Eddington bias. In Moffett et al.~the data is sub-divided by
component and the disc and little blue spheroid (LBS) components
exhibit steep low-mass end slopes which when extrapolated yields
additional mass. Similarly the dataset shown here is the only one of
the three which includes a formal Eddington bias.  The range therefore
reflects some of the systematic uncertainty inherent in the
methodology.

The black data points within the light grey/grey/dark grey shaded
region represent the full GAMA/G10-COSMOS/3D-HST combined dataset
which span almost the entire timeline of the Universe. These show good
agreement with the existing literature values but with less scatter,
and a relatively smooth behaviour in which mass builds very rapidly at
early epochs and then more slowly over later epochs (although beware
the logarithmic scaling). Very slightly noticeable is the close
agreement at high-redshift combined with a slight tendency towards low
stellar mass densities at lower-redshift. The 50 per cent point is
reached at approximately $9 \pm 1$ Gyr lookback time. We also note
that slump in data at $z \approx 0.5$ which cannot be physical (i.e.,
the stellar mass density cannot actually decline and rise this
quickly), but is most likely due to cosmic (sample) variance and an
underdensity in the G10-COSMOS at this redshift.

The main advantage of the combined dataset comes from two principle
factors: the homogeneity of stellar mass estimates across the three
datasets, and the size of the samples, bringing the errors to a
significantly narrower distribution than the assembled literature
values.

Finally we also overlay a number of recent high-redshift literature
values (Duncan et al.~2014; Grazian et al.~2015 and Song et
al.~2016). These align very closely to our high-redshift data showing
a complete record of the build-up of stellar mass from $z=8$ to the
present day (i.e, from when the Universe was 1 Gyr old to the present
time).

It is also worth noting that our dominant error at high-z is due to
uncertainty in the AGN identification (dark grey shading), of course at
some point this becomes moot as galaxies are neither AGN or
star-forming but both and ultimately effort is needed to separate the
AGN-light from the stellar emission prior to determining masses.

\begin{figure*}
\includegraphics[width=0.99\textwidth]{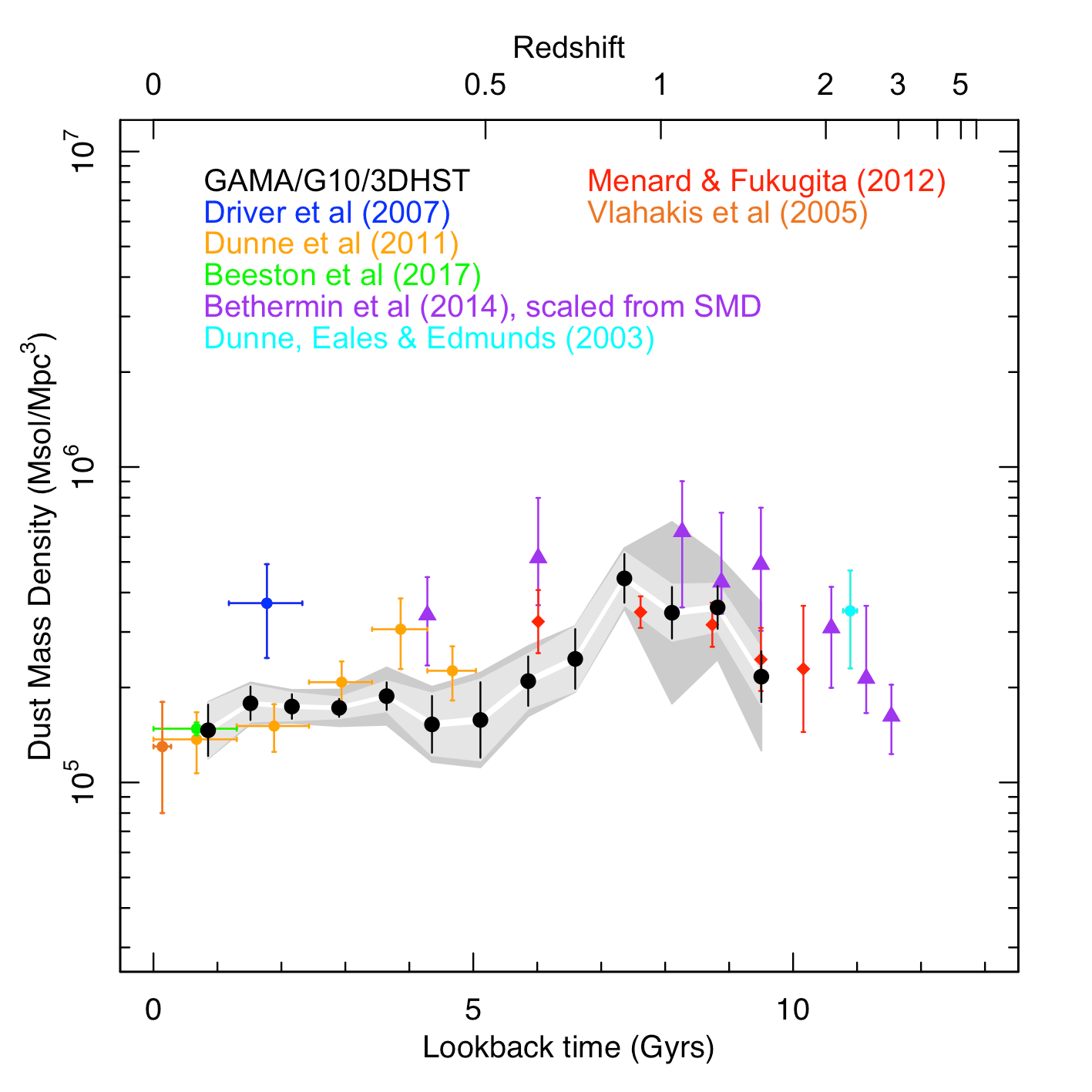}
\caption{ {{{ The dust mass density versus lookback time. Shown are
        various measurements from the literature along with our
        measurements from the combined GAMA/G10 sample (black
        dots). The error bands are shown in grey, are additive in
        indicate the statistical error (light grey), the cosmic
        variance error (grey), and the error introduced from including
        objects with or without far-IR measurements (dark grey). }}}
\label{fig:data_dmd}}
\end{figure*}

\subsection{The recent decline in dust mass density}
Fig.~\ref{fig:data_dmd} shows our recovered dust mass density against
lookback time. Initially this trend is flat then rising slowly to $z
\sim 1$ with a hint of a decline at our G10-COSMOS redshift limit of
$z=1.75$. However, we do not place any significance in this turn-down
at $z=1.75$ given the associated errors indicated by the grey shading.

One of the problems in establishing the veracity of this result is
that fairly little previous data exists at any redshift. Driver et
al.~(2008) inferred an estimate from optical data combined with a
radiative transfer model, while Dunne et al.~(2011) derived
measurements from the Herschel-ATLAS Science Definition
Phase. Concurrently to this work, an updated measurement of the low-z
dust-mass function was obtained by Beeston et al.~(2017), also based
on the GAMA database and the MAGPHYS data presented here.

Compared to the Driver et al. data we find a marginally (1.5$\sigma$)
lower dust mass density than they reported. This is likely due to
MAGPHYS recovering significantly lower than expected opacities when
compared to the Driver study. In that study a single opacity value was
derived and adopted for the entire population ($\tau_v^o=3.8$ ; Driver
et al.~2007). Compared to Dunne et al. our data agree well (all values
are within the 1$\sigma$-errors, but rather than seeing a rapidly
declining dust mass density we find a relatively flat dust mass
density. We note that Dunne et al. raise some concern and caution in
using their last data point, without which they concluded a rapidly
declining dust density. From an orthogonal analysis of dust lanes in
galaxies, Holwerda et al.~(2012) also concluded fairly flat evolution
of the dust mass density (and its planar distribution) out to $z=1$,
supporting our results.

Using mean dust-to-stellar mass ratios for the main sequence reported
by B\'ethermin et al.~(2014) we can determine rough values (purple
triangles) by scaling our stellar mass densities by these
ratios. These were derived from data at wavelengths $>1$mm (i.e., on
the Raleigh-Jeans tail to very high redshift) but do implicitly assume
that the global dust mass density is very much dominated by main
sequence systems rather than extreme burst systems. Also shown are
estimates based on MgII absorbers in Quasar sight-lines reported by
Menard \& Fukugita (2012). Note that these data implicitly assume an
SMC-like extinction curve (correcting to a MW-like extinction curve
would scale up the data by $\times 1.8$, Gergely Popping
priv. comm). The B\'ethermin and Menard data both agree reasonably
well or exceptionally well respectively, with our results painting a
consistent picture of the dust density slowly declining over a 10 Gyr
period.  Finally we include the earlier estimate by Dunne, Eales \&
Edmunds (2003) based on sub-mm constraints, provide a direct estimate
giving a high-z anchor point.  From our combined distribution and these
supporting data we can make the following statements:

(1) The dust mass density appears to peak around 8 Gyrs ago ($z \sim 1$),
suggesting that dust formation is either concurrent with
star-formation or lags no more than a few Gyrs behind the
star-formation peak (see also Cucciati et al.~2012 and Burgarella et
al.~2013). At the current time there is no clear consensus on dust
production with advocates for the dominant pathway being SN, AGB, or
ISM grain growth (see for example Gall et al.~2014; Sargent et al.~2010;
and Rowlands et al.~2014 respectively and references therein). Most
likely all pathways are relevant. A Dust Density peak coincident with
the CSFH could argue more for the former (i.e., SN which are
immediate), over AGB (which would be slightly delayed by as much as
1-2Gyrs), or ISM grain growth which would expect to be a continuous
process.

(2) We also see that the total dust mass density declines relatively
smoothly over the past 8 Gyrs, implying that in the latter half of the
Universe dust is destroyed faster than it is formed (presumably
through astration), and the Universe is becoming more
transparent. This is also consistent with the reported $A_{FUV}$
evolution in Cucciati et al.~(2012) and Burgarella et al.~(2013), see
also figure 9 of Andrews et al.~(2017a).  This is not particularly new
and unfortunately the current data cannot constrain the key question
which is what fraction of the dust is destroyed, whether it is
destroyed locally or globally and what fraction might be ejected into
the halo or even IGM.

 ~

\noindent
We will return to discuss dust evolution further in Section 6 where we build a simple toy model.

\begin{table*}
\caption{Derived cosmic star-formation densities from our combined GAMA/G10-COSMOS/3D-HST sample. \label{tab:sfr}}
\begin{tabular}{ccccccc} \hline
Age$\dagger$ & Redshift & \multicolumn{5}{c}{Star-formation rate} \\
(Gyr) & interval & \multicolumn{5}{c}{$\log_{10}(M_{\odot} yr^{-1} h_{0.7} Mpc^{-3})$}  \\ \cline{3-7}
- & - & Value$^\ddagger$ & Edd. Bias & $\Delta_{\rm possion}$ & $\Delta_{\rm cv}$ & $\Delta_{\rm AGN}$ \\ \hline
$0.85$ &  $0.02$---$0.08$ &  $-1.95$ & $0.03$ & $\pm0.00$ &  $\pm0.07$ &  $\pm0.00$ \\ 
$1.52$  & $0.06$---$0.14$ &  $-1.82$ & $0.03$ & $\pm0.01$ &  $\pm0.05$ &  $\pm0.01$ \\ 
$2.16$  & $0.14$---$0.20$ &  $-1.90$ & $0.02$ & $\pm0.00$ &  $\pm0.04$ &  $\pm0.00$ \\ 
$2.90$  & $0.20$---$0.28$ &  $-1.77$ & $0.01$ & $\pm0.00$ &  $\pm0.05$ &  $\pm0.00$ \\ 
$3.65$  & $0.28$---$0.36$ &  $-1.75$ & $0.01$ & $\pm0.00$ &  $\pm0.06$ &  $\pm0.01$ \\ 
$4.35$  & $0.36$---$0.45$ &  $-1.79$ & $0.01$ & $\pm0.01$ &  $\pm0.06$ &  $\pm0.01$ \\ 
$5.11$  & $0.45$---$0.56$ &  $-1.73$ & $0.04$ & $\pm0.01$ &  $\pm0.09$ &  $\pm0.03$ \\ 
$5.86$  & $0.56$---$0.68$ &  $-1.56$ & $0.05$ & $\pm0.00$ &  $\pm0.07$ &  $\pm0.02$ \\ 
$6.59$  & $0.68$---$0.82$ &  $-1.42$ & $0.06$ & $\pm0.01$ &  $\pm0.06$ &  $\pm0.04$ \\ 
$7.36$  & $0.82$---$1.00$ &  $-1.29$ & $0.05$ & $\pm0.00$ &  $\pm0.07$ &  $\pm0.01$ \\ 
$8.11$  & $1.00$---$1.20$ &  $-1.31$ & $0.04$ & $\pm0.00$ &  $\pm0.05$ &  $\pm0.01$ \\ 
$8.82$  & $1.20$---$1.45$ &  $-1.27$ & $0.03$ & $\pm0.00$ &  $\pm0.06$ &  $\pm0.02$ \\ 
$9.50$  & $1.45$---$1.75$ &  $-1.17$ & $0.02$ & $\pm0.00$ &  $\pm0.06$ &  $\pm0.03$ \\ 
$10.21$ & $1.75$---$2.20$ &  $-1.30$ & $0.04$ & $\pm0.01$ &  $\pm0.07$ &  $\pm0.06$ \\ 
$10.78$ & $2.20$---$2.60$ &  $-1.29$ & $0.04$ & $\pm0.01$ &  $\pm0.04$ &  $\pm0.09$ \\ 
$11.29$ & $2.60$---$3.25$ &  $-1.28$ & $0.04$ & $\pm0.01$ &  $\pm0.04$ &  $\pm0.11$ \\ 
$11.69$ & $3.25$---$3.75$ &  $-1.33$ & $0.03$ & $\pm0.01$ &  $\pm0.03$ &  $\pm0.08$ \\ 
$11.95$ & $3.75$---$4.25$ &  $-1.42$ & $0.04$ & $\pm0.04$ &  $\pm0.05$ &  $\pm0.02$ \\ 
$12.19$ & $4.25$---$5.00$ &  $-1.45$ & $0.03$ & $\pm0.04$ &  $\pm0.04$ &  $\pm0.04$ \\ 
\hline
\end{tabular}

\noindent
$\dagger$ The age of the Universe at the volume midpoint of the redshift interval.

\noindent
$\ddagger$ Note that these values have had the Eddington Bias (given in
Col. 4) {\it subtracted} from the initial measurement, i.e., they are
Eddington Bias corrected.
\end{table*}

\begin{table*}
\caption{Derived cosmic stellar mass densities from our combined GAMA/G10-COSMOS/3D-HST sample. \label{tab:smd}}
\begin{tabular}{ccccccc} \hline
Age$\dagger$ & Redshift & \multicolumn{5}{c}{Stellar mass density} \\
(Gyr) & interval & \multicolumn{5}{c}{$\log_{10}(M_{\odot} h_{0.7} Mpc^{-3})$} \\ \cline{3-7}
- & - & Value$\ddagger$ & Edd. Bias & $\Delta_{\rm possion}$ & $\Delta_{\rm cv}$ & $\Delta_{\rm AGN}$ \\ \hline
$0.85$ &  $0.02$---$0.08$ &  $8.30$ & $0.01$ & $\pm0.01$ &  $\pm0.08$ &  $\pm0.00$ \\ 
$1.52$  & $0.06$---$0.14$ &  $8.33$ & $0.01$ & $\pm0.00$ &  $\pm0.05$ &  $\pm0.00$ \\ 
$2.16$  & $0.14$---$0.20$ &  $8.27$ & $0.02$ & $\pm0.00$ &  $\pm0.03$ &  $\pm0.00$ \\ 
$2.90$  & $0.20$---$0.28$ &  $8.28$ & $0.00$ & $\pm0.00$ &  $\pm0.04$ &  $\pm0.00$ \\ 
$3.65$  & $0.28$---$0.36$ &  $8.30$ & $-0.01$ & $\pm0.00$ &  $\pm0.04$ &  $\pm0.00$ \\ 
$4.35$  & $0.36$---$0.45$ &  $8.18$ & $-0.01$ & $\pm0.00$ &  $\pm0.06$ &  $\pm0.01$ \\ 
$5.11$  & $0.45$---$0.56$ &  $8.09$ & $-0.01$ & $\pm0.00$ &  $\pm0.07$ &  $\pm0.02$ \\ 
$5.86$  & $0.56$---$0.68$ &  $8.29$ & $-0.03$ & $\pm0.01$ &  $\pm0.09$ &  $\pm0.01$ \\ 
$6.59$  & $0.68$---$0.82$ &  $8.23$ & $0.01$ & $\pm0.01$ &  $\pm0.07$ &  $\pm0.02$ \\ 
$7.36$  & $0.82$---$1.00$ &  $8.29$ & $0.01$ & $\pm0.00$ &  $\pm0.07$ &  $\pm0.00$ \\ 
$8.11$  & $1.00$---$1.20$ &  $8.08$ & $0.01$ & $\pm0.00$ &  $\pm0.07$ &  $\pm0.01$ \\ 
$8.82$  & $1.20$---$1.45$ &  $8.02$ & $0.01$ & $\pm0.00$ &  $\pm0.06$ &  $\pm0.02$ \\ 
$9.50$  & $1.45$---$1.75$ &  $7.97$ & $0.01$ & $\pm0.00$ &  $\pm0.05$ &  $\pm0.02$ \\ 
$10.21$ & $1.75$---$2.20$ &  $7.92$ & $0.01$ & $\pm0.01$ &  $\pm0.07$ &  $\pm0.07$ \\ 
$10.78$ & $2.20$---$2.60$ &  $7.75$ & $0.01$ & $\pm0.01$ &  $\pm0.05$ &  $\pm0.15$ \\ 
$11.29$ & $2.60$---$3.25$ &  $7.54$ & $0.04$ & $\pm0.03$ &  $\pm0.05$ &  $\pm0.23$ \\ 
$11.69$ & $3.25$---$3.75$ &  $7.33$ & $0.05$ & $\pm0.03$ &  $\pm0.04$ &  $\pm0.13$ \\ 
$11.95$ & $3.75$---$4.25$ &  $7.33$ & $0.01$ & $\pm0.08$ &  $\pm0.04$ &  $\pm0.09$ \\ 
$12.19$ & $4.25$---$5.00$ &  $7.11$ & $0.02$ & $\pm0.02$ &  $\pm0.03$ &  $\pm0.12$ \\ 
\hline
\end{tabular}

\noindent
$\dagger$ The age of the Universe at the volume midpoint of the redshift interval.

\noindent
$\ddagger$ Note that these values have had the Eddington Bias (given in
Col. 4) {\it subtracted} from the initial measurement, i.e., they are
Eddington Bias corrected.
\end{table*}

\begin{table*}
\caption{Derived cosmic dust mass densities from our combined GAMA/G10 sample. \label{tab:dmd}}
\begin{tabular}{ccccccc} \hline
Age$\dagger$ & Redshift & \multicolumn{5}{c}{Dust mass density} \\
(Gyr) & interval & \multicolumn{5}{c}{$\log_{10}(M_{\odot} h_{0.7} Mpc^{-3})$}  \\ \cline{3-7}
- & - & Value$\ddagger$ & Edd. Bias & $\Delta_{\rm possion}$ & $\Delta_{\rm cv}$ & $\Delta_{\rm far-IR}$ \\ \hline
$0.85$ &  $0.02$---$0.08$ &  $5.17$ & $0.05$ & $\pm0.01$ &  $\pm0.08$ &  $\pm0.00$ \\ 
$1.52$  & $0.06$---$0.14$ &  $5.25$ & $0.06$ & $\pm0.00$ &  $\pm0.05$ &  $\pm0.01$ \\ 
$2.16$  & $0.14$---$0.20$ &  $5.24$ & $0.07$ & $\pm0.00$ &  $\pm0.04$ &  $\pm0.01$ \\ 
$2.90$  & $0.20$---$0.28$ &  $5.24$ & $0.06$ & $\pm0.00$ &  $\pm0.03$ &  $\pm0.02$ \\ 
$3.65$  & $0.28$---$0.36$ &  $5.27$ & $0.04$ & $\pm0.00$ &  $\pm0.04$ &  $\pm0.04$ \\ 
$4.35$  & $0.36$---$0.45$ &  $5.18$ & $0.12$ & $\pm0.01$ &  $\pm0.09$ &  $\pm0.02$ \\ 
$5.11$  & $0.45$---$0.56$ &  $5.20$ & $0.12$ & $\pm0.01$ &  $\pm0.12$ &  $\pm0.02$ \\ 
$5.86$  & $0.56$---$0.68$ &  $5.32$ & $0.15$ & $\pm0.01$ &  $\pm0.08$ &  $\pm0.02$ \\ 
$6.59$  & $0.68$---$0.82$ &  $5.39$ & $0.18$ & $\pm0.01$ &  $\pm0.09$ &  $\pm0.00$ \\ 
$7.36$  & $0.82$---$1.00$ &  $5.65$ & $0.14$ & $\pm0.01$ &  $\pm0.08$ &  $\pm0.01$ \\ 
$8.11$  & $1.00$---$1.20$ &  $5.54$ & $0.14$ & $\pm0.01$ &  $\pm0.08$ &  $\pm0.20$ \\ 
$8.82$  & $1.20$---$1.45$ &  $5.55$ & $0.11$ & $\pm0.01$ &  $\pm0.07$ &  $\pm0.09$ \\ 
$9.50$  & $1.45$---$1.75$ &  $5.34$ & $0.14$ & $\pm0.01$ &  $\pm0.08$ &  $\pm0.14$ \\ 
\hline
\end{tabular}

\noindent
$\dagger$ The age of the Universe at the volume midpoint of the redshift interval.

\noindent
$\ddagger$ Note that these values have had the Eddington Bias (given in
Col.~4) {\it subtracted} from the initial measurement, i.e., they are
Eddington Bias corrected.
\end{table*}

\begin{figure*}
\includegraphics[width=0.99\textwidth]{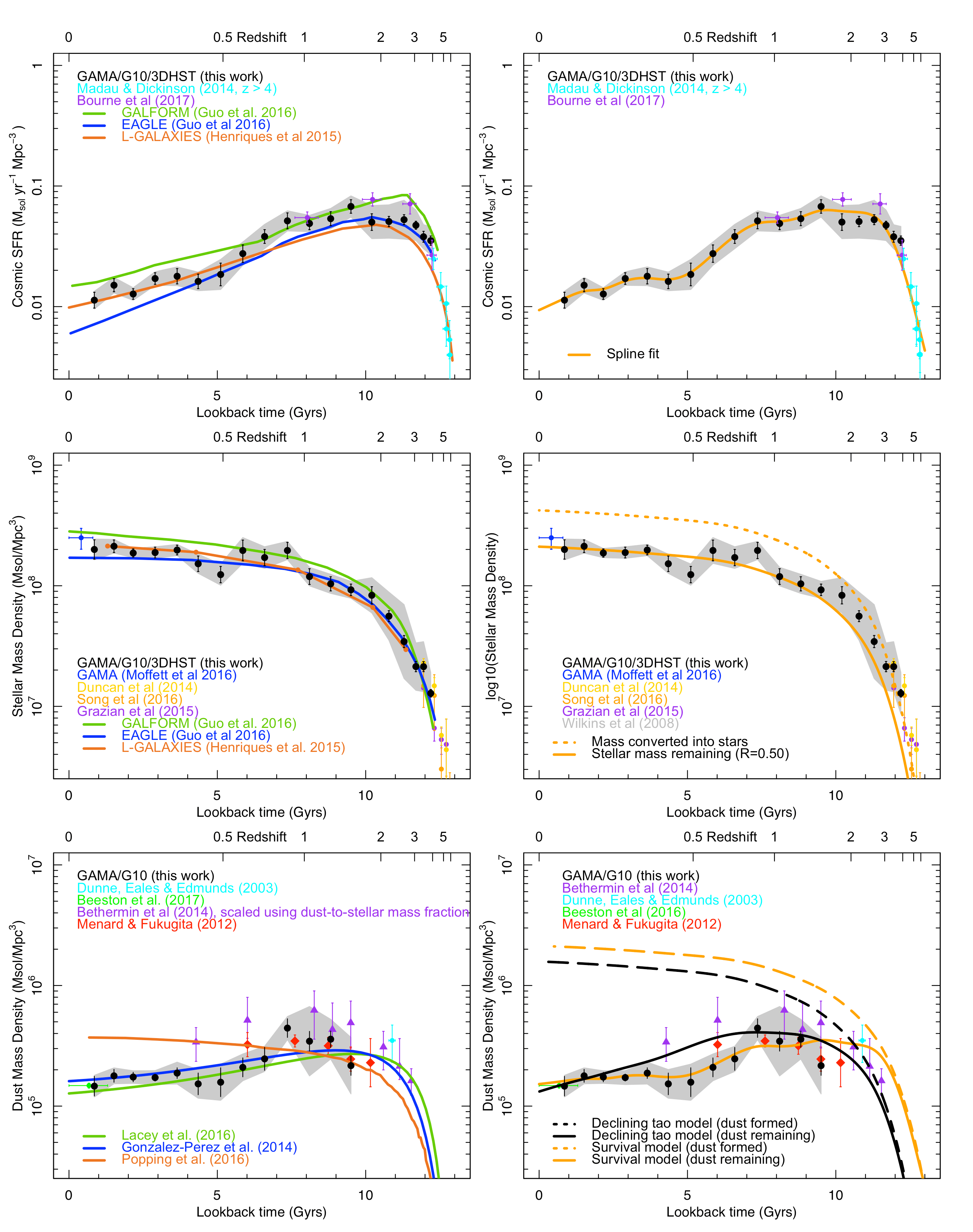}
\caption{ {{{ Our derived cosmic star-formation history (upper panels),
  stellar-mass build-up (middle panels) and dust mass build-up (lower
  panels) versus lookback time. Overlain are model lines as described
  in the text with predictions from numerical simulations (left-side
  panels), and our phenomenological model (right-side panels).
\label{fig:sims} }}} }
\end{figure*}

\section{Discussion}
{{{ The primary goal of this paper is to provide homogenous data of
      the cosmic star-formation history, the stellar mass density and
      dust mass density as shown in
      Figs.~\ref{fig:data_sfr},\ref{fig:data_smd}, \&
      \ref{fig:data_dmd} and Tables
      \ref{tab:sfr},~\ref{tab:smd},~\&\ref{tab:dmd}. Here we briefly
      compare these data to the outputs of simulations, as well
      attempt to build a phenomenological model of the dust evolution,
      and complete by putting the new data into the context of the
      evolution of the baryon budget.}}}

\subsection{Comparison to numerical, hydro and semi-analytic simulations}
Fig.~\ref{fig:sims} (left side panels) shows our results and selected
data (for clarity), but now compared to various curves produced from
semi-analytic or hydrodynamical models.

We show in Fig.~\ref{fig:sims} the CSFH adopted by GALFORM (Guo et
al.~2016, which uses the Gonzalez-Peres et al.~2014 version of
GALFORM; green line), EAGLE (Guo et al.~2016; mauve line), and
L-GALAXIES (Henriques et al.~2015; orange line). We see that EAGLE
tends to under-predict our CSFH at very late times but by relatively
modest amounts.  GALFORM, on the other-hand agrees very well at
intermediate ages just slightly over-predicting the CSFH in the 1-5Gyr
lookback time range. Also shown (orange line) is the latest version of
the Munich semi-analytic model (L-GALAXIES) by Henriques et
al.~(2015). This includes the Henriques et al.~(2013) prescription to
re-incorporate gas ejected from SN feedback in order to delay
star-formation in low mass galaxies, and avoid excessive build-up of
these objects at early times. In addition, it implements a slight
modified version of AGN radio model feedback, a lower star-formation
threshold and ram-pressure stripping only in clusters. Combined, these
modifications ensure that massive galaxies are predominately quenched
at earlier times while low mass galaxies have star-formation histories
more extended towards the local Universe. As can be seen this results
in a slight under-prediction at very high-redshift, compared to our
data, but a better fit locally.

Overlain on Fig.~\ref{fig:sims} (middle left side panel) are the same
three models from the GALFORM (Guo et al.~2016), EAGLE (Guo et
al.~2016), and L-GALAXIES (Henriques et al.~2015) simulations,
compared to our measured evolution of the stellar mass
density. Generally as for the CSFH, these mostly provide a good
representation to the data with the GALFORM data slightly
over-predicting the total stellar mass content at late times ---
inline with their enhanced late-time star-formation and EAGLE slightly
under-predicts. Nevertheless the agreement between the models and data
is noteworthy particularly as while the simulations are calibrated to
the $z=0$ galaxy stellar mass function, they have not been tailored to
reproduce the cosmic star-formation history or the stellar-mass
build-up in detail.

Finally overlain on Fig~\ref{fig:sims} (lower left side panel) are two
prescriptions from GALFORM for dust evolution (Lacey et al.~2016 and
Gonzalez-Perez et al.~2014). The GALFORM models presented here have
very different assumptions for the IMF. The Gonzalez-Perez et
al. (2014) model assumes a universal MW-like IMF, while Lacey et
al. (2016) introduced a top-heavier IMF during starbursts. The latter
is done to allow for quick metal enrichment at high redshift in very
star-forming galaxies, so that the model can reproduce the abundance
and luminosities of dusty galaxies. Consequently, the Lacey et
al. model predicts a larger dust abundance at higher redshift compared
to the Gonzalez-Perez et al. model. Towards lower redshift this
inverts due to the Lacey et al. model predicting lower gas fractions
of galaxies than the Gonzalez-Perez et al. model (see Lagos et
al. 2014 for a comparison of the cosmic gas density in these two
models), and under the model assumption of a metallicity dependent
dust-to-gas ratio, this directly translates into a lower dust
abundance.  Nevertheless both curves match the data reasonably well
across the full redshift range despite the relatively sweeping
assumptions adopted in the dust prescriptions.

A more dust-focused model by Popping et al.~(2016), which advocates
dust ejection, marginally under-predicts at early times and
significantly over-predicts at later times. In their model dust is
lost into the CGM and ejected, but is still forming dust faster than
it is being ejected, resulting in an upward trend in the predicted dust
density. Our measurements should include both the ISM and close-in CGM
dust, but would miss fully ejected material. However more critical is
the decline that we see which, as it derives mainly from the more
massive systems, strongly supports the notion of dust destruction
over ejection.

Overall the simulations seem to be doing pretty well in reproducing
the cosmic star-formation history, the stellar mass build-up, and dust
density evolution with the largest concern being, curiously, the
redshift zero star-formation rates which are over or under predictions
compared to our measurements for GALFORM and EAGLE respectively, by
$\sim 50$ per cent. Conversely L-GALAXIES manage to match the
very-high, intermediate, low-z Universe extremely well but just
slightly under-predict our CSFH measurements at very early times
(i.e., $2.5 < z < 4$).

{{{ We finish this section by advocating that the data presented
      here (Tables \ref{tab:sfr},~\ref{tab:smd},~\&~\ref{tab:dmd}),
      represents the first homogeneous sample of the CSFH, SMD and DMD
      over all time, and therefore optimal to inform (calibrate)
      future numerical, hydrodynamical, and semi-analytic
      simulations.}}}

\subsection{Phenomenological modelling}
Fig.~\ref{fig:sims} (right side panels) essentially encodes the
life-cycle of stars and dust and can ultimately be used to place hard
empirical constraints on factors such as the mass-return fraction,
stripped stellar mass, dust production rate and dust destruction
rate. The key of course is the CSFH which provides an empirical-based
description of the rate at which stars form. Explicitly {\it under the
  assumption of a constant Chabrier IMF}, the stellar-mass density can
be represented by the cumulative distribution of the CSFH, modulo mass
returned to the ISM and mass-lost through ejection, stripping or other
processes. The orange line on Fig.~\ref{fig:sims} (right side panels)
represents a simple spline-fit to our CSFH where we also include the
Madau \& Dickinson (2014) data with $z>4$ and the recent Bourne et
al.~(2017) data. The spline shows a good fit to all the data
points. Integrating this spline-fit yields the cumulative stellar-mass
density excluding mass-loss.  This is shown as the dashed line in the
middle right side panel of Fig.~\ref{fig:sims} and while fitting the
data at very early lookback times clearly exceeds the measurements at
latter times. The obvious explanation is replenishment of the ISM
through stellar mass-loss, but one cannot rule out an additional
portion arising due to stellar-mass also being stripped or ejected
from the individual galaxies, i.e., that which ultimately makes up the
intra cluster- and intra-group light.

Integrating the CSFH spline to z=0.0 we find a total mass converted
into stars of $(4.2 \pm 0.2) \times 10^{8}$M$_{\odot}$Mpc$^{-3}$ (at
$z = 0.0$) compared to our measured value (see
Tables~\ref{tab:sfr},\ref{tab:smd},~\&~\ref{tab:dmd}) of $(2.1 \pm 0.4)
\times 10^{8}$M$_{\odot}$Mpc$^{-3}$ (at $z\approx0.1$). Correcting for
the slight redshift offset this implies a mass-loss factor of $(0.50
\pm 0.07)$. This is consistent with that expected for a Chabrier IMF
(see Courteau et al.~2014, figure 3) of 0.44. This agreement is
extremely good and suggests that the CSFH and stellar-mass density are
in close agreement. It also leaves a little room for stripped or
ejected stellar-mass, with implications for the optical ICL and
IHL. Pushing to the limit of the error range suggests that $<13$ per
cent of the stellar-mass is likely to be stripped, consistent with the
$<20$ per cent value determined from EBL considerations (see Driver et
al.~2016). One might also be tempted to use this as confirmation of
the Chabrier IMF but the case is not entirely clear. Switching to an
alternative IMF would result in both the CSFH and stellar-mass density
varying in slightly different ways depending exactly how the
stellar-masses were derived (i.e., via single band optical, near-IR,
colours or full SED fitting). We can, however, say that the simplistic
assumption of a universal Chabrier IMF is fully consistent with our
homogenised measurements of the CSFH and stellar-mass density based on
our MAGPHYS analysis. Adding in a mass-loss factor of 0.49 we can now
re-plot the predicted stellar-mass as the solid black line
(Fig.~\ref{fig:sims}, middle right side panel) yielding a consistent
fit throughout. The implication being $44$ per cent of this is through
normal stellar mass loss processes, and $6 \pm 7$ per cent through
stripping via merging and/or harassment.

Moving to Fig.~\ref{fig:sims} (lower right side panel) we see the dust
mass density evolution which represents relatively new territory. The
recent study by Popping, Somerville \& Galametz~(2016) argued that
dust is rapidly formed with minimal destruction and continuously
accumulates with some portion ejected into the CGM and beyond. Our
data, following on from Dunne et al.~(2011), disagrees as we see a
gradually declining dust density from early times to the present day.
While we cannot rule out a systematic upward error in our dust
measurements towards higher-z, particularly as we rely more heavily on
optical/near-IR data combined with upper-limit estimates, our data
also appear to agree reasonable well with literature constraints.  At
low redshift the data of Dunne et al. (2003, 2011), Vlahakis et
al.~(2005) appear fully consistent, as do the data based on MgII
absorbers from Menard \& Fukugita (2012). In comparison to the Bethermin
data, rescaled using a constant dust to stellar mass ratio, we do see some
discrepancy but this does include the adoption of a universal
dust-to-stellar mass ratio.

Given that we see a steady decline in the dust mass density, dust
destruction is clearly critical. This is non-controversial given the
known lack of dust in old massive Elliptical systems where ejection
processes are unlikely to be effective (because of the halos ability
to retain ejected mass). We therefore consider two simplistic toy
models. One where the dust is destroyed globally (by assigning some
dust half-life), and one where it is destroyed locally (by assigning
some survival fraction). The toy models are intended to convey the
notion that in the first scenario 

dust is formed and escapes the star-forming region but ultimately
depleted through astration or expulsion from the galaxy by either
radiation pressure or galactic winds. The second assumes that while
dust is formed in star-forming regions, the majority of it might also
be destroyed in the same location by supernova shocks. Most likely
both processes are occurring but it is convenient to ask if both
scenarioes are independently viable.

For the declining tao model
we can introduce the idea
of dust destruction via a simple exponential decay:
\begin{equation}
\mbox{i.e.,} M_D(t_z) = \sum_{t=0}^{t=tz} \epsilon \psi_*(t)e^{-\frac{t_z-t}{\tau}}
\end{equation}
where $\epsilon$ is the fraction of dust formed per unit stellar mass,
$\psi_*$ is the cosmic star-formation rate at time $t$, $\tau$ is the
dust-folding time and $M_D(t_z)$ is the dust mass remaining at time
$t_z$. A simple match to the literature data (red and mauve points),
then provides best fit parameters of:

~

$\epsilon$ = $0.004^{+0.001}_{-0.001}$

$\tau$ = $2.25^{+0.5}_{-0.5}$ Gyr

~

Note that we fit the literature data, as this simple model struggles
to match the shape of our data at around 5Gyr look back time where
there is an obvious dip (which we assume is due to CV).  The arbitrary
fitting function we have adopted, and shown by the solid black line
on Fig.~\ref{fig:sims} (lower left), appears to trace the literature data very
well. At face value this function suggests that dust destruction (or
loss), is indeed a major factor, and while a significant amount of
dust is formed the majority is either destroyed (i.e., via astration),
or lost (i.e., ejected).

In our second model we adopt a constant dust survival fraction, i.e., a constant
fraction of dust is formed during star-formation, the majority of which
is destroyed {\it in situ}. This model is shown by the orange line and
follows our data (rather than the literature data), extremely closely.
Here we have adopted a formation rate of 0.0065 dust masses for every
unit of stellar mass formed of which 5 per cent survives indefinitely.

Note that the dotted black and orange lines show the cumulative dust
formation curves (i.e., ignoring the dust destruction mechanism).
While the two scenarioes are quite different they plausibly bracket
the extreme scenarioes where dust is destroyed globally (the declining
tau model) or locally (the constant survival model). This essentially
informs us that the full set of models that incorporate both global
and local mechanisms are also likely to be plausible. Finally we note
that of course the dust formation fraction may also evolve with time
as the ambient ISM metallicity increases. Exploring these various
effects will clearly require additional data and constraints and we
leave this for further future work.

However, it is interesting (or perhaps obvious) that the two toy
models, despite their differences, both predict similar fractions for
the total dust mass formed. From our two toy models we can now infer that at
$z=0$ the total amount of dust formed is $(2.1$---$1.6) \times
10^{6}$M$_{\odot}$Mpc$^{-3}$, yet we measure a density of $(1.5 \pm
0.2) \times 10^{5}$M$_{\odot}$Mpc$^{-3}$, implying that approximately
90 to 95 per cent of all dust formed has either been destroyed (most
likely), or ejected into the IGM (less likely given our measurements
are mostly constrained by massive systems with strong gravitational
fields).

\begin{figure*}
\includegraphics[width=0.99\textwidth]{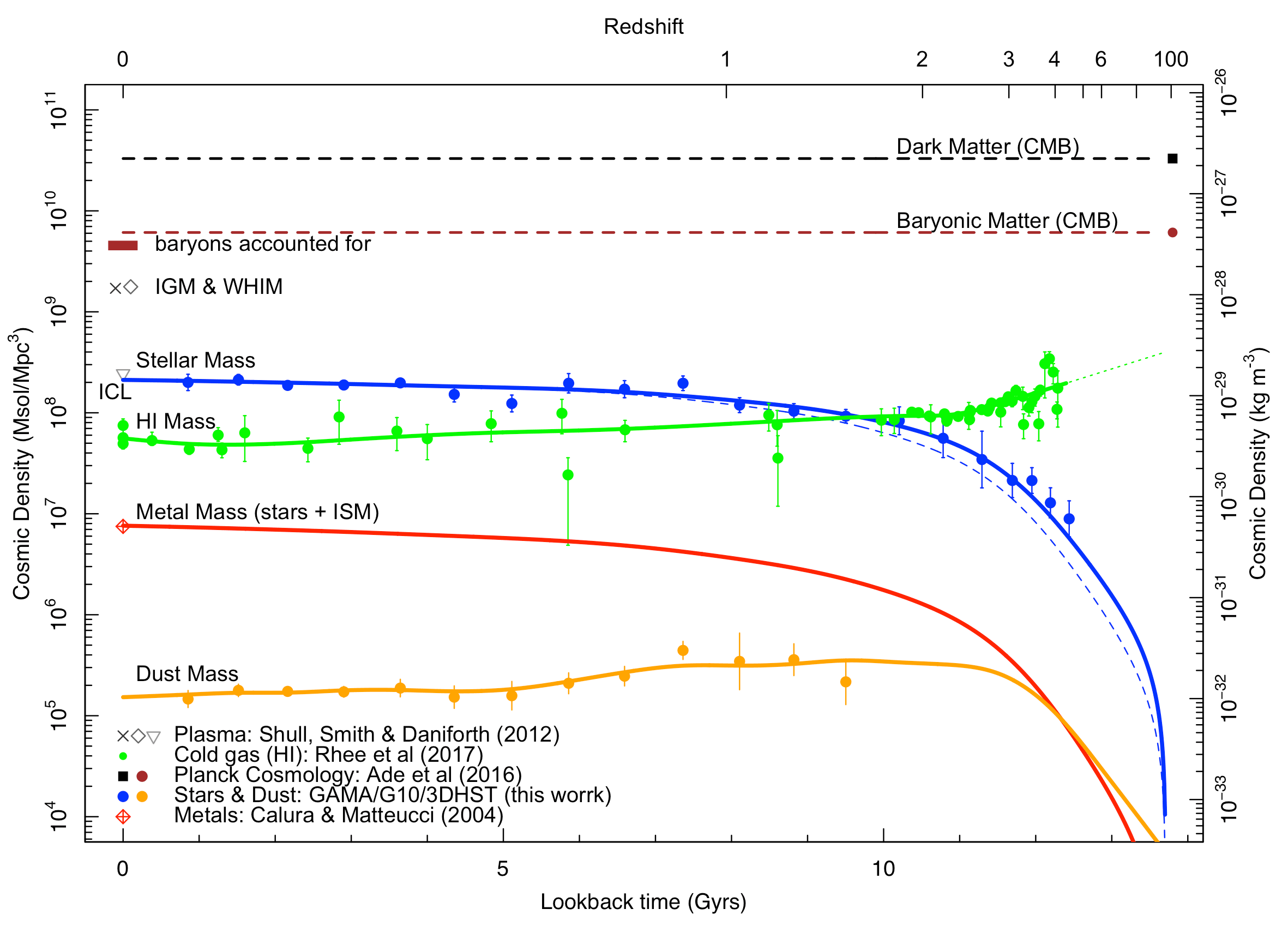}
\caption{ {{{ Our findings placed in the context of the mass budget and its evolution since the Big Bang.
\label{fig:density} }}} }
\end{figure*}

\subsection{Implications for the bound baryon budget and its evolution over time}
We return now to the opening discussion on the evolution of the
baryons and present an overview perspective in
Fig.~\ref{fig:density}. This includes the base measurements of the
dark matter density and baryon density from CMB studies (Ade et
al.~2016) as well as the IGM, WHIM and ICL values reported in Shull,
Smith \& Danforth (2012). Now we can add the stellar mass and dust
mass density derived here (blue and orange points respectively). We
can also provide a simple estimate of the metal evolution by adopting
a fixed yield and tying metal production to the total mass going into
star-formation. This we calibrate at $z=0$ by adopting the value for
the mean metallicity in the present day Universe from Calura \&
Matteucci (2005). Finally we include the cosmic HI compendium from
Rhee et al.~(2017) shown as green data. The figure paints an
interesting picture, not least of which is how the almost mirror image
between the stellar mass and cosmic HI mass trends. The simplest
interpretation is that the sum of stars and HI is constant, i.e., gas
is imply turned into stars and not replenished. However we note the
very large errors associated with the HI measurements and the need for
further measurements in the intermediate age range.

\section{Summary}
We have combined data from the GAMA, G10-COSMOS, and 3D-HST surveys to
produce a meta-catalogue which samples the stellar-mass, dust-mass and
star-formation measurements over a broad mass/star-formation range and
over all time. For the GAMA, G10-COSMOS and 3D-HST datasets we use the
popular energy-balance code MAGPHYS to determine cosmic star-formation
rates, stellar-mass, and dust-mass densities. These values compare
well with previous estimates and now constitute a homogenous set of
measurements across the full timeline of the Universe.

We construct histograms of the star-formation, stellar-mass, and
dust-mass space-densities and their associated first moments which
represent their differential contribution to the overall mass and
star-formation budgets. After identifying appropriate limits for each
dataset we use a 7-point spline to simultaneously fit all three datasets
weighting by the inverse errors squared. In general the three datasets
show good agreement and the spline fits are well behaved allowing us
to determine the cosmic star-formation rate, stellar-mass and dust-mass
densities reliably over a consistent mass and star-formation range and
over a broad range in redshift.

The resulting cosmic star-formation density, stellar-mass density, and
dust-mass density versus lookback time agree reasonably well with
previous estimates but constitute a major advancement in terms of
homogeneity, and sample-size. In particular at intermediate lookback
times the combination of GAMA with G10-COSMOS and 3D-HST enables
robust sampling of the full dynamic range of masses and star-formation
rates with GAMA defining the high mass, high star-formation rate end
with high significance and G10-COSMOS defining the low
mass/star-formation rate end. As a consequence the scatter implied by
our measurement is a significant advancement over the previous
compendium of disparate datasets. In particular the stellar-mass
density and star-formation densities are now known over a 12 Gyr period
with an uncertainty at any time interval of $< \sim \pm 30$ per cent. The
data are provided in Tables~\ref{tab:sfr},~\ref{tab:smd},~\&~\ref{tab:dmd}.

Our measurement of the dust mass density over lookback time represents
the first tentative study of dust over such a broad range of time and
builds on the earlier HAtlas-SDP study of Dunne et al.~(2011). While
our data formally agree over the lookback range in common within the
quoted errors we do not agree with the conclusion of a rapid declining
dust mass density. Instead we find a steady but shallower decline in
the dust density from a peak somewhere between 7 --- 10 Gyrs ago. We
note that the hydro and semi-analytic models tested here also
reproduce reasonably well the cosmic star-formation history, the
stellar-mass dust build-up, and the evolution in the dust mass
density.

We attempt to explore consistency across the three density
distributions. Starting with a spline-fit to our cosmic star-formation
history we find that we predict the stellar mass density at later
times in agreement with the findings of Madau \& Dickinson~(2014)
recovering a stellar-mass replenishment factor of ($0.50\pm0.07$), consistent
with our adopted Chabrier IMF and some addition stellar-mass loss
through stripping ($0.06 \pm 0.07$).

Again starting from our CSFH spline-fit we show two simple toy models
where dust formation is closely linked to the cosmic star-formation
history and dust destruction follows either an exponential decline or
a constant survival fraction. From our models, that bracket the range
of data, we conclude that for every unit of stellar mass that is
formed 0.0065 --- 0.0040 units of dust mass is formed and that over
the lifetime of the Universe approximately 90 --- 95 per cent of all
dust which has formed is unaccounted for: either it is destroyed,
ejected or a combination of both.

{{{
Finally we show how our results mesh with the unfolding picture of the
evolution of the bound baryons from the early Universe to the present
day with the rise in stellar mass appearing to mirror the evolution of
the cosmic HI density, but with low significance due to the inherent
uncertainties.}}}

This work is a preliminary step which demonstrates the power of
linking comprehensive high quality datasets which each sample distinct
regions of the mass-redshift plane. Obvious improvements to this work
include improved measurements of far-IR fluxes, particularly for the
G10-COSMOS and the 3D-HST fields, both through improved analysis but
also improved observations such as might become available through the
proposed future NASA Origins mission.

\section*{acknowledgments}
GAMA is a joint European-Australasian project based around a
spectroscopic campaign using the Anglo-Australian Telescope. The GAMA
input catalogue is based on data taken from the Sloan Digital Sky
Survey and the UKIRT Infrared Deep Sky Survey. Complementary imaging
of the GAMA regions is being obtained by a number of independent
survey programmes including GALEX MIS, VST KiDS, VISTA VIKING, WISE,
Herschel-ATLAS, GMRT and ASKAP providing UV to radio coverage. GAMA is
funded by the STFC (UK), the ARC (Australia), the AAO, and the
participating institutions. The GAMA website is
http://www.gama-survey.org/ . Based on observations made with ESO
Telescopes at the La Silla Paranal Observatory under programme ID
179.A-2004.
 
The G10-COSMOS redshift catalogue, photometric catalogue and cutout
tool uses data acquired as part of the Cosmic Evolution Survey
(COSMOS) project and spectra from observations made with ESO
Telescopes at the La Silla or Paranal Observatories under programme ID
175.A-0839. The G10-COSMOS cutout tool is hosted and maintained by funding
from the International Centre for Radio Astronomy Research (ICRAR) at
the University of Western Australia. Full details of the data,
observation and catalogues can be found in Davies et al. (2015) and
Andrews et al. (2017a), or on the G10-COSMOS website:
cutout.icrar.org/G10/dataRelease.php

This work is based on observations by the 3D-HST Treasury Program (GO
12177 and 12328) with the NASA/ESA HST, which is operated by the
Association of Universities for Research in Astronomy, Inc. under NASA
contract NAS5-26555.

This work was supported by resources provided by the Pawsey
Supercomputing Centre with funding from the Australian Government and
the Government of Western Australia.

SKA is supported by an Australian Postgraduate Award. LD and SJM
acknowledge support from European Research Council (ERC) in the form
of Advanced Investigator grant COSMICISM, and Consolidator grant Cosmic
Dust.

\section*{References}

\reference Planck Collaboration: Ade P.A.R. et al. 2016, A\&A, 594, 13 

\reference Ahn C.P., et al., 2014, ApJS, 211, 17

\reference Andrews S.K., Driver S.P., Davies L.J.M., Kafle P.R.,
Robotham A.S.G., Wright A.H., 2017a, MNRAS, 464, 1579

\reference Andrews S.K., et al., 2017b, MNRAS, in press

\reference Andrews S.K., et al., 2017c, MNRAS, submitted

\reference Baldry I.K., Glazebrook K., 2003, ApJ, 593, 258

\reference Baldry I.K., Glazebrook \& Driver, 2008, MNRAS, 388, 945

\reference Baldry I.K., et al., 2010, MNRAS, 404, 86

\reference Baldry I.K., et al., 2012, MNRAS, 412, 621

\reference Baldry I.K., et al., 2014, MNRAS, 441, 2440

\reference B\'ethermin M., et al., 2014, MNRAS, 567, 103

\reference Beeston R., et al., 2017, MNRAS, in prep.

\reference Bertin E., 2011, ASPC, 442, 435

\reference Bonamente M., Nevalainen J.M., Tilton E., Liivam\"age, Tempel E., Hein\"am\"aki, Fang T., MNRAS, 457, 4236

\reference Bondi M., Ciliegi P., Schinnerer R., Smolcic V., Jahnke K., Carilli C., Zamorani G., et al., 2008, ApJ, 681, 1129

\reference Bourne N., et al. 2016, MNRAS, 462, 1714

\reference Bourne N., et al. 2017, MNRAS, in press (arXiv:1607.04283)

\reference Bouwens R.J., et al. 2012a, ApJ, 752, 5

\reference Bouwens R.J., et al. 2012b, ApJ, 754, 83

\reference Burgarella D., et al., 2013, A\&A, 554, 70

\reference Brammer G., et al., 2012, ApJ, 758, 17

\reference Calura F., Matteucci F., 2004, MNRAS, 350, 351

\reference Capak P., et al., 2007, ApJS, 172, 99

\reference Chabrier G., 2003, PASP, 115, 763

\reference Clemens M.S., et al., 2013, MNRAS, 433, 695 

\reference Colless M., et al., 2001, MNRAS, 328, 1039

\reference Courteau, S et a., 2014, RvMP, 86, 47

\reference Cool R.J., et al., 2013, ApJ, 767, 118

\reference Cucciati O., et al., 2012, A\&A, 539, 31

\reference Cyburt R.H., Fields B.D., Olive K.A., Yeh T-H., 2016, 88, 5004

\reference da Cunha E., Charlot S., Elbaz, D., 2008, MNRAS, 388, 1595

\reference Danforth C.W., et al., 2016, ApJ, 817, 111

\reference Dariush A., et al., 2016, MNRAS, 456, 2221

\reference Davies L.J.M., Driver S.P., Robotham A.S.G., Baldry I.K.,
Lange R., Liske J., Meyer M., Popping A., Wilkins S.M., Wright A.H.,
2015, MNRAS, 447, 1014

\reference Davies L.J.M., et al., 2016, MNRAS, 461, 458

\reference Decarli R., et al., 2016, ApJ, 833, 69

\reference Delhaize J., Meyer M.J., Staveley-Smith L., Boyle B.J., 2013, MNRAS, 433, 1398

\reference Donley J.L., et al., 2012, ApJ, 748, 142

\reference Driver S.P., Robotham A.S.G., 2010, MNRAS, 407, 2131

\reference Driver S.P., et al., 2009, A\&G, 50, 12

\reference Driver S.P., Popescu C.C., Tuffs R., Liske J., Graham A.W., Allen P.D., de Propris R., 2007, MNRAS, 379, 1022

\reference Driver S.P., et al., 2012, MNRAS, 427, 3244

\reference Driver S.P., et al., 2011, MNRAS, 413, 971

\reference Driver S.P., et al., 2016, ApJ, 827, 108

\reference Duncan et al, 2014, MNRAS, 444, 2960

\reference Dunne, L., Eales, S.A., Edmunds M.G., 2003, MNRAS< 341, 589

\reference Eales S., et al., 2010, PASP, 122, 499 

\reference Eckert D., et al., 2015, Nature, 528, 105

\reference Fukugita, Hogan \& Peebles, 1998, ApJ, 503, 518

\reference Gall C., Hjorth J., Watson D., Dwek E., Manaund J.R., Fox, O., Leloudas G., Malesani D., Day-Jones A.C., 2014, Nature, 511, 326

\reference Gonzalez-Perez V., Lacey C.G., Baugh C.M., Helly C.D.P., Campbell D.J.R., Mitchell P.D., 2014, MNRAS, 439, 264

\reference Graham A., Driver S.P., Allen P.D., Liske J., 2007, MNRAS, 378, 198

\reference Grazian A., et al., 2015, A\&A, 575, 96

\reference Grootes M., et al., 2013, ApJ, 766, 59

\reference Grootes M., et al., 2016, MNRAS, submitted

\reference Guo Q., et al, 2016, MNRAS, 461, 3457

\reference Gunawardhana M., et al., 2013, MNRAS, 433, 2764

\reference Gunawardhana M., et al., 2011, MNRAS, 415, 1647

\reference Hayward C.C., Smith D.J.B., 2015, MNRAS, 446, 1512 

\reference Henriques B., White S.D.M., Thomas P.A., Angulo R.E., Guo
Q., Lemson G., Springer V., 2013, MNRAS, 431, 3373

\reference Henriques B., White S.D.M., Thomas P.A., Angulo R., Guo Q.,
Lemson G., Springel V., Overzier R., 2015, MNRAS, 451, 2663

\reference Hernandez-Monteagudo C., Ma, Y-Z., Kitaura F.S., Wang, W., Genova-Santos R., Macias-Perez J, Herranz, D., 2015, PhRvL, 115, 1301

\reference Hinshaw G.F., et al., 2013, ApJS, 208, 19

\reference Holwerda B., Dalcanton J.J., Radburn-Smith D., de Jong R.S., Guhathakurta P., Koekemoer A., Allen R.J., B\"oker T., 2012, ApJ, 753, 25

\reference Hopkins A.M., Beacom  J.F., 2006, ApJ, 651, 142

\reference Hopkins A.M., et al., 2013, MNRAS, 430, 2047

\reference Hurley P.D., et al., 2016, MNRAS, in press (arXiv:1606.05770)

\reference Keres D., Yun M.S., Young J.S., 2003, ApJ, 582, 659

\reference Kriek M., et al., 2009.~2009, ApJ, 700, 221

\reference Kroupa P., 2001, MNRAS, 322, 231

\reference Lagos C., Baugh C.M., Lacey C.G., Benson A.J., Kim H-S., Power C., 2011, MNRAS, 418, 1649

\reference Lacey C., et al., 2016, MNRAS, 462, 3854

\reference Laigle C., et al., 2016, ApJS, 224, 24

\reference Le Fevre O., et al., 2013, A\&A, 559, 14

\reference Lilly S.J., Le Fevre O., Hammer F., Crampton D., 1996, ApJ, 460, 1

\reference Liske J., et al., 2015, MNRAS, 452, 2087

\reference Madau P., Dickinson M., 2014, ARA\&A, 52, 415

\reference Martin A.M., Papastergis E., Giovanelli R., Haynes M.P., Springob C.M., Stierwalt S., 2010, ApJ, 723, 1359

\reference Martin C., et al. 2005, ApJ, 619, 1

\reference Menard B., Fukugita M., 2012, ApJ, 754, 116 

\reference Moffett A., et al., 2016, MNRAS, 462, 4336

\reference Momcheva I., et al., 2016, ApJS, 225, 27

\reference Mutlu Pakdil M., Seigar M.S., David B.L., 2016, in press (arXiv:1607.07325)
 
\reference Oliver S., et al., 2012, MNRAS, 424, 1614 

\reference Peng Y., et al., 2010, ApJ, 721, 193

\reference Popping G., Somerville R.S., Galametz M., 2016, MNRAS, submitted (arXiv:1609.08622)

\reference Rhee, J., et al., 2017, MNRAS, in press

\reference Robotham A.S.G., Driver S.P., 2011, MNRAS, 413, 2570

\reference Rowlands K., Gomez H.L., Dunne L., Aragon-Salamanca A., Dye
S., Maddox S., da Cunha E., van der Werf P., 2014, MNRAS, 441, 1040

\reference Sargent B., et al., 2010, ApJ, 878

\reference Schenker M.A., et al., 2012, ApJ, 744, 179

\reference Schinnerer R., et al., 2007, ApJS, 172, 46

\reference Scoville N., et al., 2007a, ApJS, 172, 1

\reference Scoville N., et al., 2007b, ApJS, 172, 38

\reference Seymour N., et al., 2008, MNRAS, 386, 1695

\reference Shankar F., Salucci P., Granato G.L., De Zotti G., Danese L., 2004, MNRAS, 354, 1020

\reference Skelton R.E., et al., 2014, ApJS, 214, 24
 
\reference Shull J.M., Smith B.D., Danforth C.W., 2015, ApJ, 811, 3

\reference Smith D.J.B., et al., 2012, MNRAS, 427, 703

\reference Song M., et al., 2016, ApJ, 825, 5

\reference Stocke J.T. et al., 2013, ApJ, 763, 148

\reference Sutherland W.J., 2015, A\&A, 575, 25

\reference Taniguchi Y., et al., 2007, ApJS, 172, 9

\reference Valinate E., et al., 2016, MNRAS, 462, 3146

\reference van Dokkum P., et al., 2013, ApJ, submitted (arXiv:13052140)

\reference Vlahakis C., Dunne L., Eales D., 2005, MNRAS, 364, 1253

\reference Vika, M., Driver S.P., Graham A.W., Liske J., 2009, MNRAS, 400, 1451

\reference Walter F., et al., 2014, ApJ, 782, 79

\reference Weingartner J.C., Draine B.T., 2001, ApJS, 134, 263

\reference Whitaker K.E., 2014, ApJ, 795, 104

\reference Wilkins S.M., Trentham N., Hopkins A.M., 2008, MNRAS, 385, 687

\reference Wright A.H., et al., 2016, MNRAS, 460, 765

\reference Wright A.H., et al., 2017a, MNRAS, submitted

\reference Wright A.H., et al., 2017b, MNRAS, in prep

\reference Wright E.L., et al., 2010, AJ, 140, 1868

\reference York D., et al., AJ, 2000, 120, 1579

\reference Zwaan M., Meyer M.J., Staveley-Smith L., Webster R.L., 2005, MNRAS, 359, 30

\pagebreak

\appendix

\section{Expanding the G10-COSMOS far-IR sampling}
The original G10-COSMOS catalogue described in Andrews et al.~(2016)
used a fairly aggressive cascade in the far-IR to pre-determine which
objects should be measured.  Essentially an attempt to measure a flux
using the LAMBDAR code was only made for a small fraction of the
objects. This is because of the much lower comparative signal-to-noise
of the HerMES PACS and SPIRE data as compared to the very deep Subaru
data. In earlier attempts to measure all 170,000 systems from the
HerMES data we found that the density of faint objects led to numerous
instances of overlapping apertures (in fact every pixel has on average
10 overlapping apertures). The net result is that flux sharing codes
such as LAMBDAR, when confronted with very poor resolution data and
multiple targets within a single pixel, end up averaging the flux
across all the objects and eroding the bright systems by
redistributing their flux to the fainter objects. While LAMBDAR has a
process to mitigate this, like any code, it has limits and when
flooded with an excessive number of faint objects in very poor
resolution data LAMBDAR simply fails. The way this was addressed in
Andrews et al.~(and also with XID+ in Hurley et al.~2016) is to
introduce a prior via a cascade process, whereby only objects with
likely measurable fluxes were measured. In Andrews et al.~this was
implemented via a cascade process whereas only objects with clear
MIPS24$\mu$m data were passed forward for measurement. Those objects
which had solid detections in PACS 100 $\mu$m or 160 $\mu$m were then
passed on for SPIRE250 $\mu$m measurements and only those with
detections in SPIRE250 $\mu$m passed on for SPIRE350 measurements
etc. The net result is that of the initial 170,000 objects in the full
G10-COSMOS i-band selected catalogue only 11925 had attempted measurements in
MIPS24, PACS100 and PACS160 bands reducing to 7178, 3446 and 2636 in
the SPIRE bands. Hence ultimately measurements were only attempted for
1.55 per cent of the original input catalogue. In review we believe
this cascade was overly aggressive and in our first pass MAGPHYS
catalogue many systems were found with predicted dust masses which
should be detectable. We therefore decided to revisit the inputs to
our LAMBDAR analysis and remeasure PACS and SPIRE fluxes for all
objects with an initial $\log_{10}[\frac{M_d}{d_l^2}] > 0.0, 0.35$ or
$0.5$ resulting in samples of the dustiest 24k, 12k or 6k objects. We
ran all three catalogues through our LAMBDAR code on the PACS and
SPIRE PEPS and HerMES data and reviewed the outputs. Comparing the
output aperture masks to the input image we established that the 24k
catalogue was appropriate for PACS and SPIRE 250, the 12k catalogue
appropriate for SPIRE 350 and the 6k catalogue appropriate for
SPIRE500.  These revised output catalogues were then spliced into
G10CosmosLAMBDARCatv05 (Andrews et al.~2016) to make
G10CosmosLAMBDARCatv06 which was used in this work.

\section{Extended analysis plots}
\begin{figure}
\begin{center}
\includegraphics[width=\columnwidth]{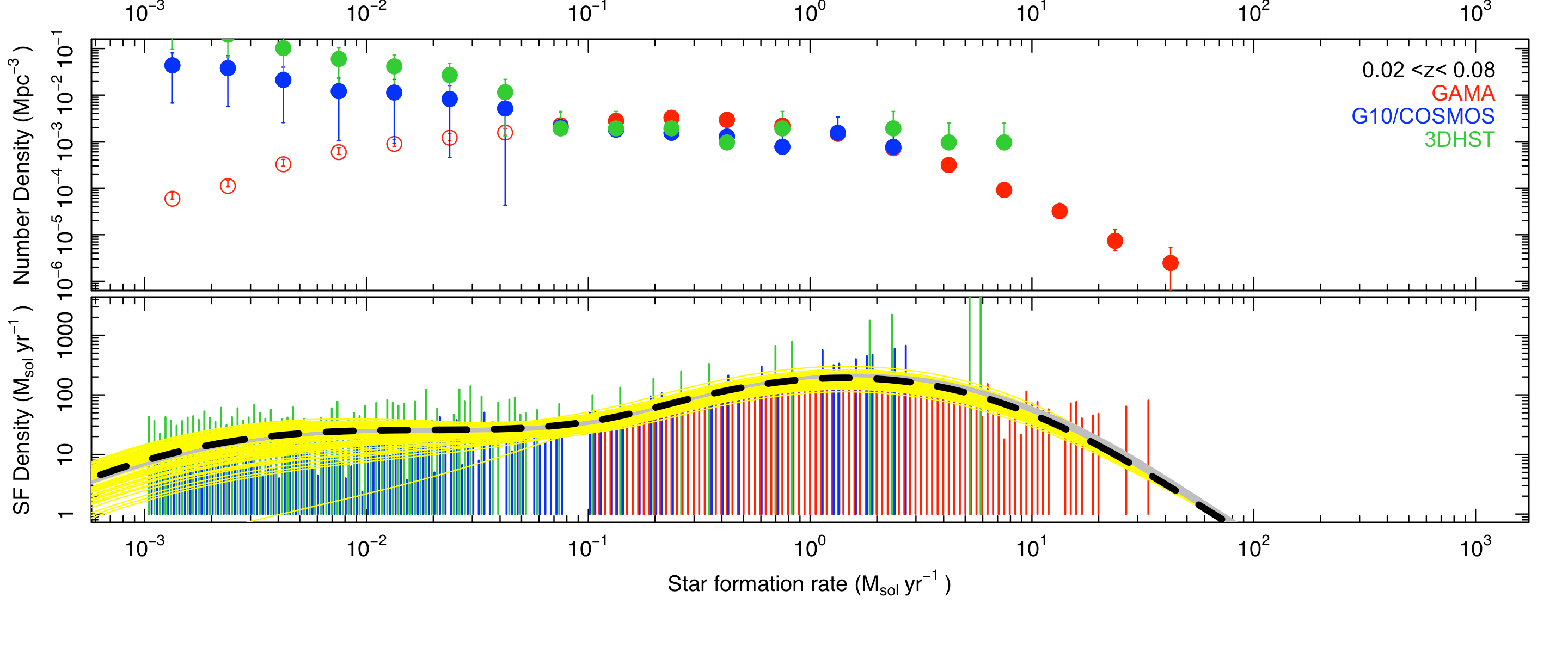}
\vspace{-0.5cm}
\includegraphics[width=\columnwidth]{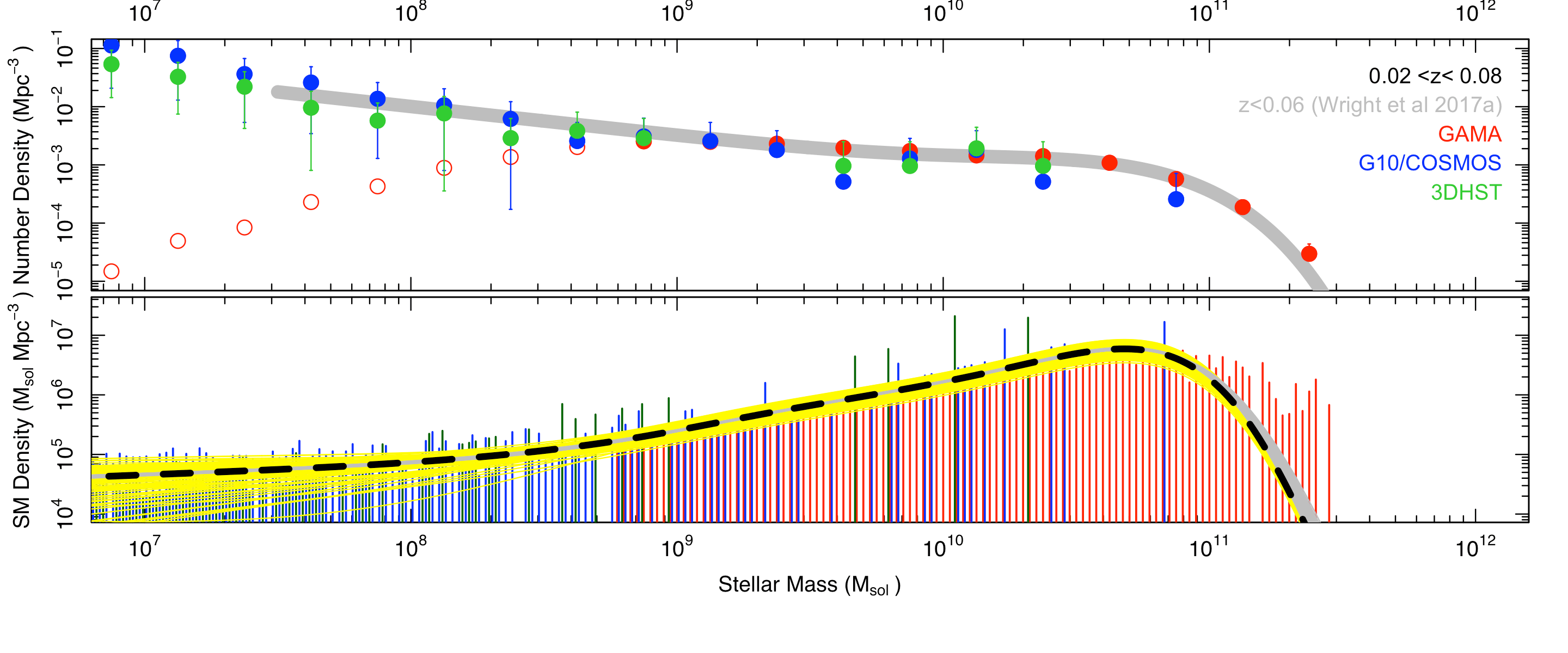}
\vspace{-0.5cm}
\includegraphics[width=\columnwidth]{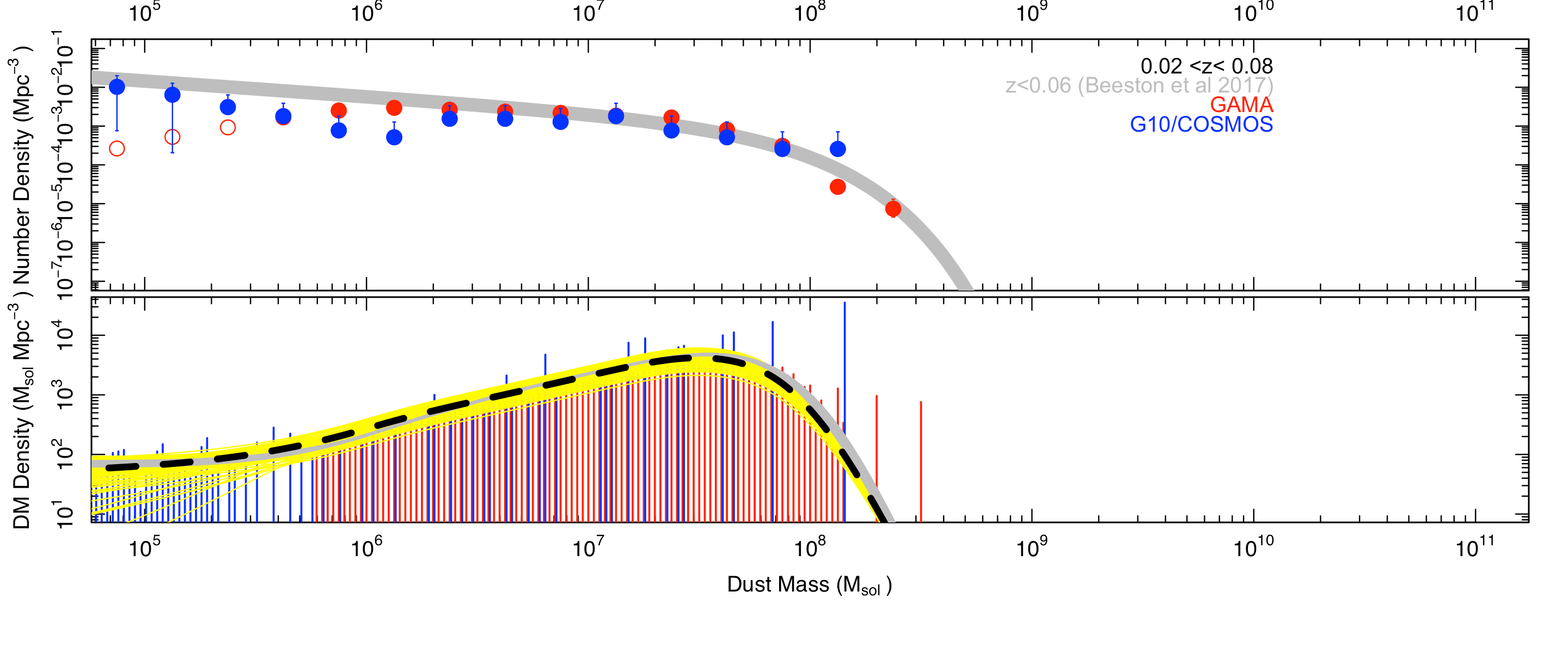}
\vspace{-0.5cm}
\caption{As for Fig.~\ref{fig:methods} except for the redshift range indicated. \label{fig:methods1}}
\end{center}
\end{figure}

Figs~\ref{fig:methods1} to B18 show the analysis fits in all redshift bands for the cosmic star-formation (upper panels), stellar mass (middle panels), and dust mass (lower panels) distributions. 

\begin{figure}
\begin{center}
\includegraphics[width=\columnwidth]{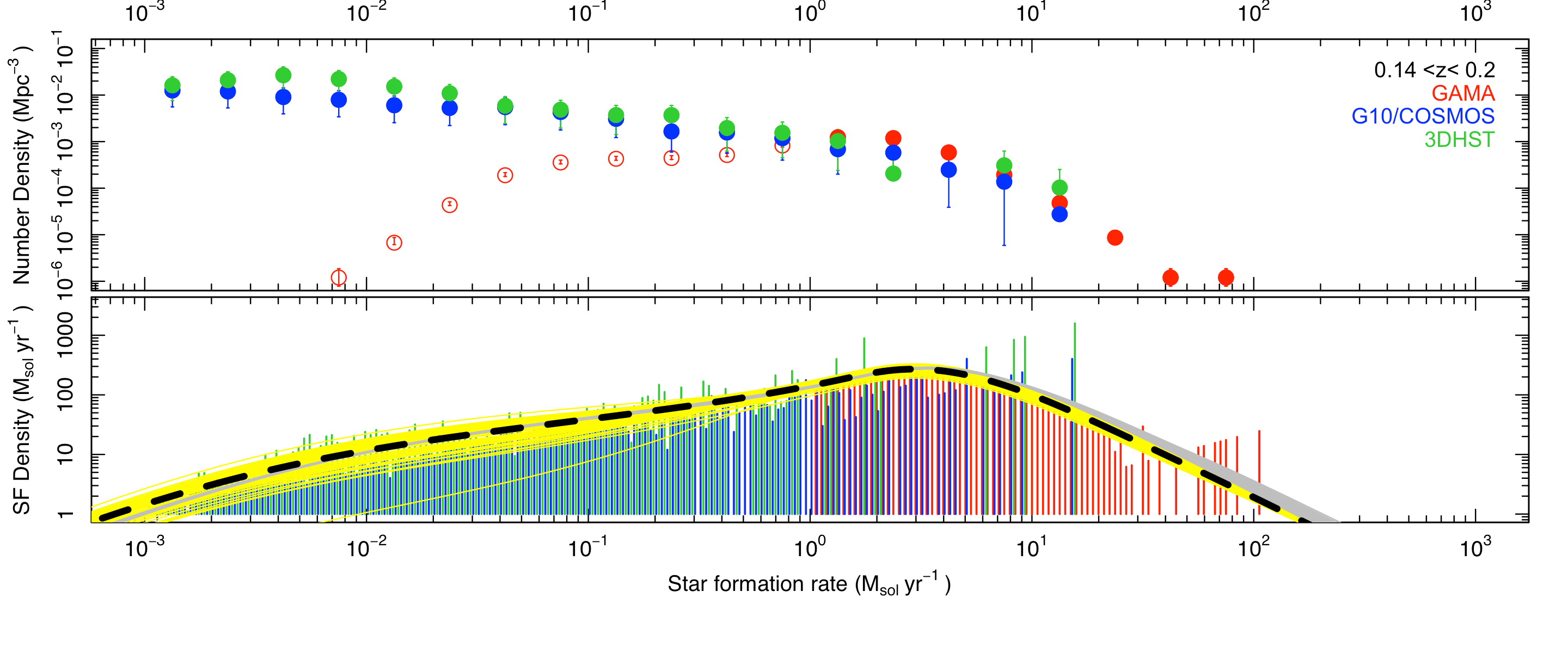}
\vspace{-0.5cm}
\includegraphics[width=\columnwidth]{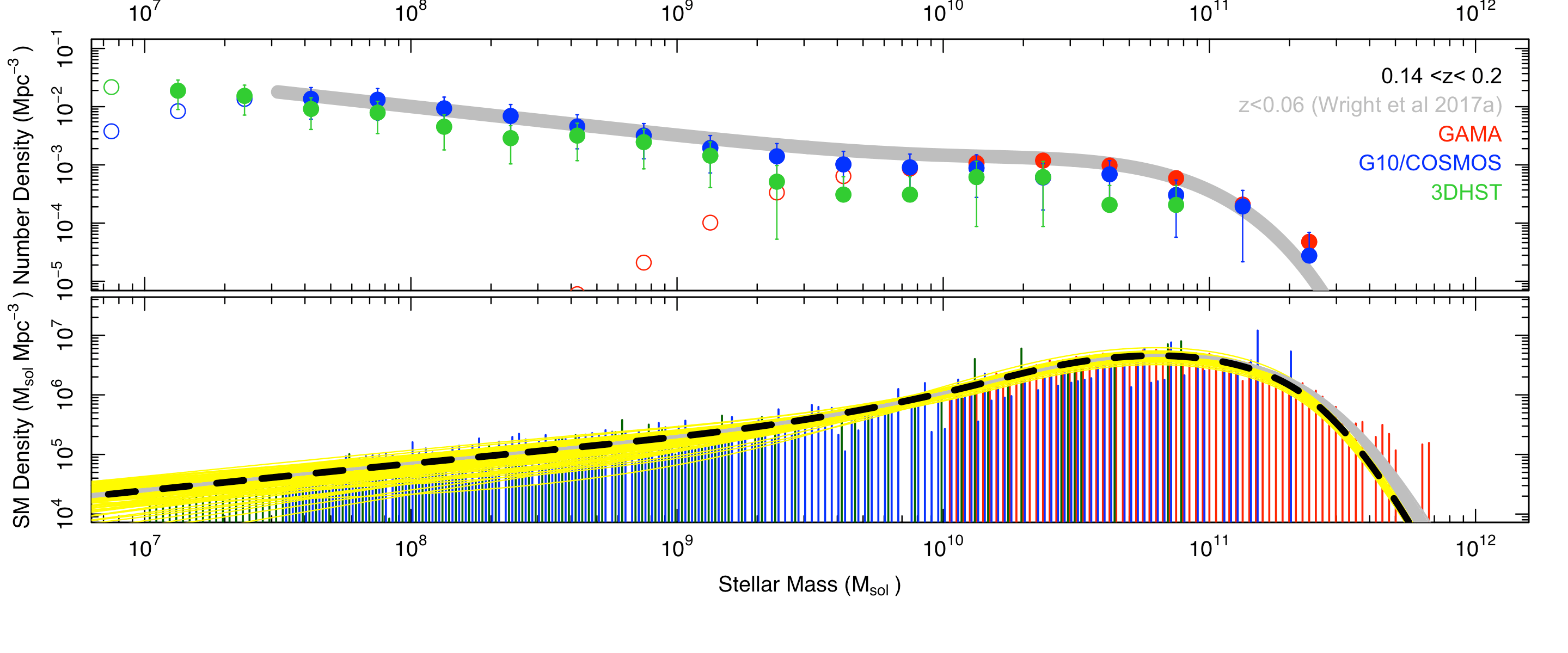}
\vspace{-0.5cm}
\includegraphics[width=\columnwidth]{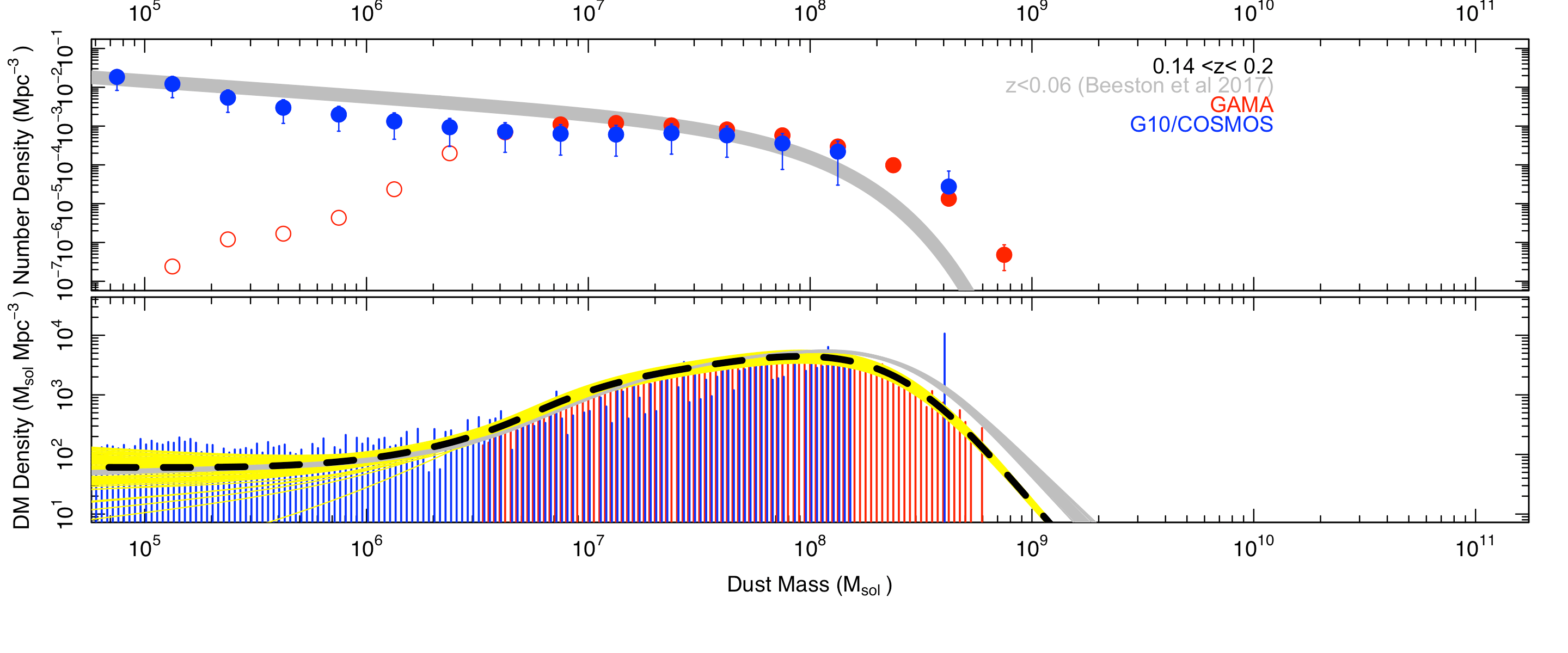}
\vspace{-0.5cm}
\caption{As for Fig.~\ref{fig:methods} except for the redshift range indicated.}
\end{center}
\end{figure}

\clearpage

\begin{figure}
\begin{center}
\includegraphics[width=\columnwidth]{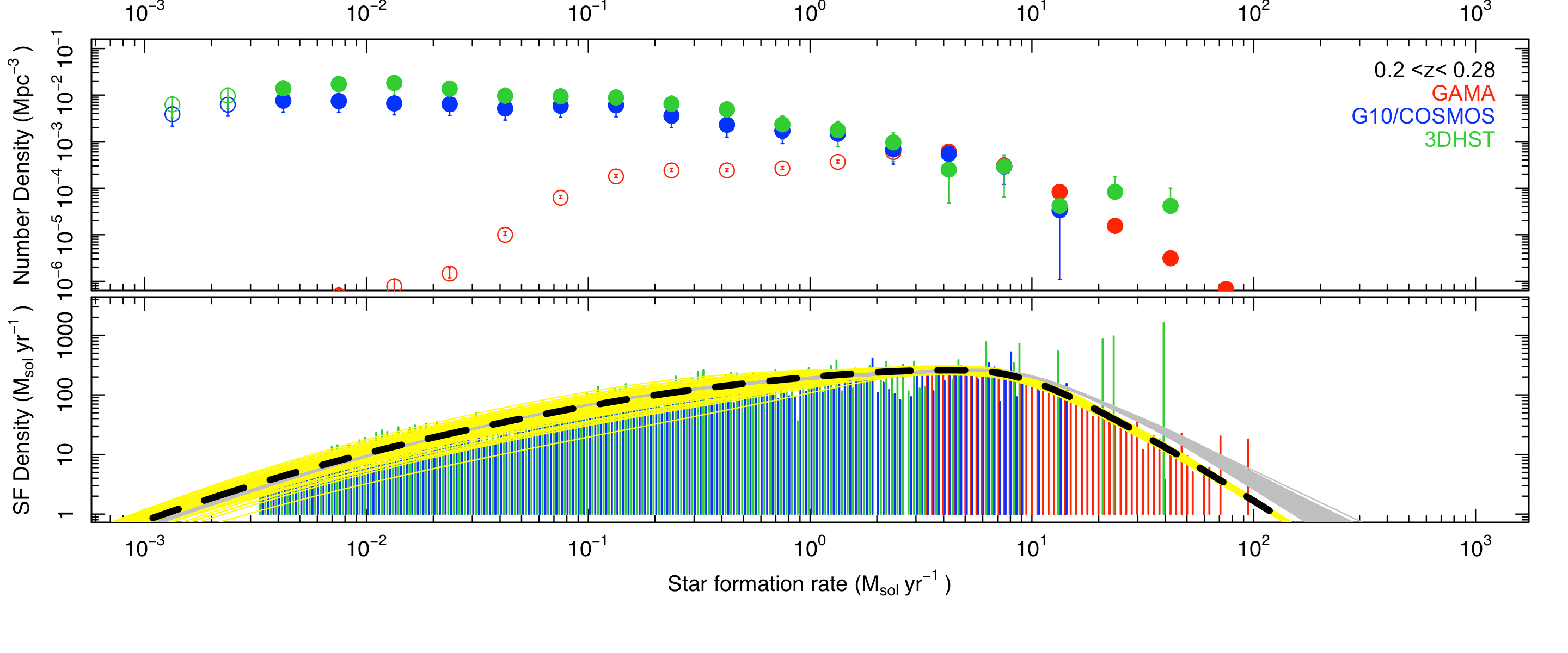}
\vspace{-0.5cm}
\includegraphics[width=\columnwidth]{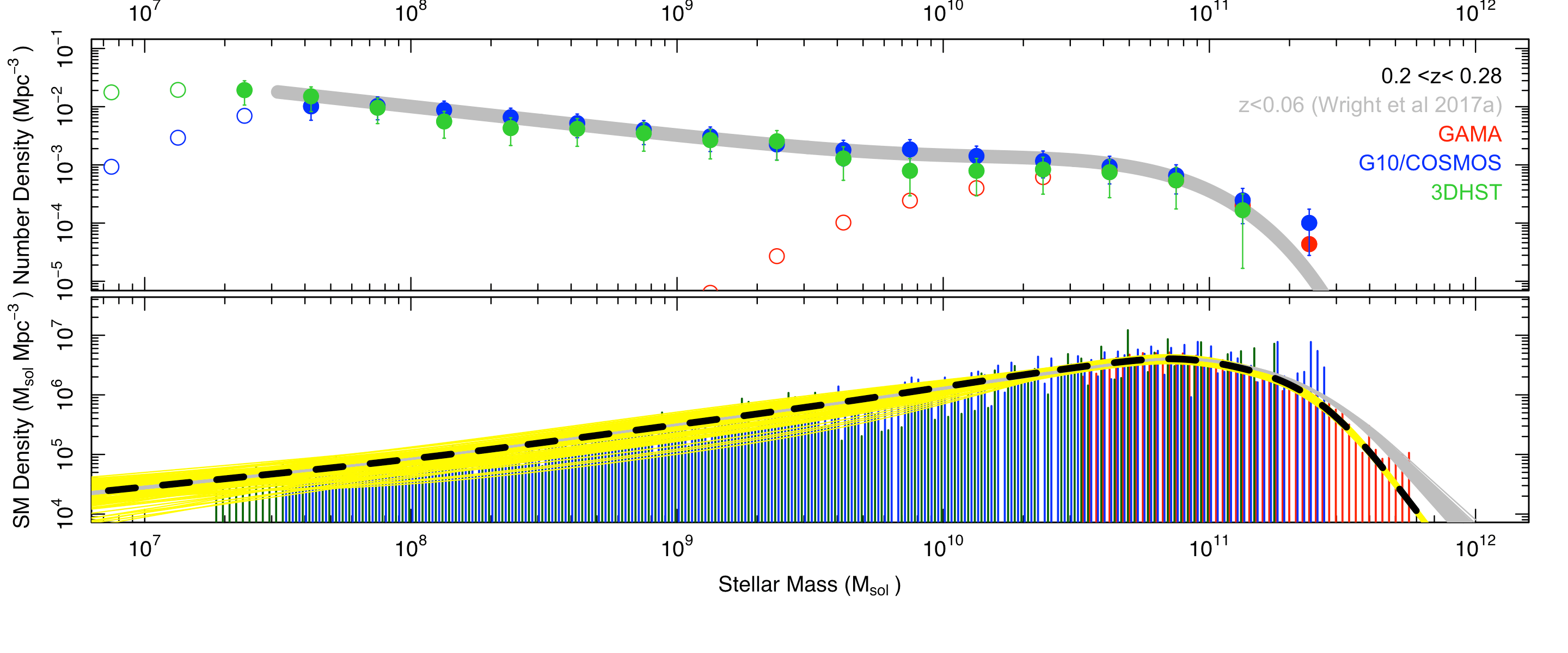}
\vspace{-0.5cm}
\includegraphics[width=\columnwidth]{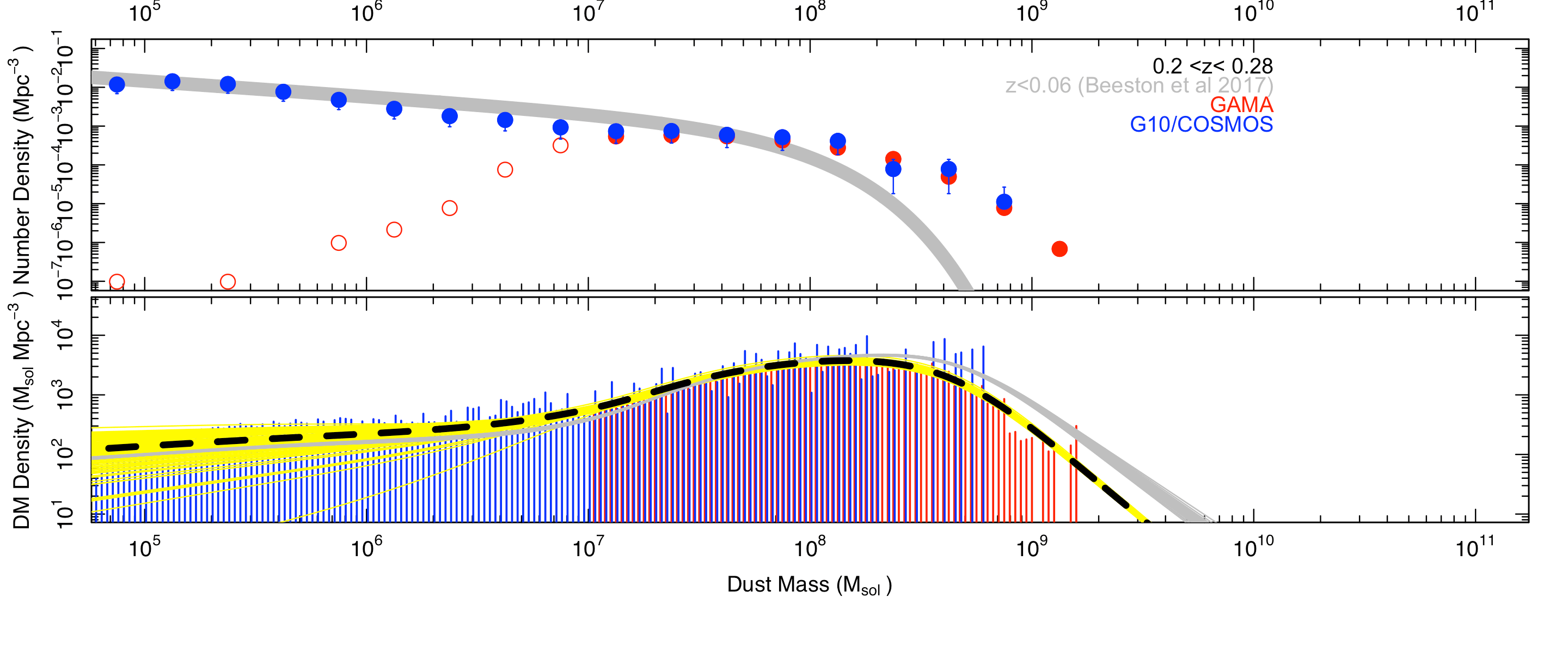}
\vspace{-0.5cm}
\caption{As for Fig.~\ref{fig:methods} except for the redshift range indicated.}
\end{center}
\end{figure}


\begin{figure}
\begin{center}
\includegraphics[width=\columnwidth]{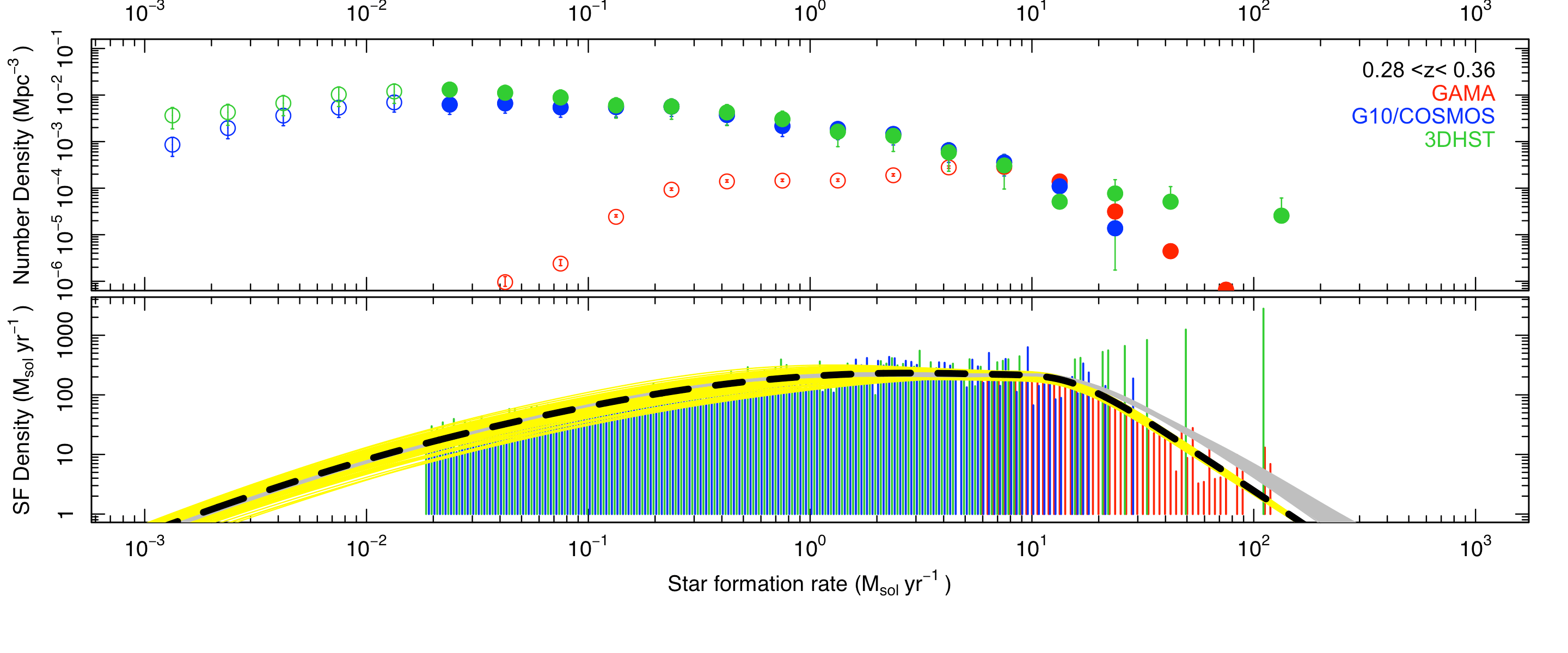}
\vspace{-0.5cm}
\includegraphics[width=\columnwidth]{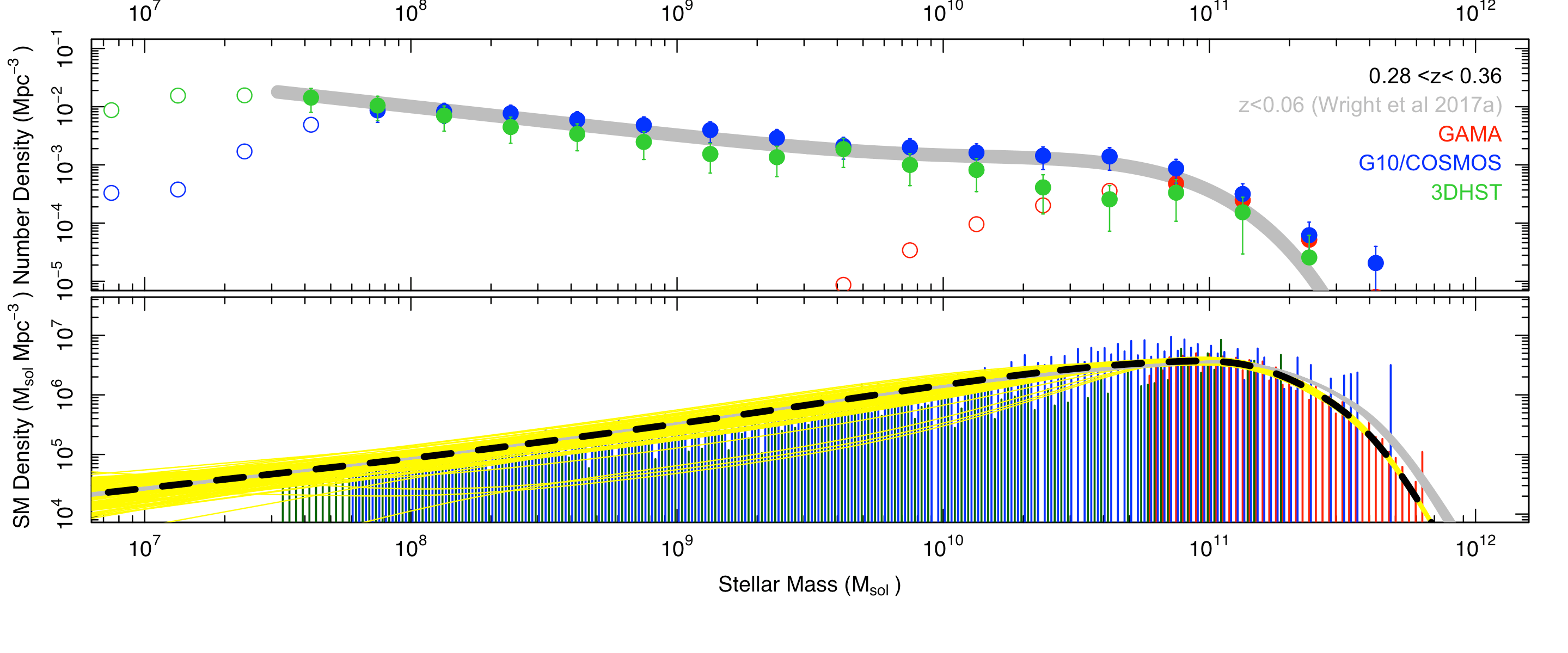}
\vspace{-0.5cm}
\includegraphics[width=\columnwidth]{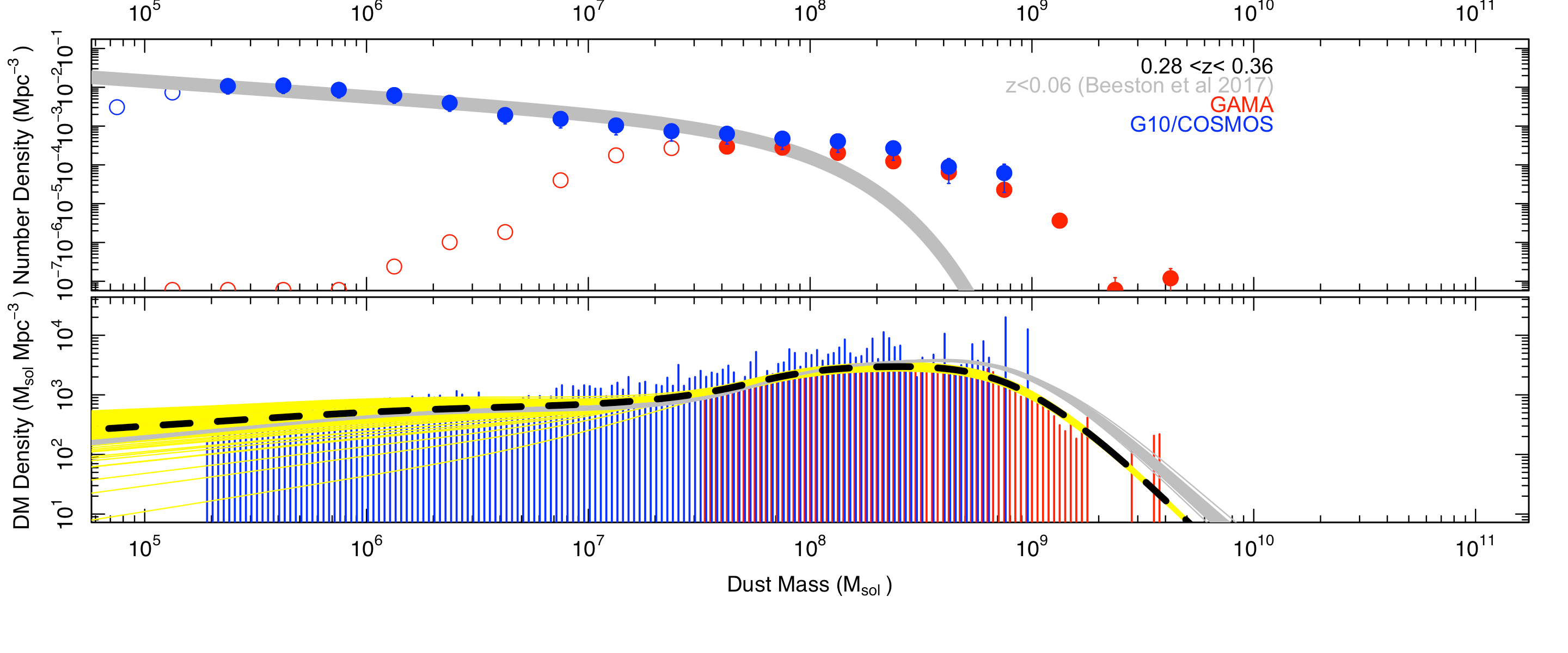}
\vspace{-0.5cm}
\caption{As for Fig.~\ref{fig:methods} except for the redshift range indicated.}
\end{center}
\end{figure}


\begin{figure}
\begin{center}
\includegraphics[width=\columnwidth]{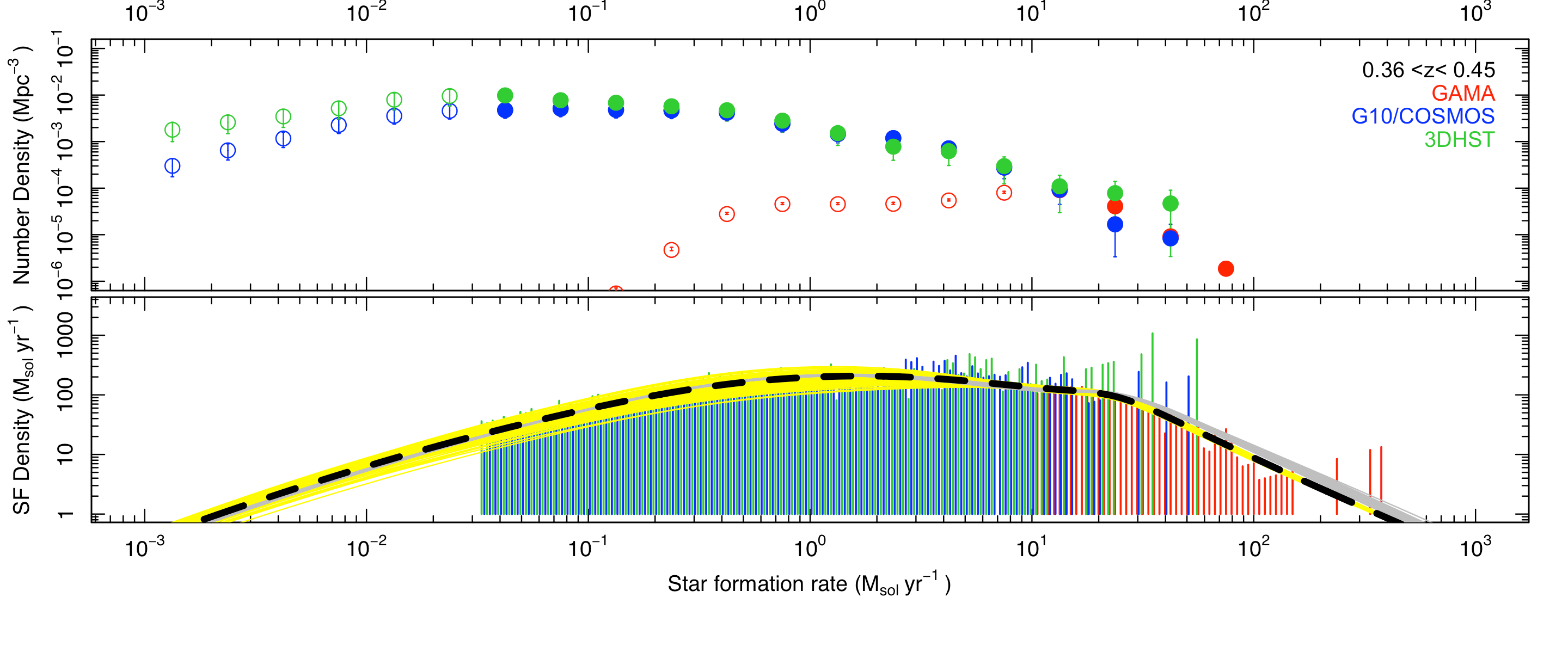}
\vspace{-0.5cm}
\includegraphics[width=\columnwidth]{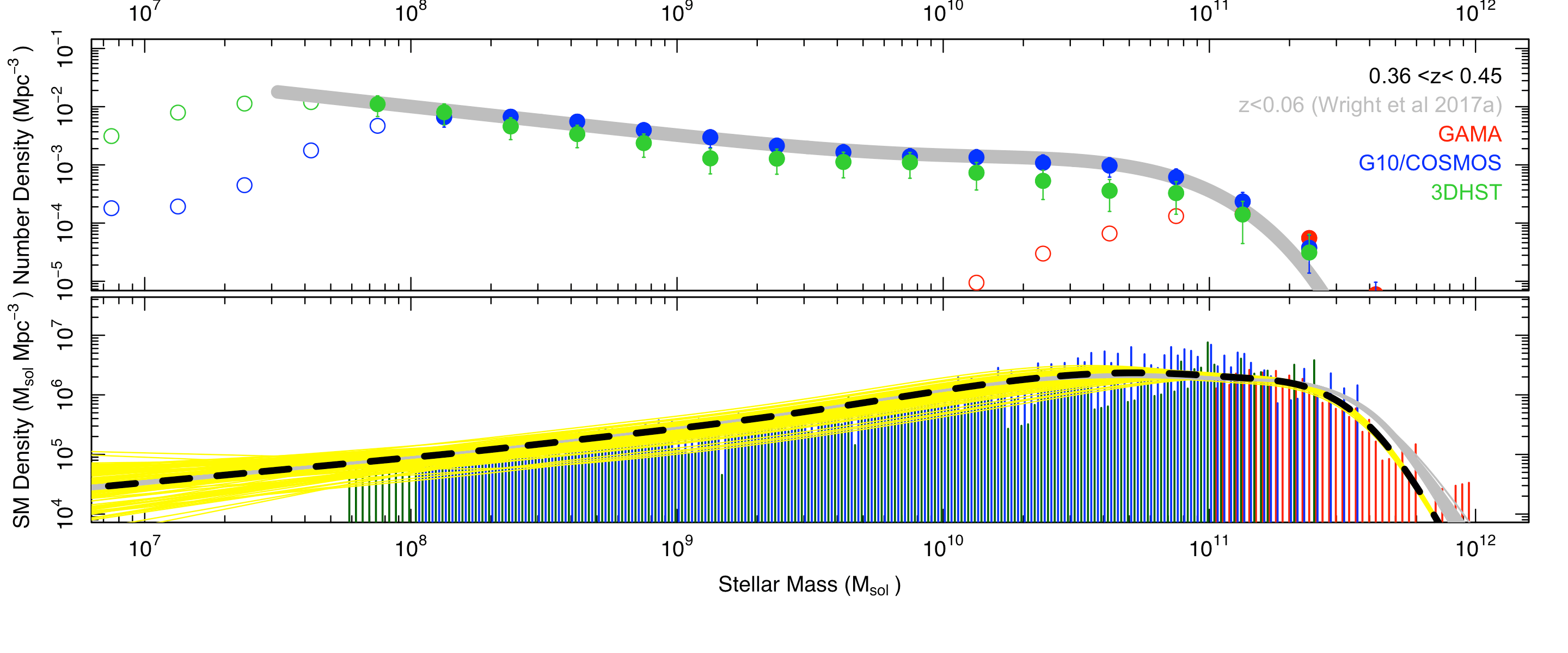}
\vspace{-0.5cm}
\includegraphics[width=\columnwidth]{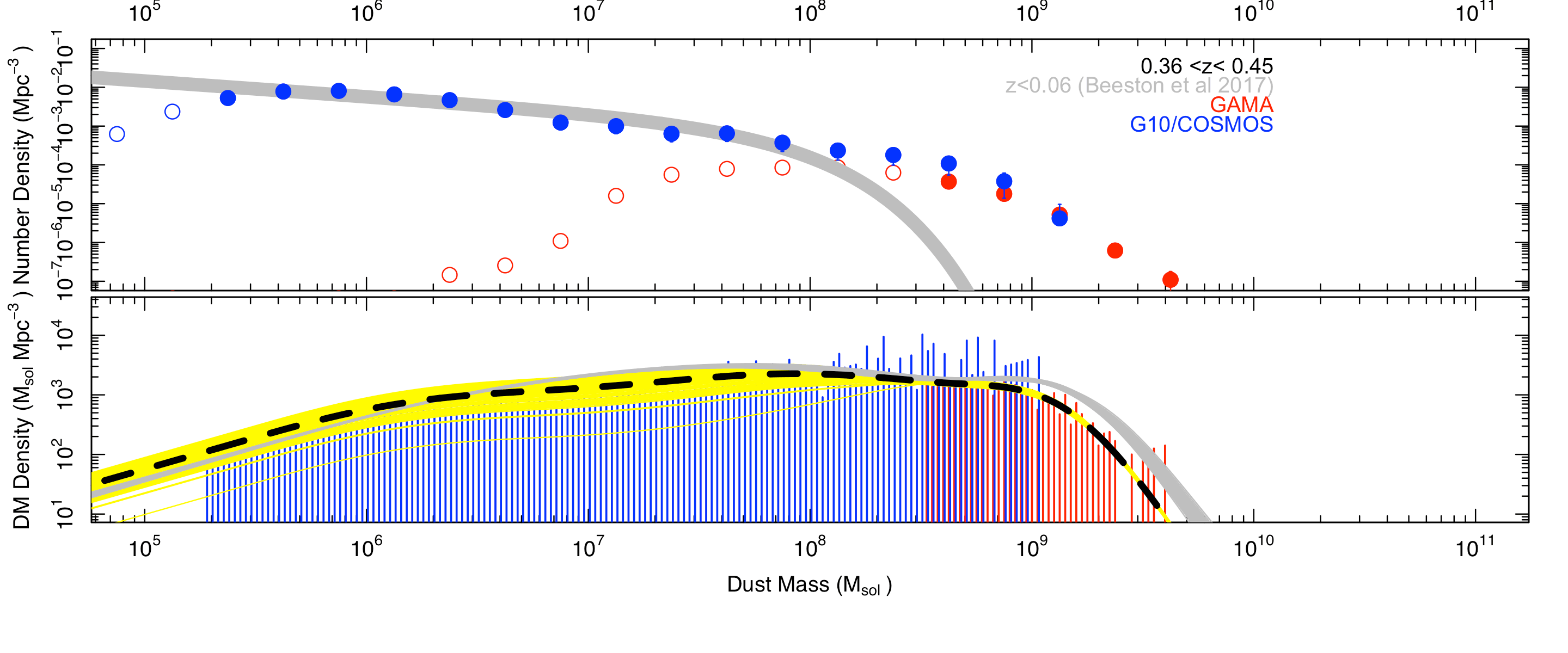}
\vspace{-0.5cm}
\caption{As for Fig.~\ref{fig:methods} except for the redshift range indicated.}
\end{center}
\end{figure}


\begin{figure}
\begin{center}
\includegraphics[width=\columnwidth]{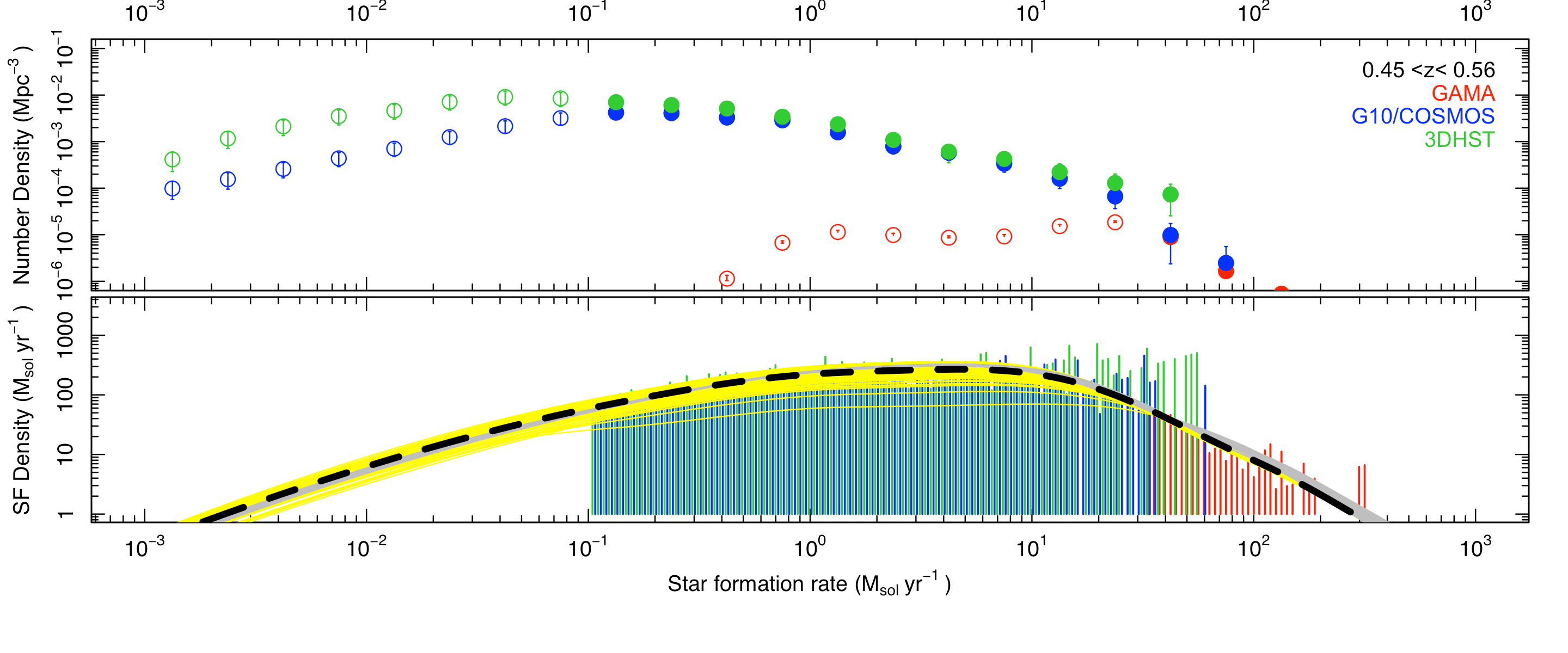}
\vspace{-0.5cm}
\includegraphics[width=\columnwidth]{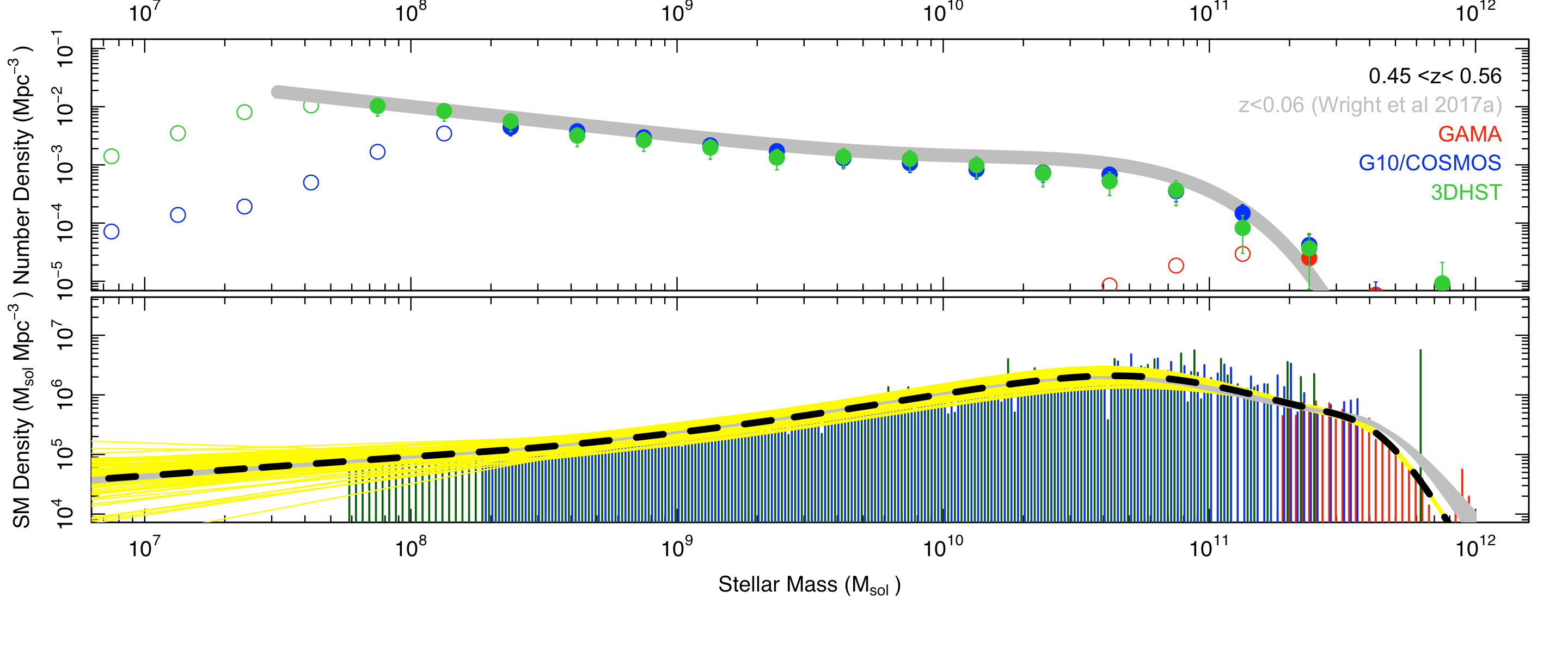}
\vspace{-0.5cm}
\includegraphics[width=\columnwidth]{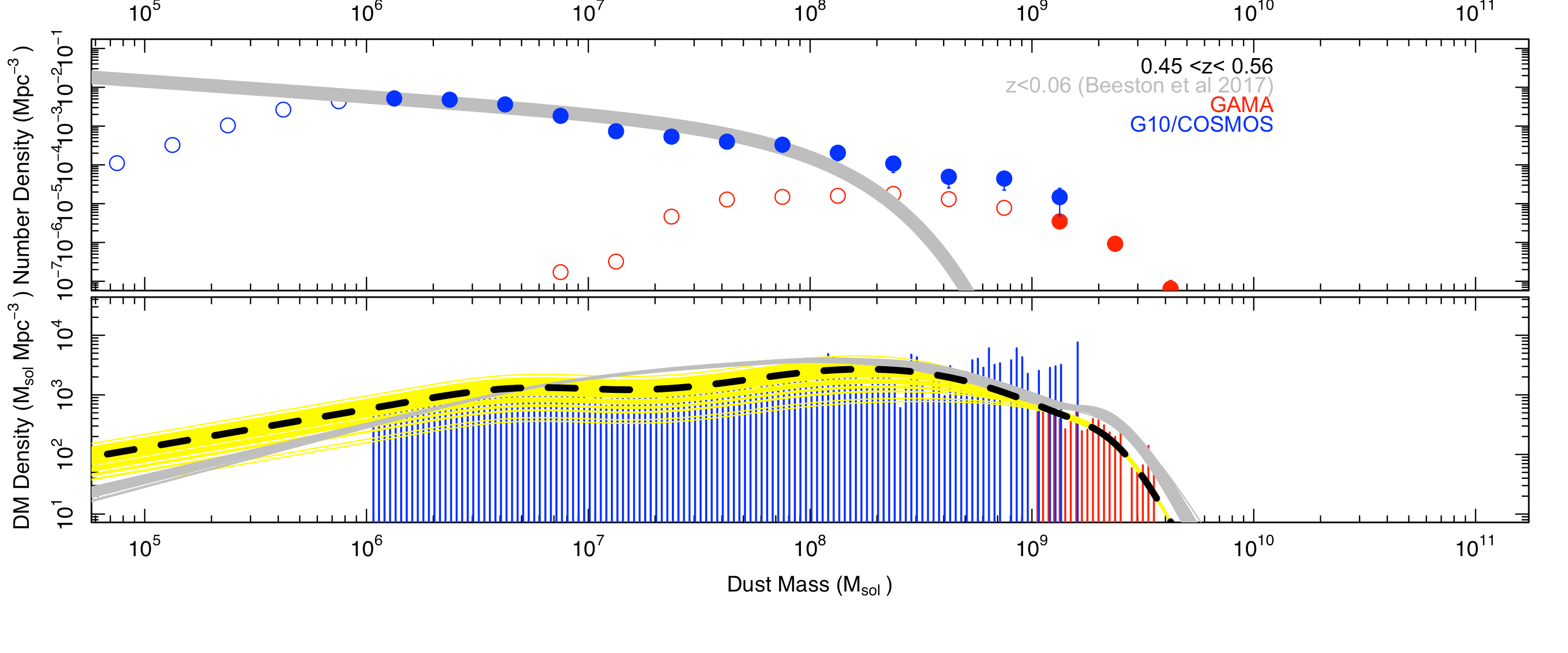}
\vspace{-0.5cm}
\caption{As for Fig.~\ref{fig:methods} except for the redshift range indicated.}
\end{center}
\end{figure}

\clearpage

\begin{figure}
\begin{center}
\includegraphics[width=\columnwidth]{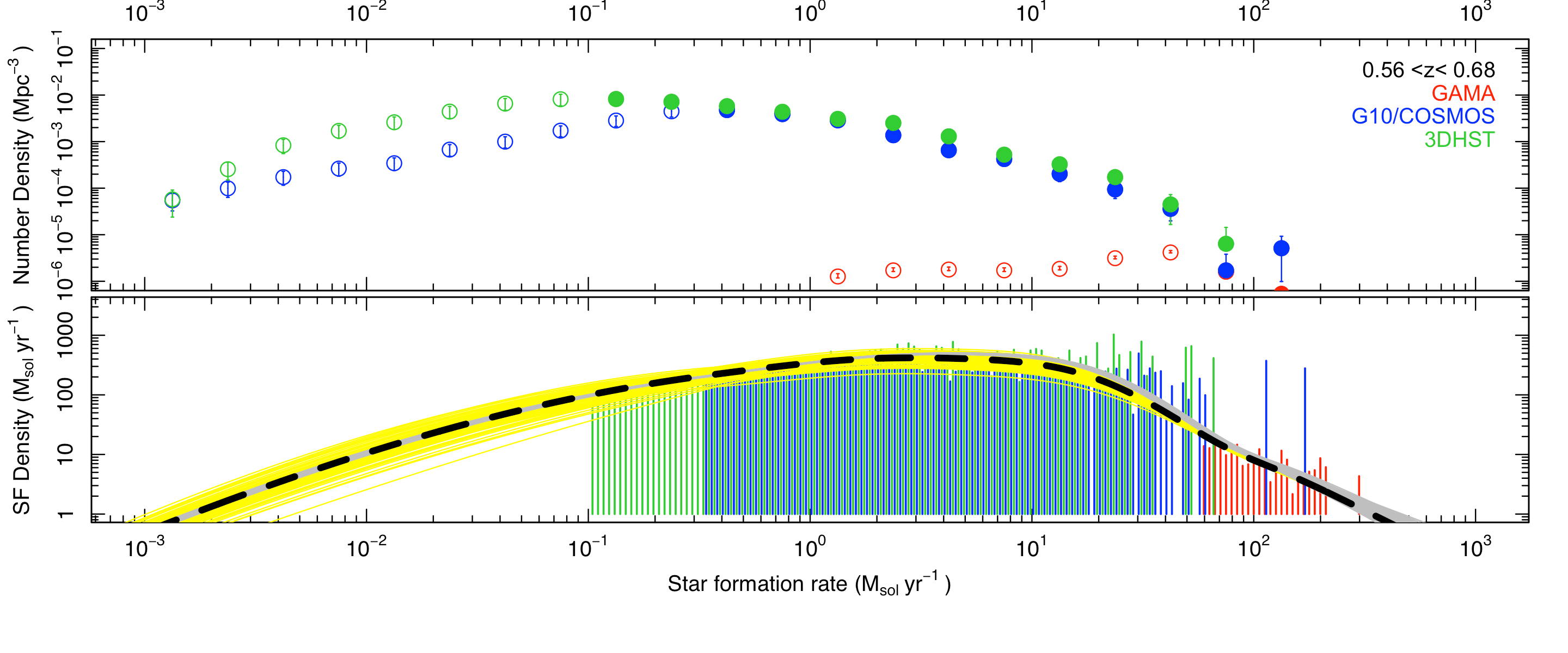}
\vspace{-0.5cm}
\includegraphics[width=\columnwidth]{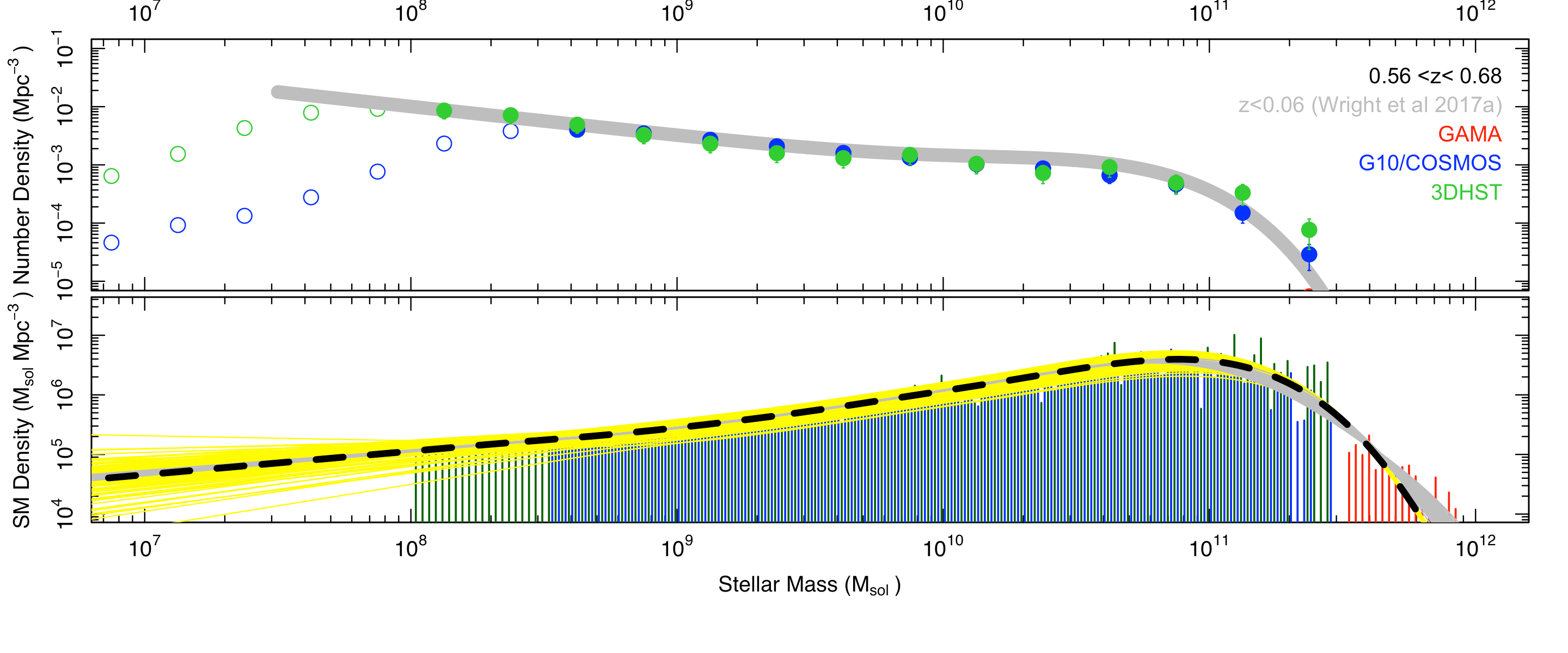}
\vspace{-0.5cm}
\includegraphics[width=\columnwidth]{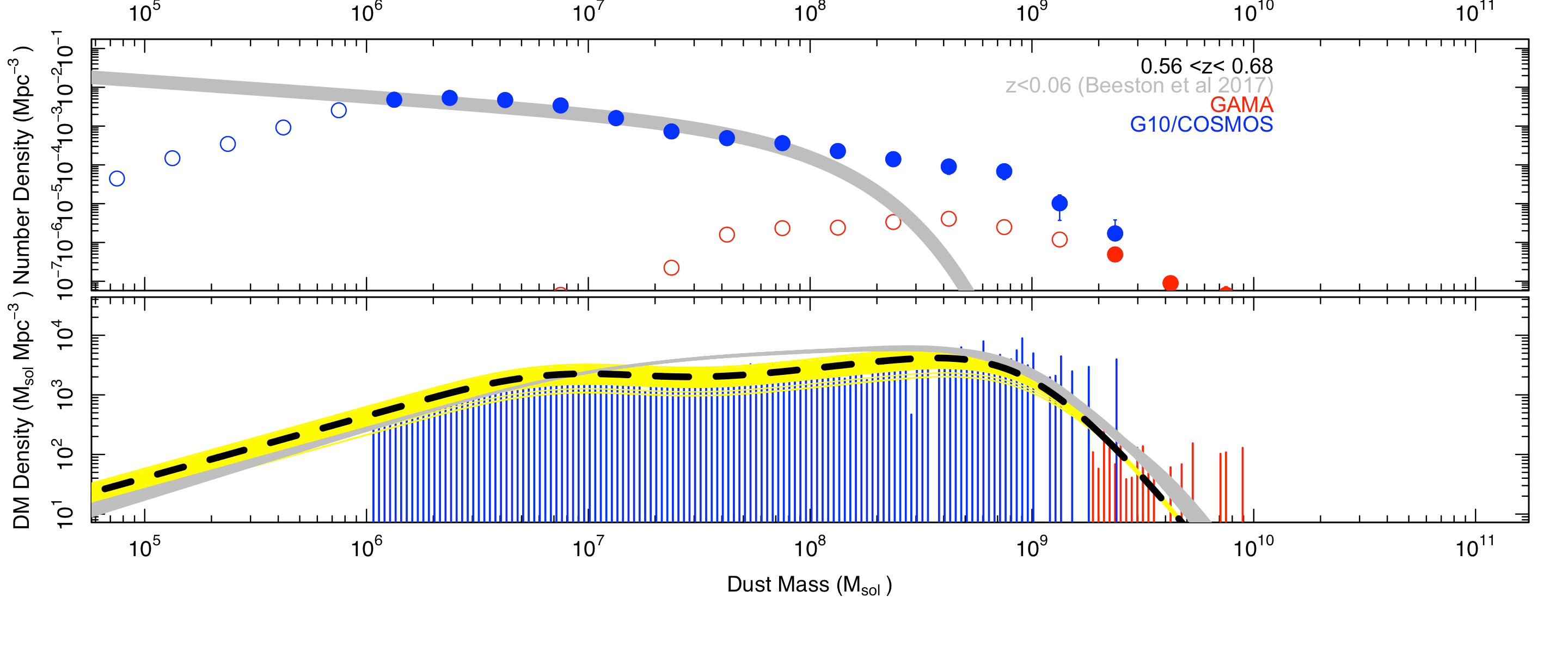}
\vspace{-0.5cm}
\caption{As for Fig.~\ref{fig:methods} except for the redshift range indicated.}
\end{center}
\end{figure}


\begin{figure}
\begin{center}
\includegraphics[width=\columnwidth]{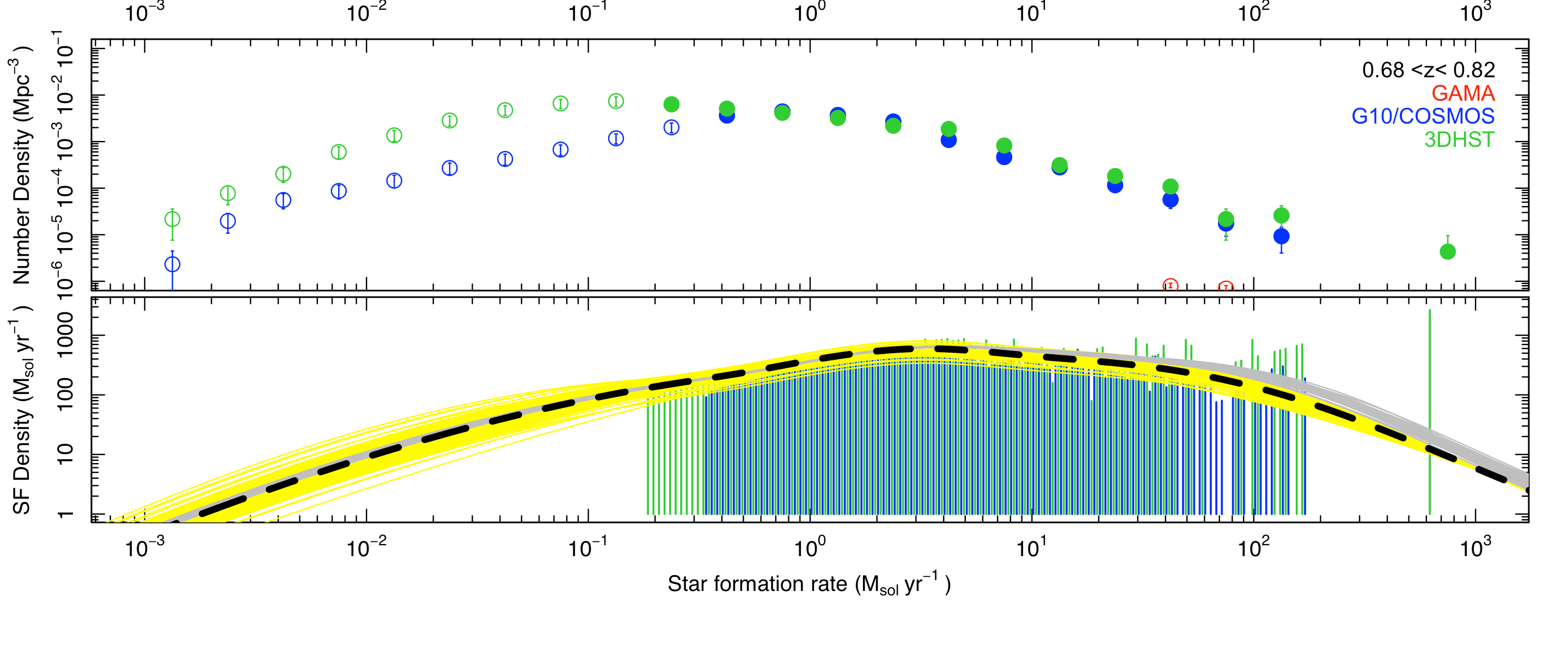}
\vspace{-0.5cm}
\includegraphics[width=\columnwidth]{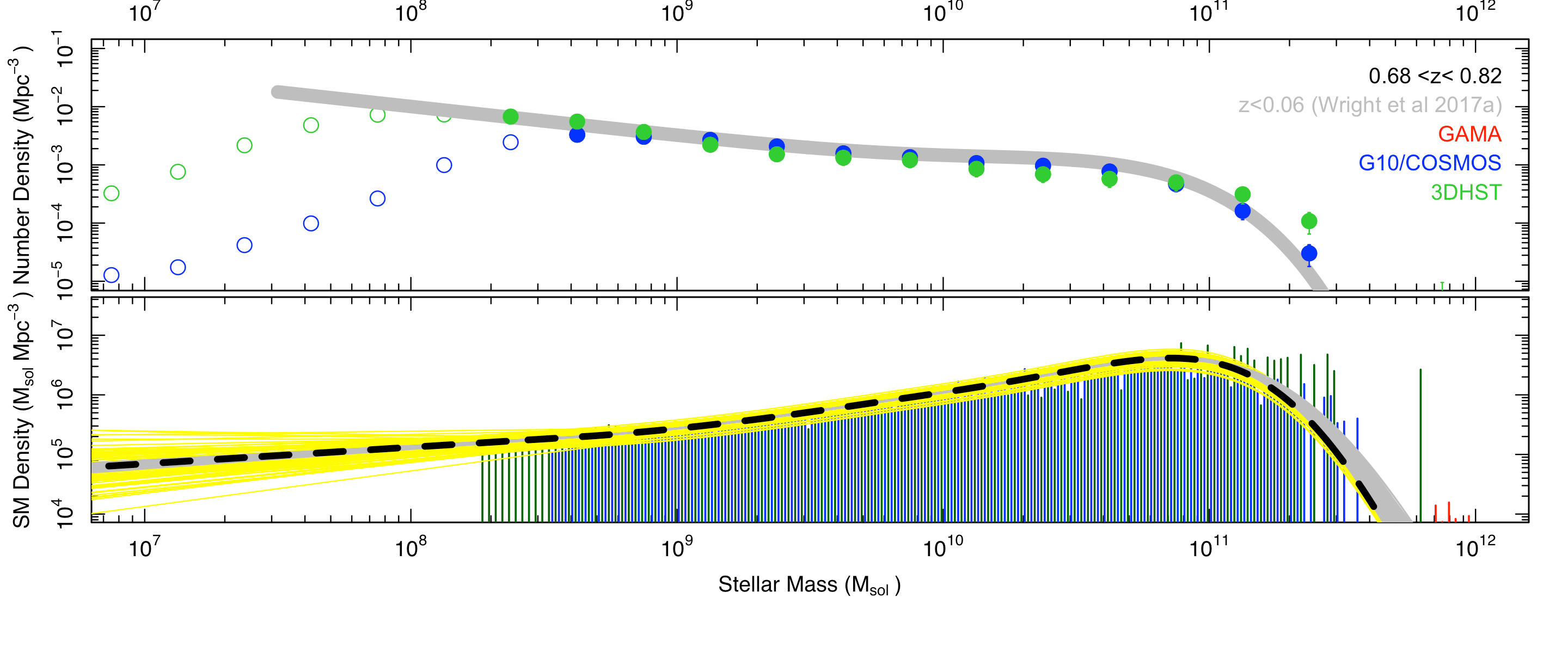}
\vspace{-0.5cm}
\includegraphics[width=\columnwidth]{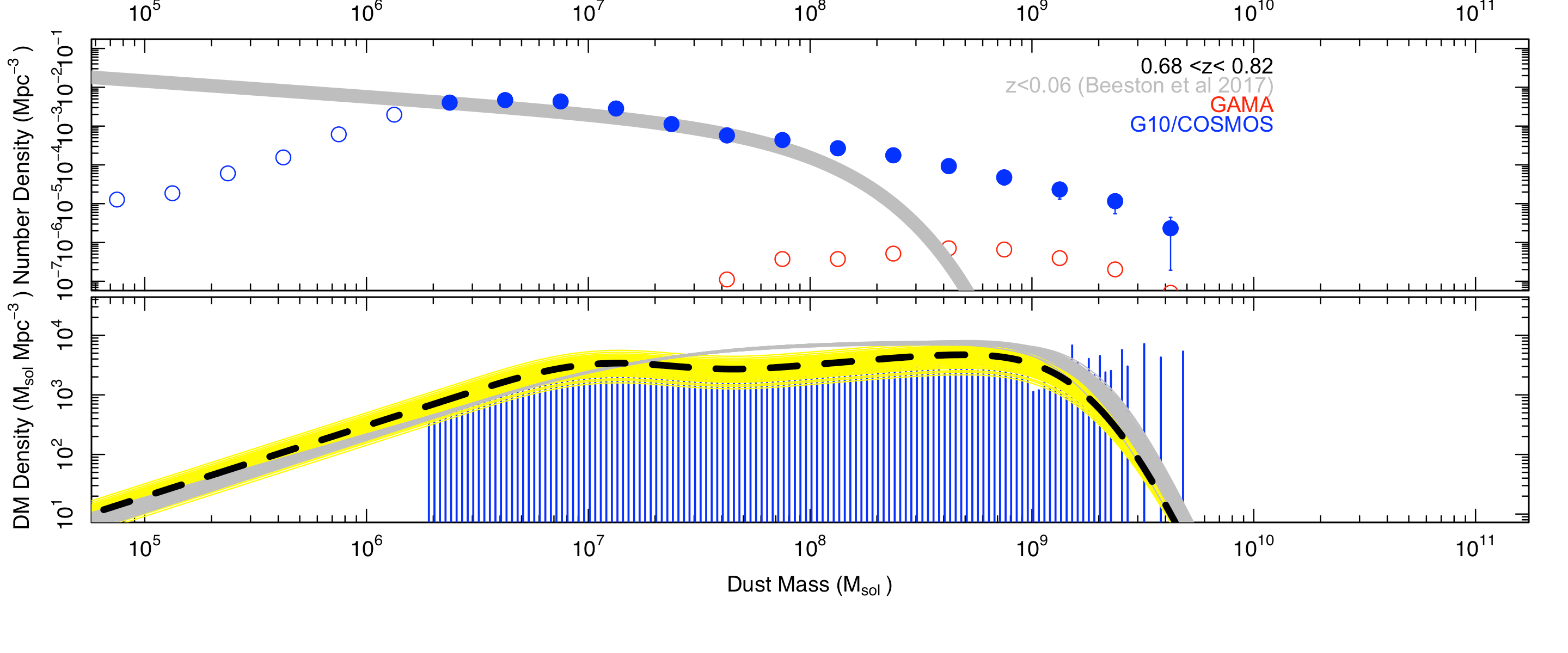}
\vspace{-0.5cm}
\caption{As for Fig.~\ref{fig:methods} except for the redshift range indicated.}
\end{center}
\end{figure}


\begin{figure}
\begin{center}
\includegraphics[width=\columnwidth]{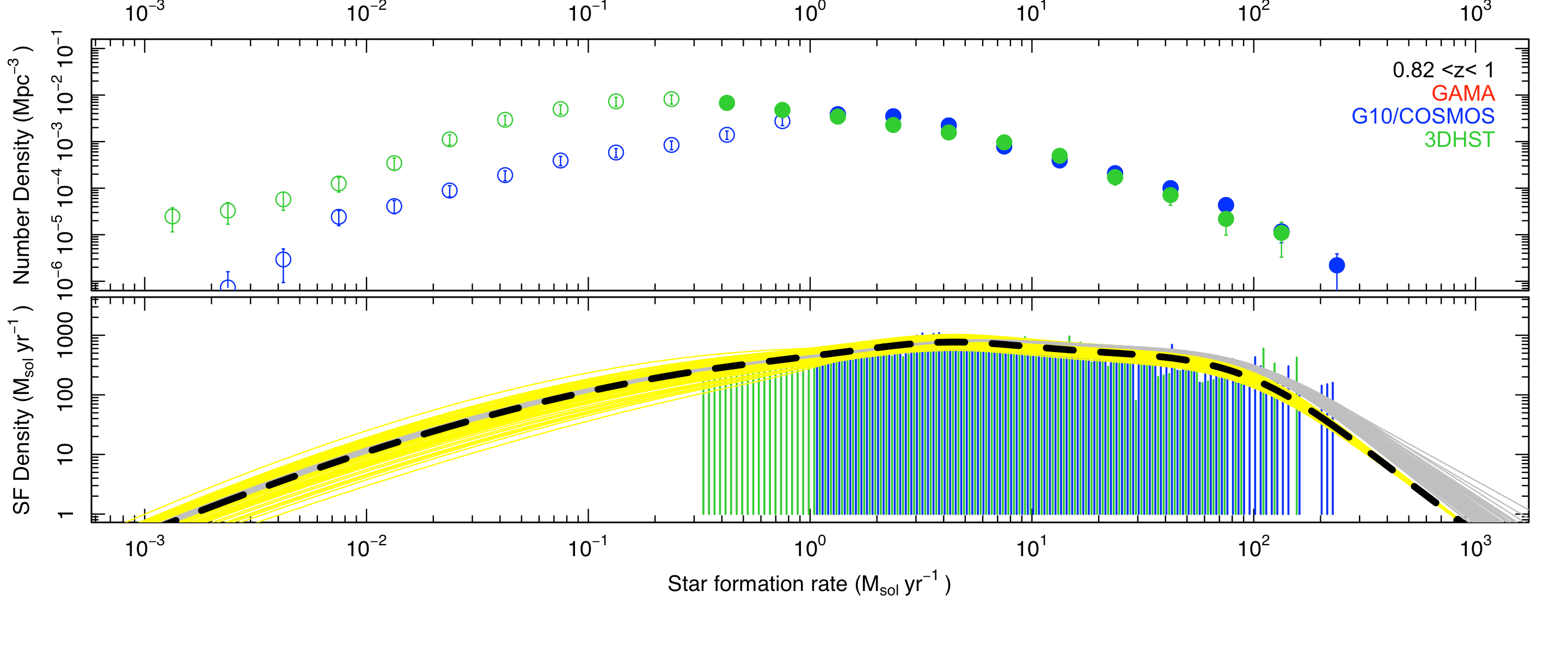}
\vspace{-0.5cm}
\includegraphics[width=\columnwidth]{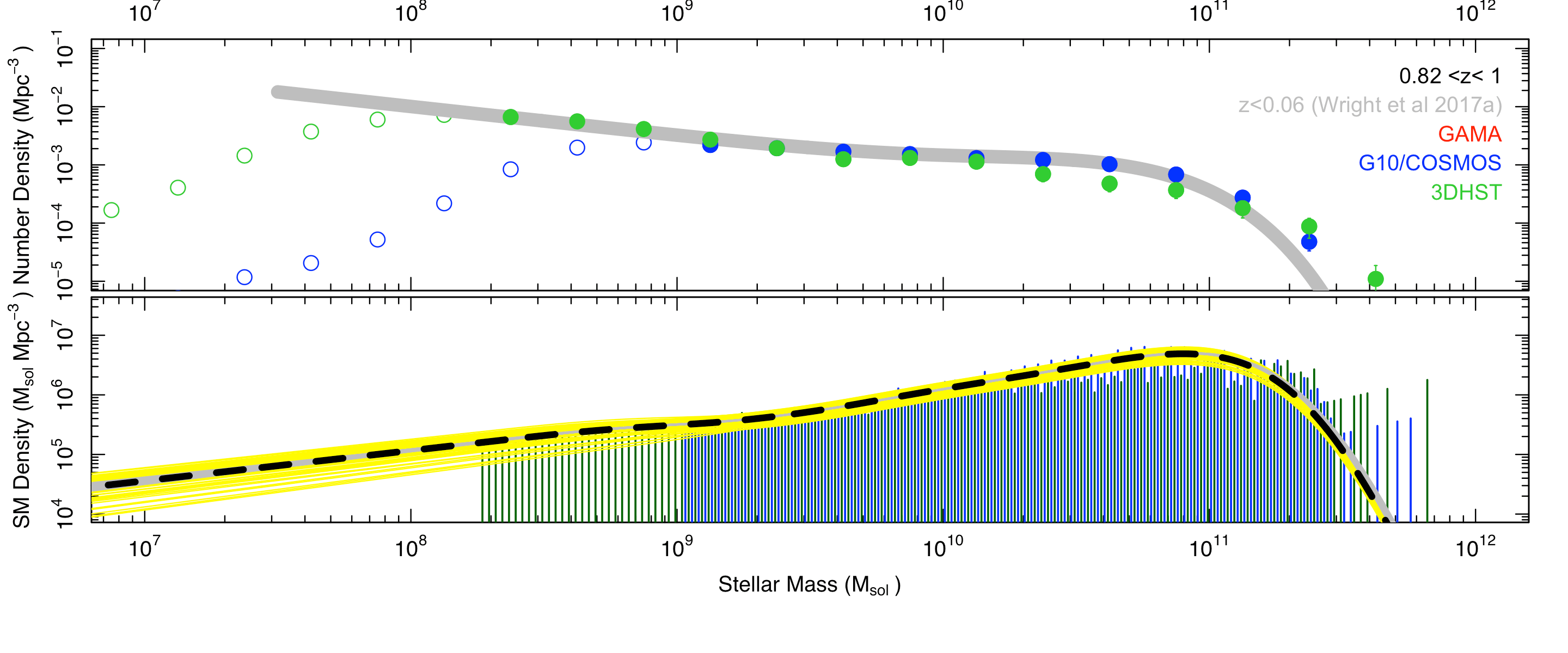}
\vspace{-0.5cm}
\includegraphics[width=\columnwidth]{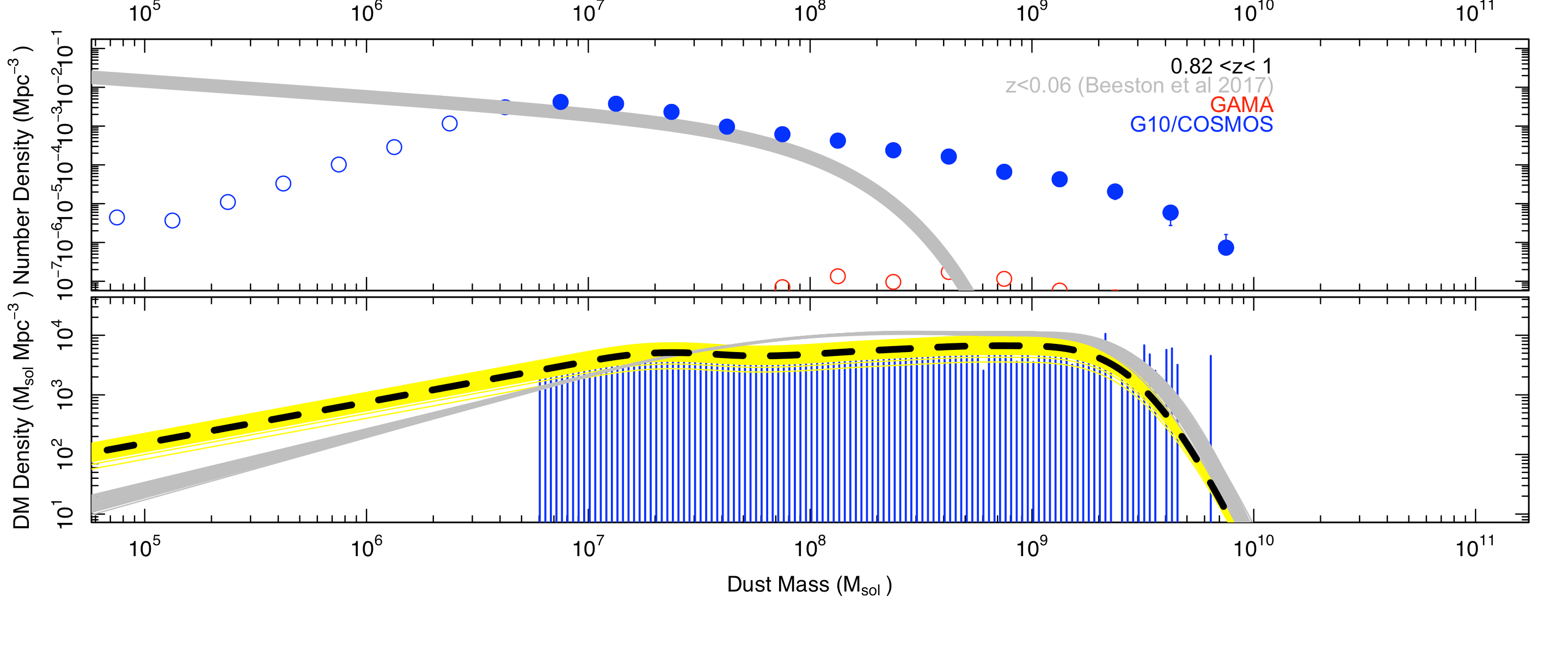}
\vspace{-0.5cm}
\caption{As for Fig.~\ref{fig:methods} except for the redshift range indicated.}
\end{center}
\end{figure}


\begin{figure}
\begin{center}
\includegraphics[width=\columnwidth]{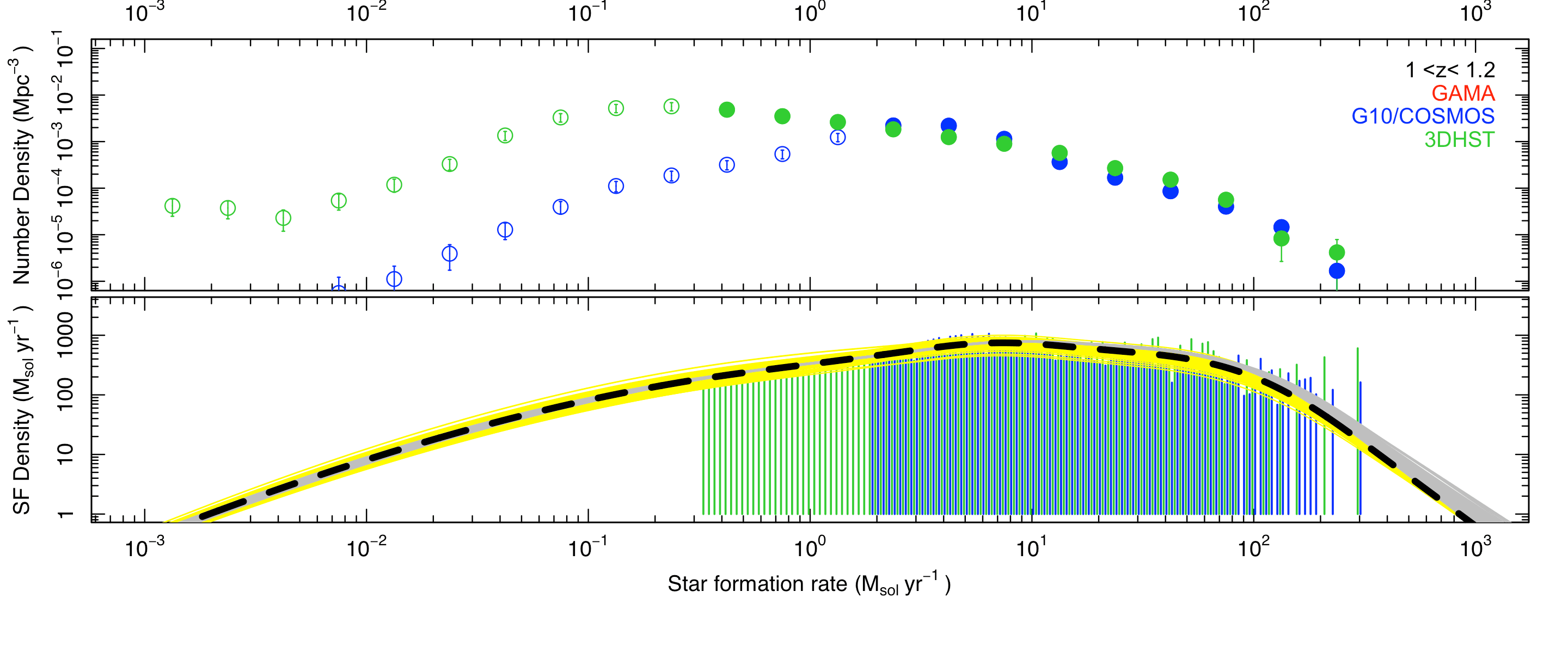}
\vspace{-0.5cm}
\includegraphics[width=\columnwidth]{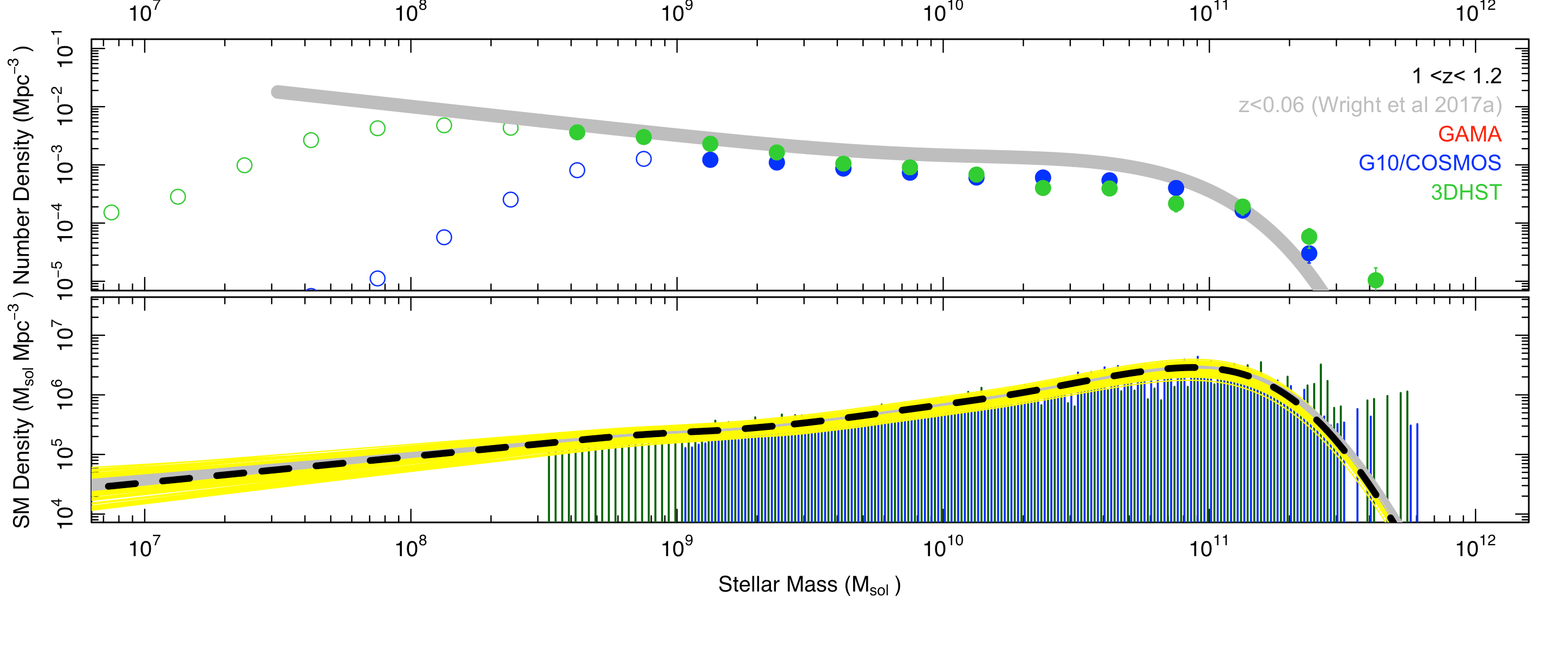}
\vspace{-0.5cm}
\includegraphics[width=\columnwidth]{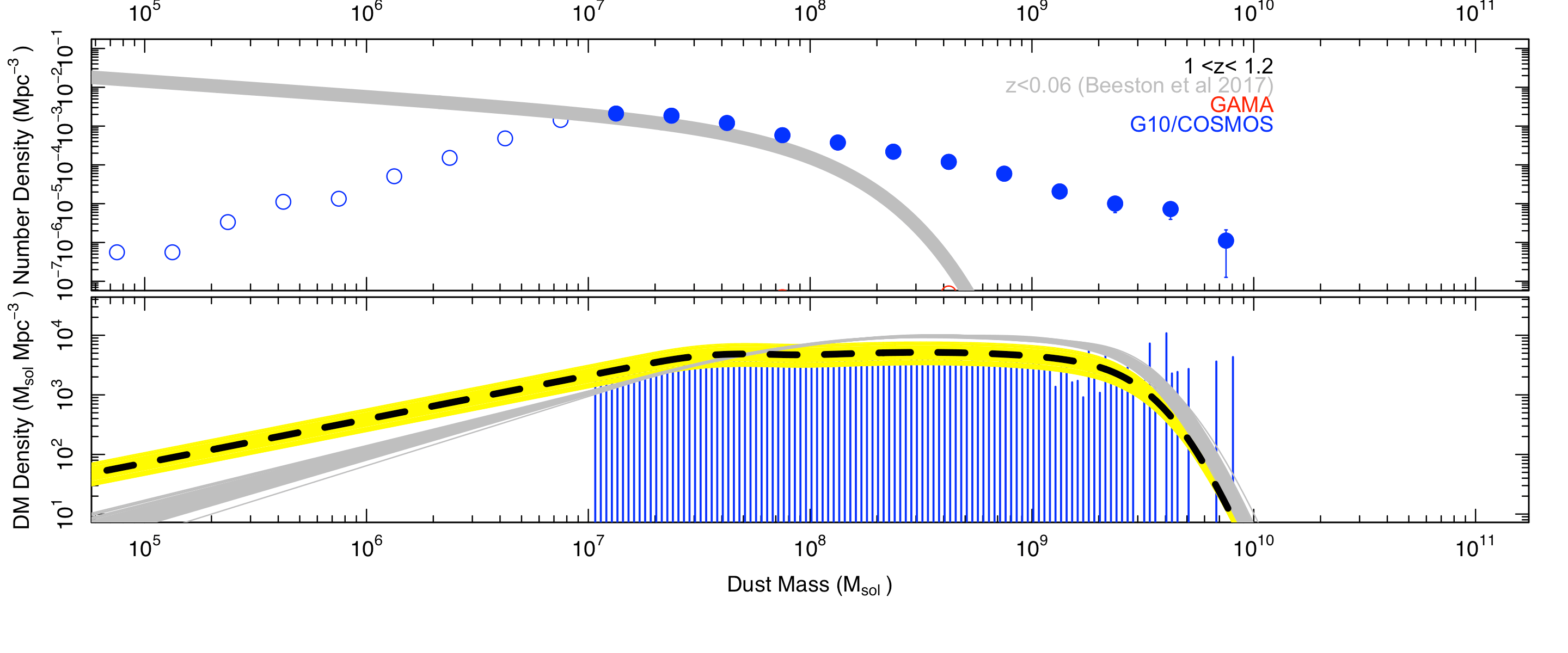}
\vspace{-0.5cm}
\caption{As for Fig.~\ref{fig:methods} except for the redshift range indicated.}
\end{center}
\end{figure}

\clearpage

\begin{figure}
\begin{center}
\includegraphics[width=\columnwidth]{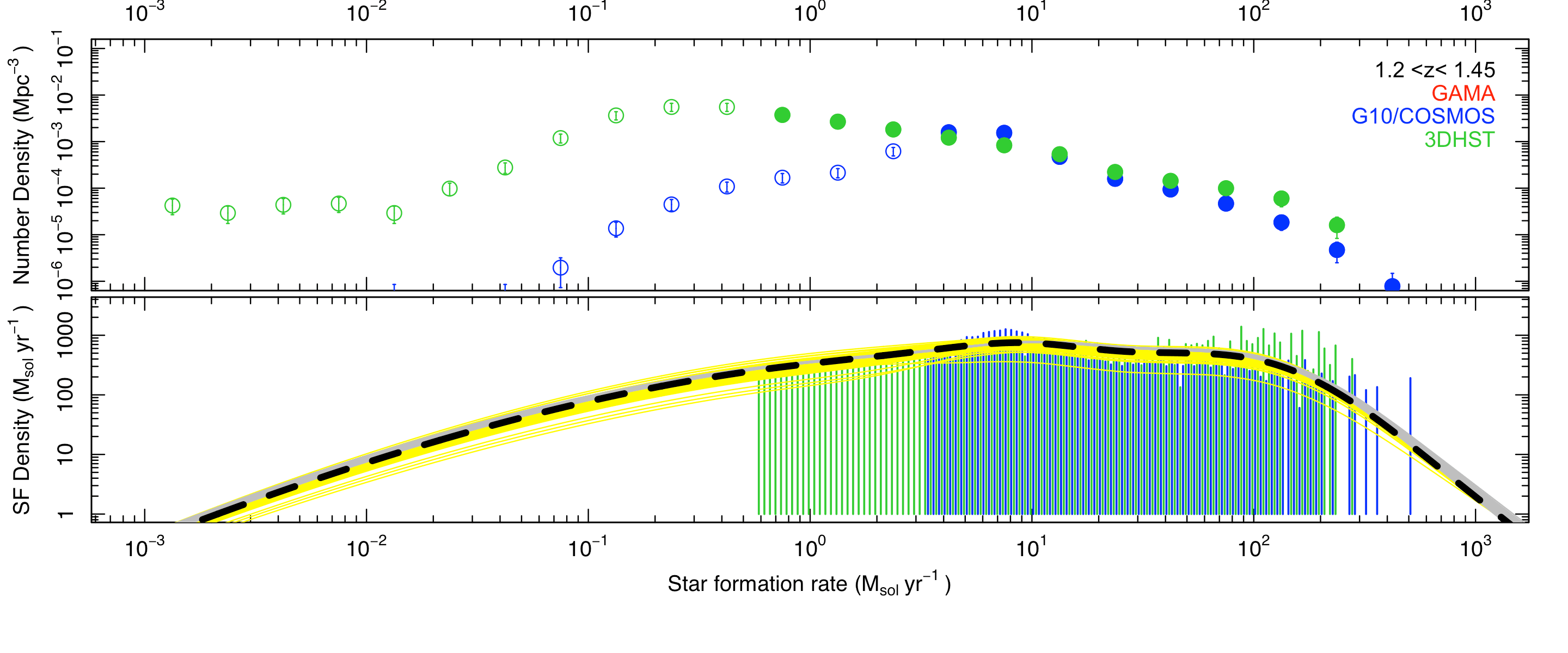}
\vspace{-0.5cm}
\includegraphics[width=\columnwidth]{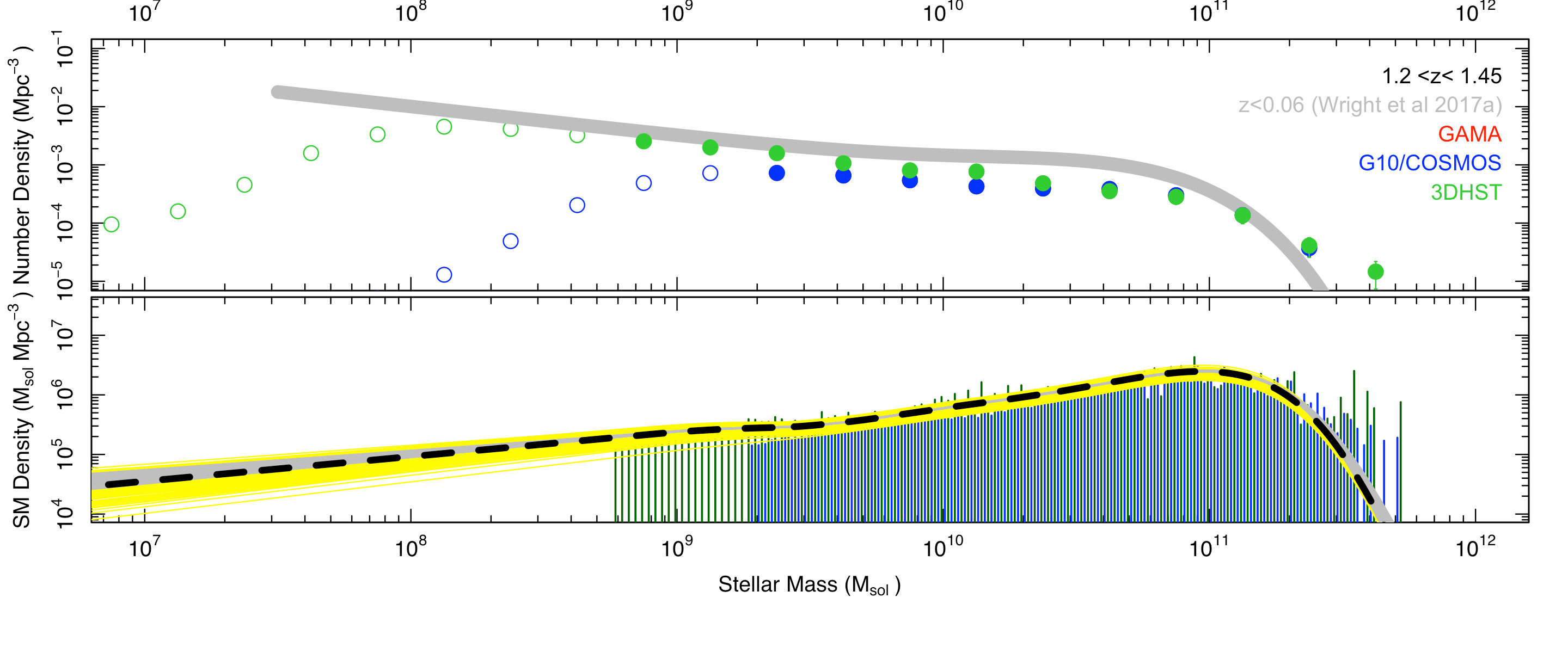}
\vspace{-0.5cm}
\includegraphics[width=\columnwidth]{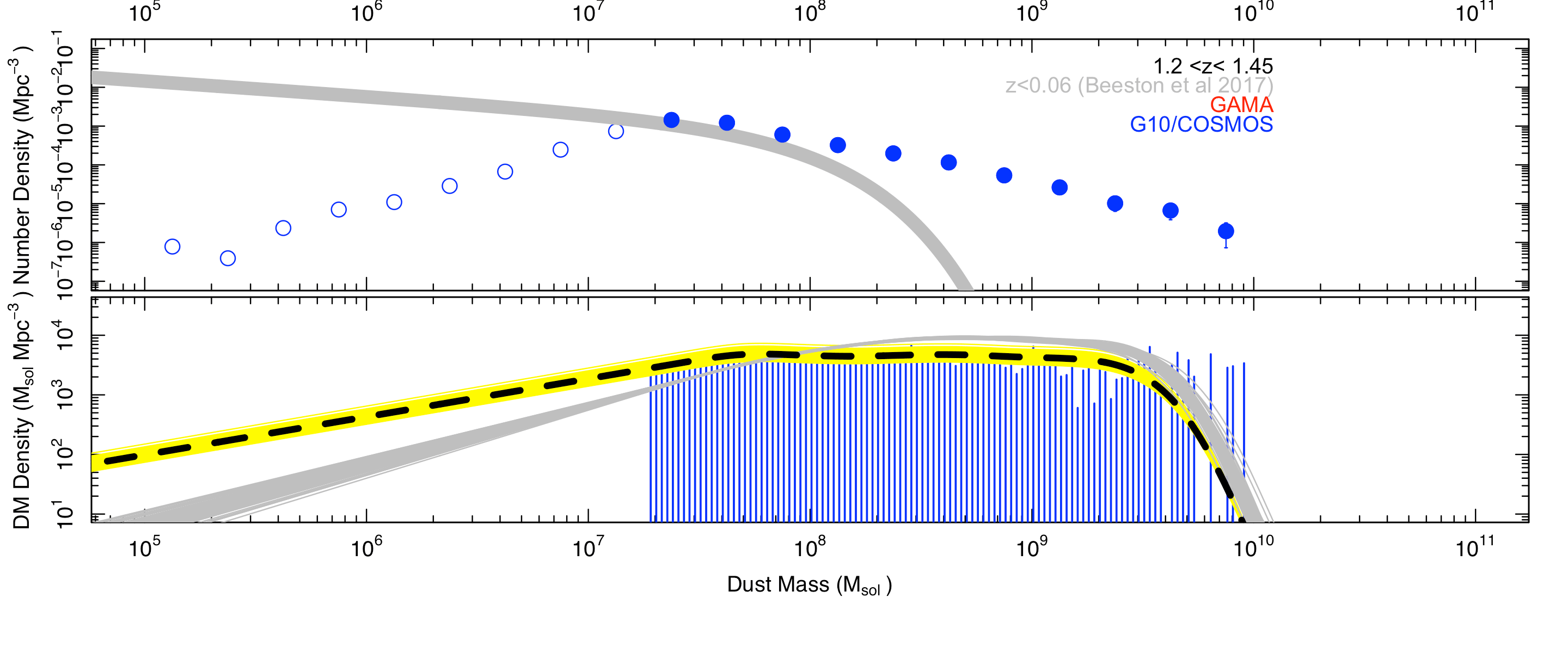}
\vspace{-0.5cm}
\caption{As for Fig.~\ref{fig:methods} except for the redshift range indicated.}
\end{center}
\end{figure}


\begin{figure}
\begin{center}
\includegraphics[width=\columnwidth]{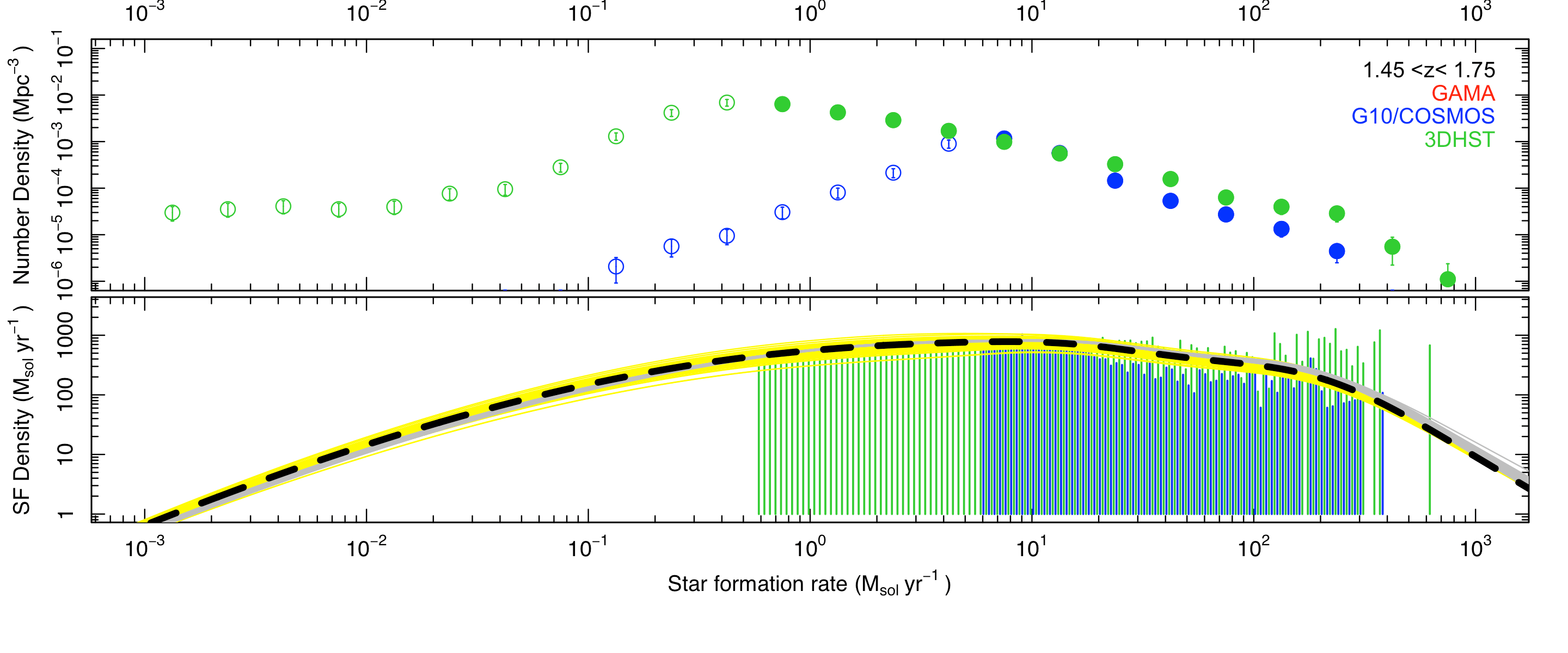}
\vspace{-0.5cm}
\includegraphics[width=\columnwidth]{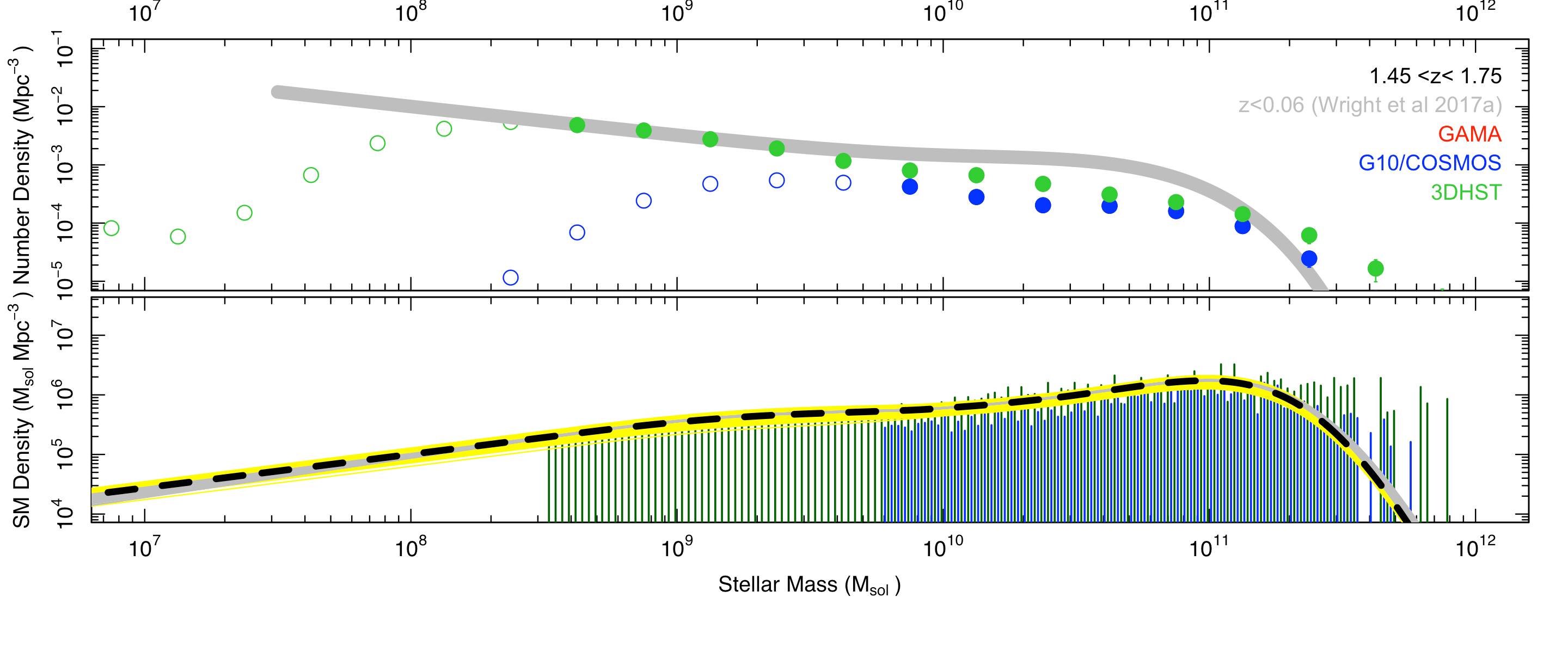}
\vspace{-0.5cm}
\includegraphics[width=\columnwidth]{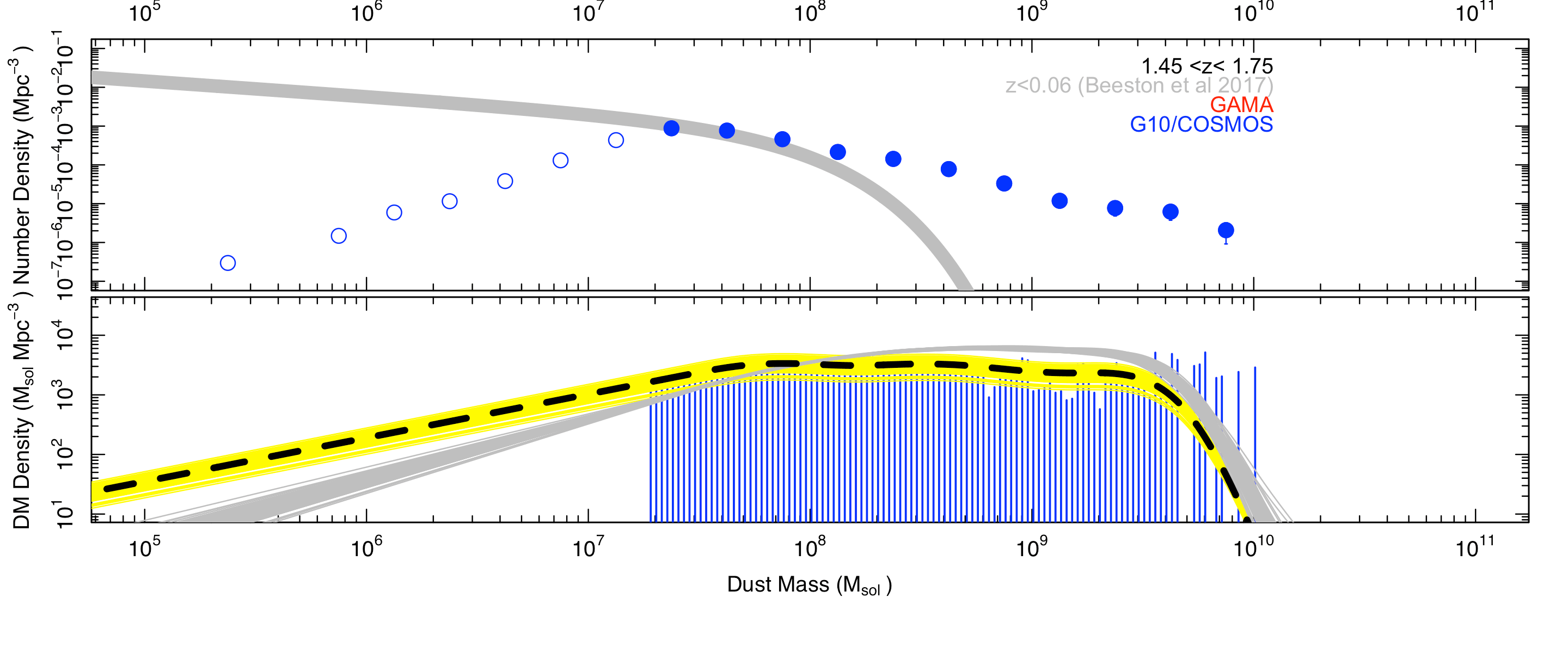}
\vspace{-0.5cm}
\caption{As for Fig.~\ref{fig:methods} except for the redshift range indicated.}
\end{center}
\end{figure}


\begin{figure}
\begin{center}
\includegraphics[width=\columnwidth]{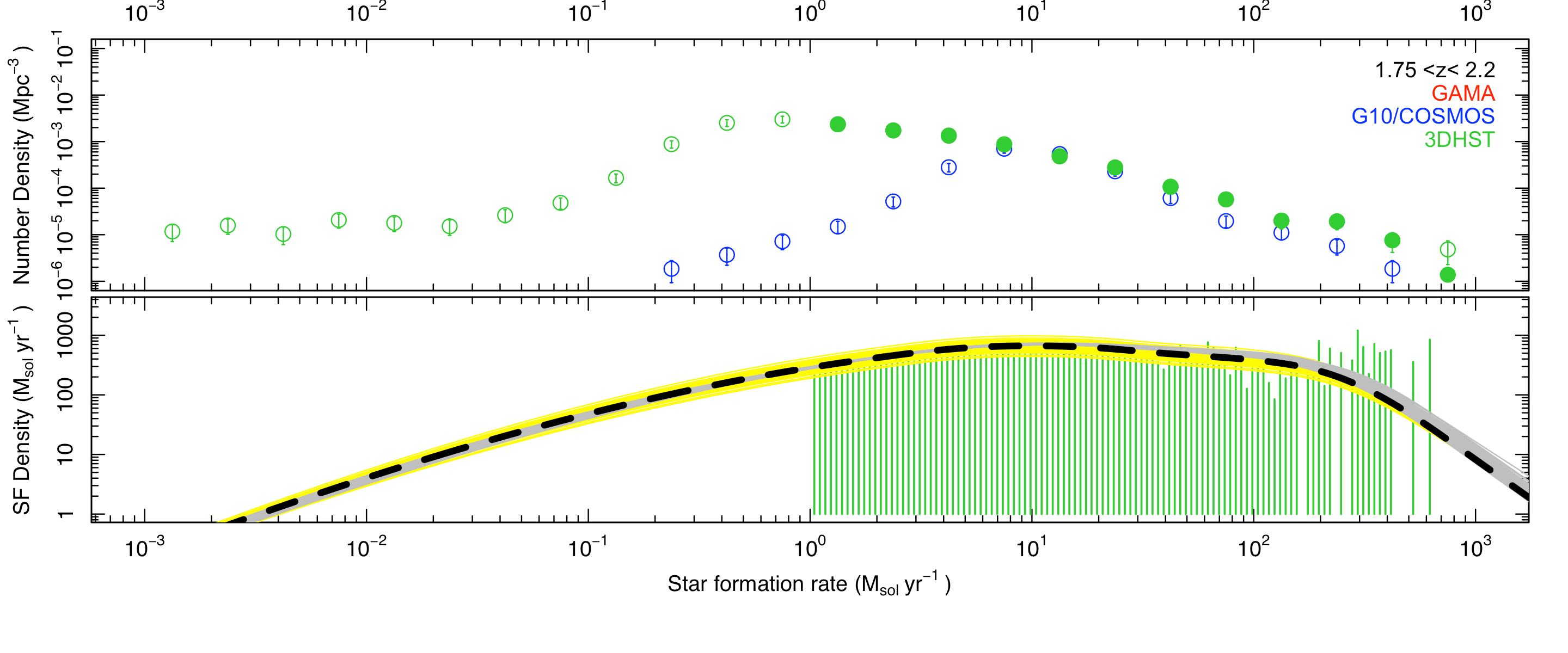}
\vspace{-0.5cm}
\includegraphics[width=\columnwidth]{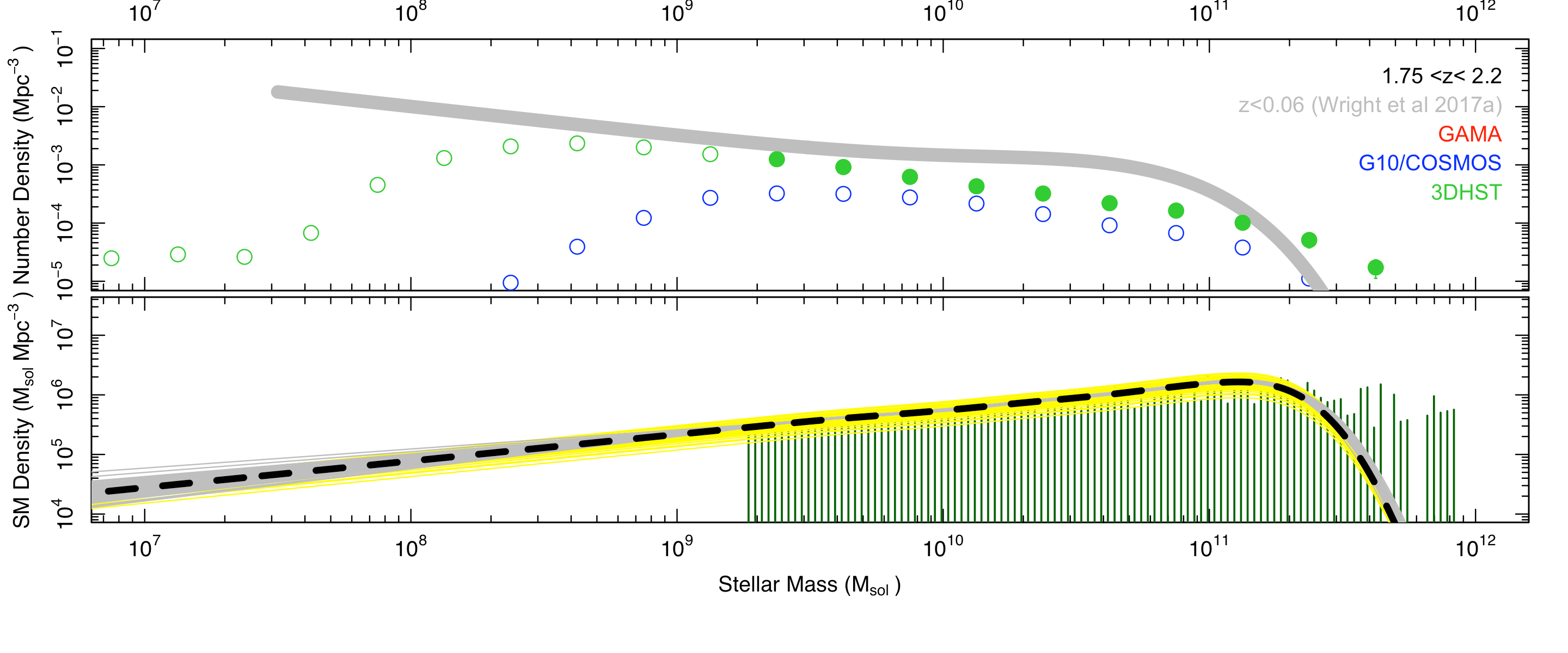}
\vspace{-0.5cm}
\caption{As for Fig.~\ref{fig:methods} except for the redshift range indicated.}
\end{center}
\end{figure}


\begin{figure}
\begin{center}
\includegraphics[width=\columnwidth]{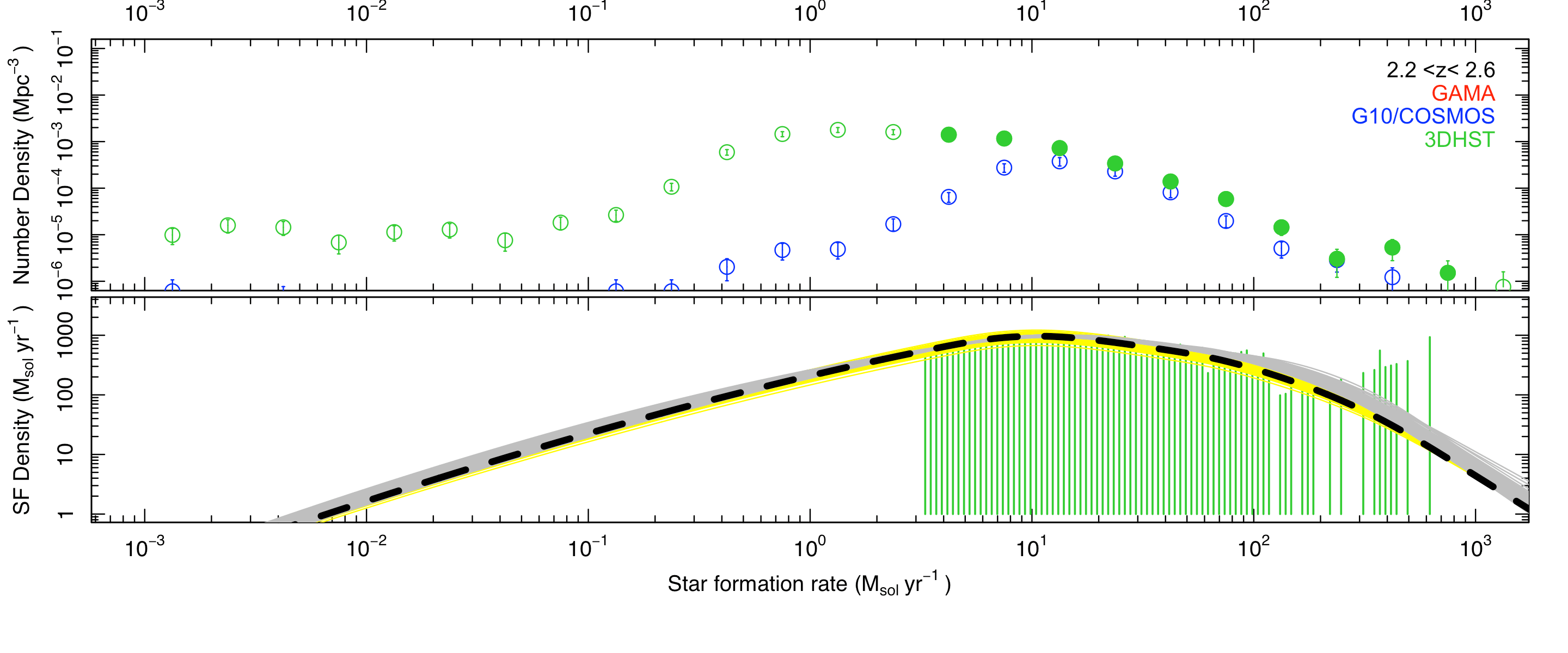}
\vspace{-0.5cm}
\includegraphics[width=\columnwidth]{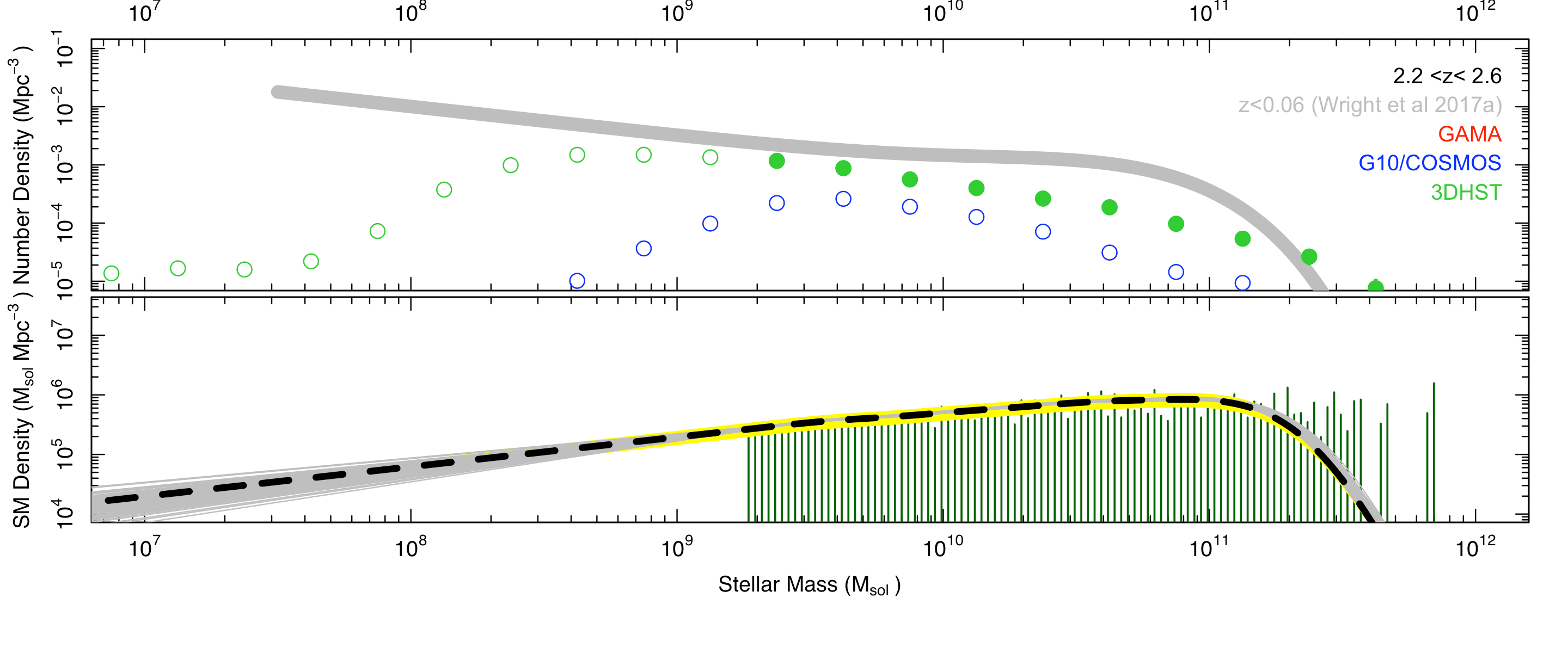}
\vspace{-0.5cm}
\caption{As for Fig.~\ref{fig:methods} except for the redshift range indicated.}
\end{center}
\end{figure}

\clearpage

\begin{figure}
\begin{center}
\includegraphics[width=\columnwidth]{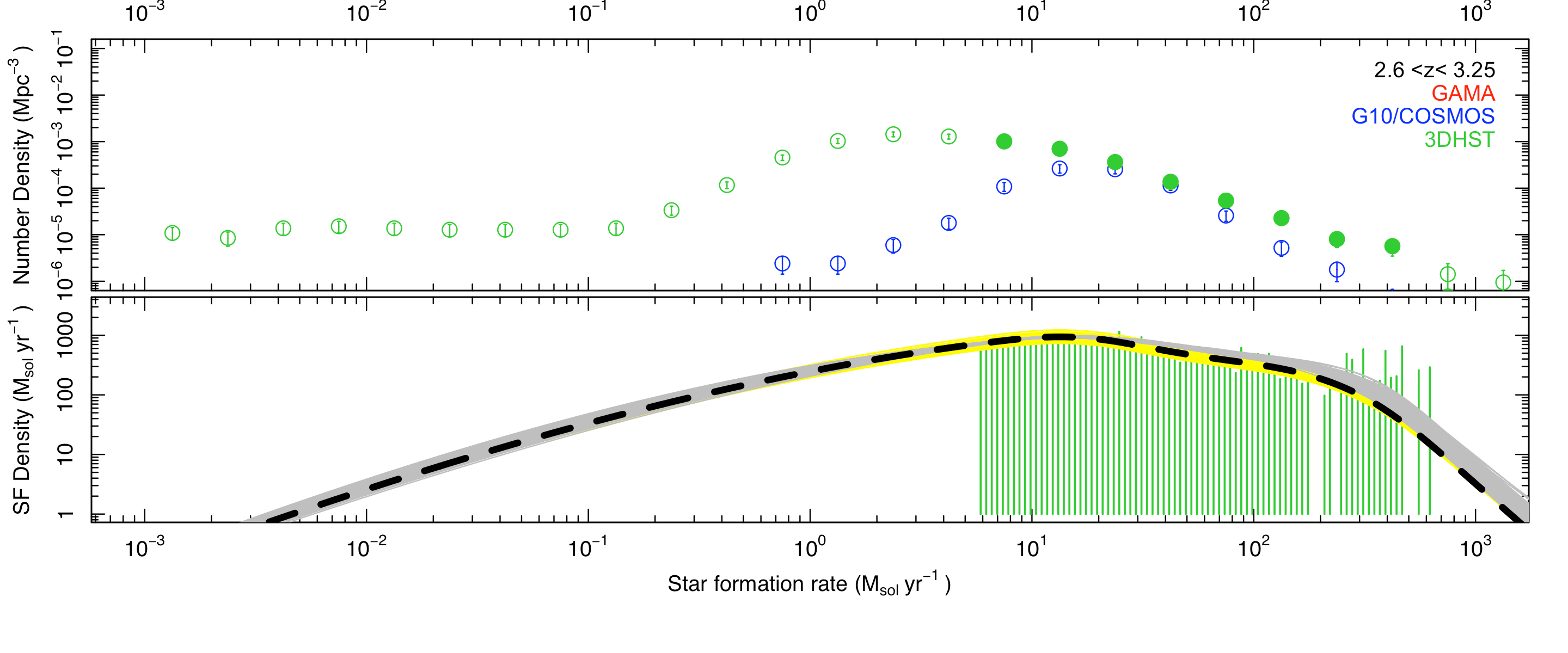}
\vspace{-0.5cm}
\includegraphics[width=\columnwidth]{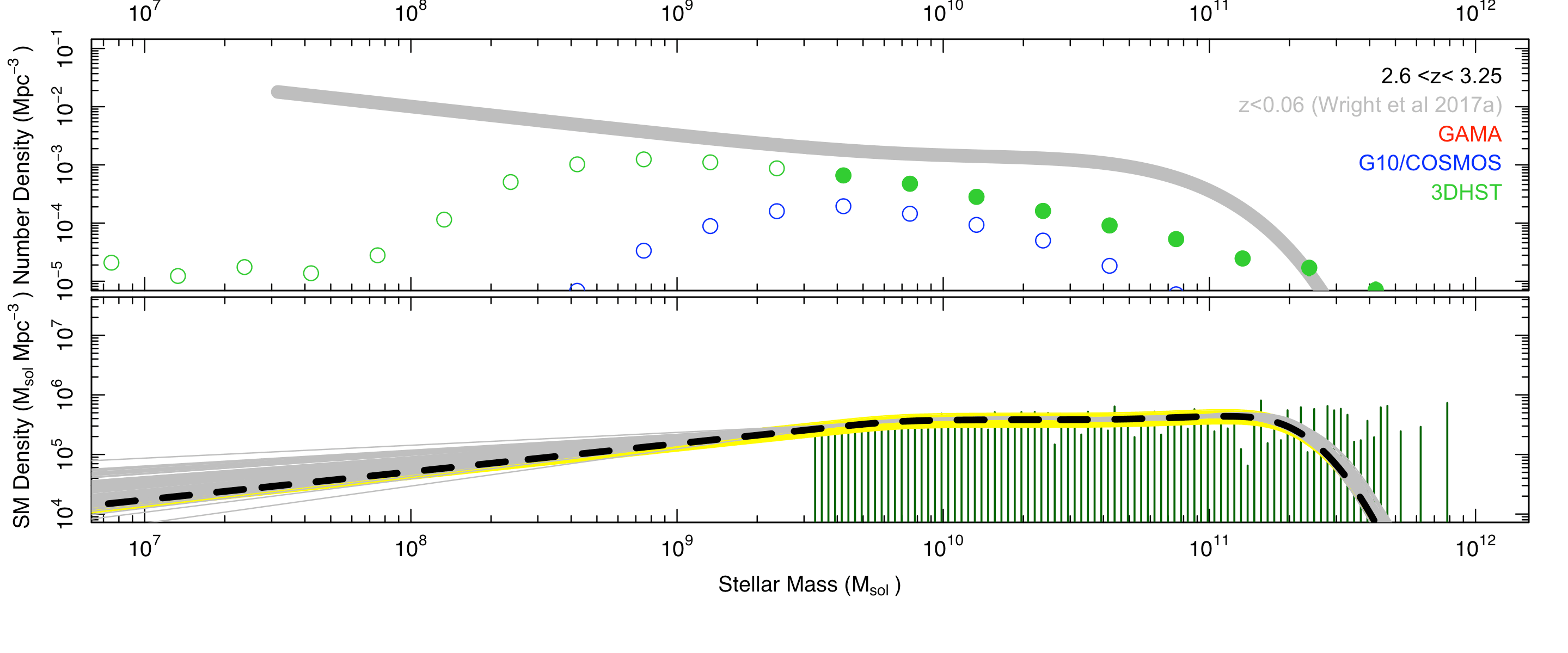}
\vspace{-0.5cm}
\caption{As for Fig.~\ref{fig:methods} except for the redshift range indicated.}
\end{center}
\end{figure}


\begin{figure}
\begin{center}
\includegraphics[width=\columnwidth]{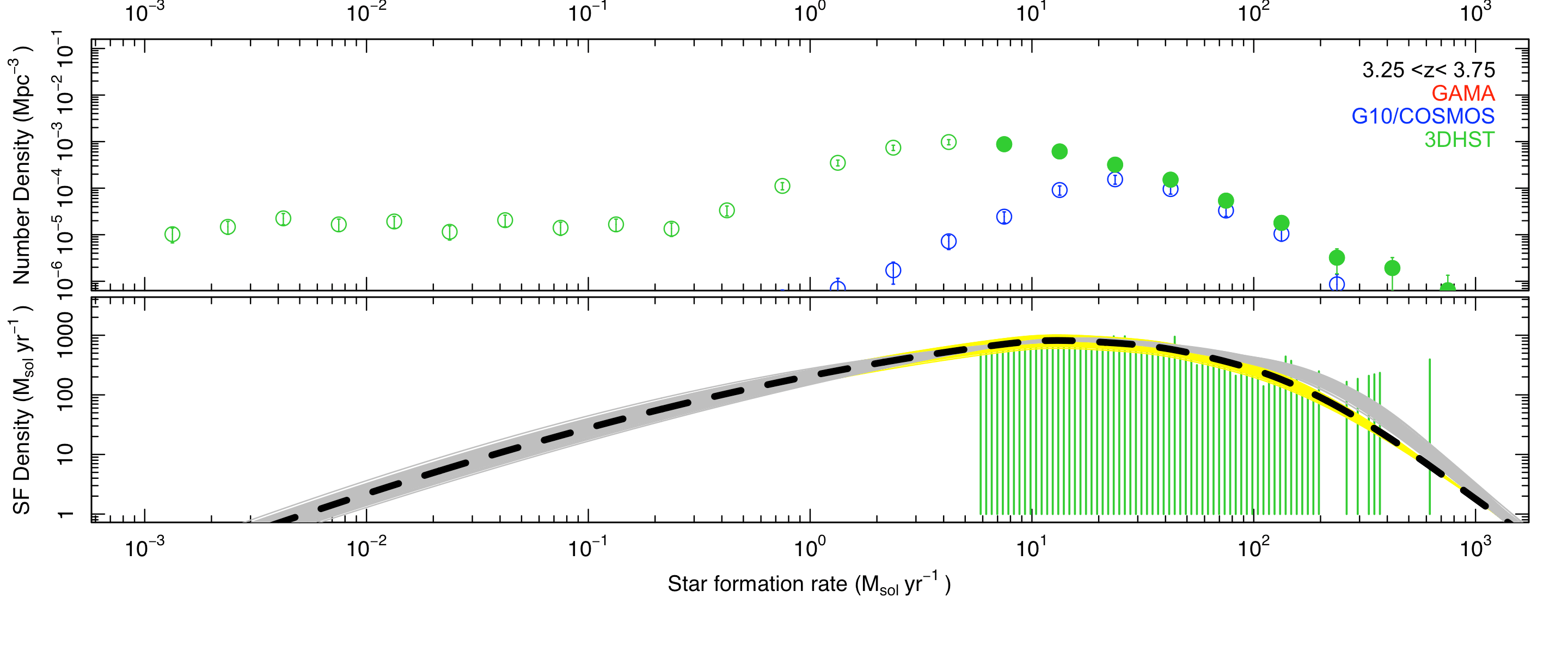}
\vspace{-0.5cm}
\includegraphics[width=\columnwidth]{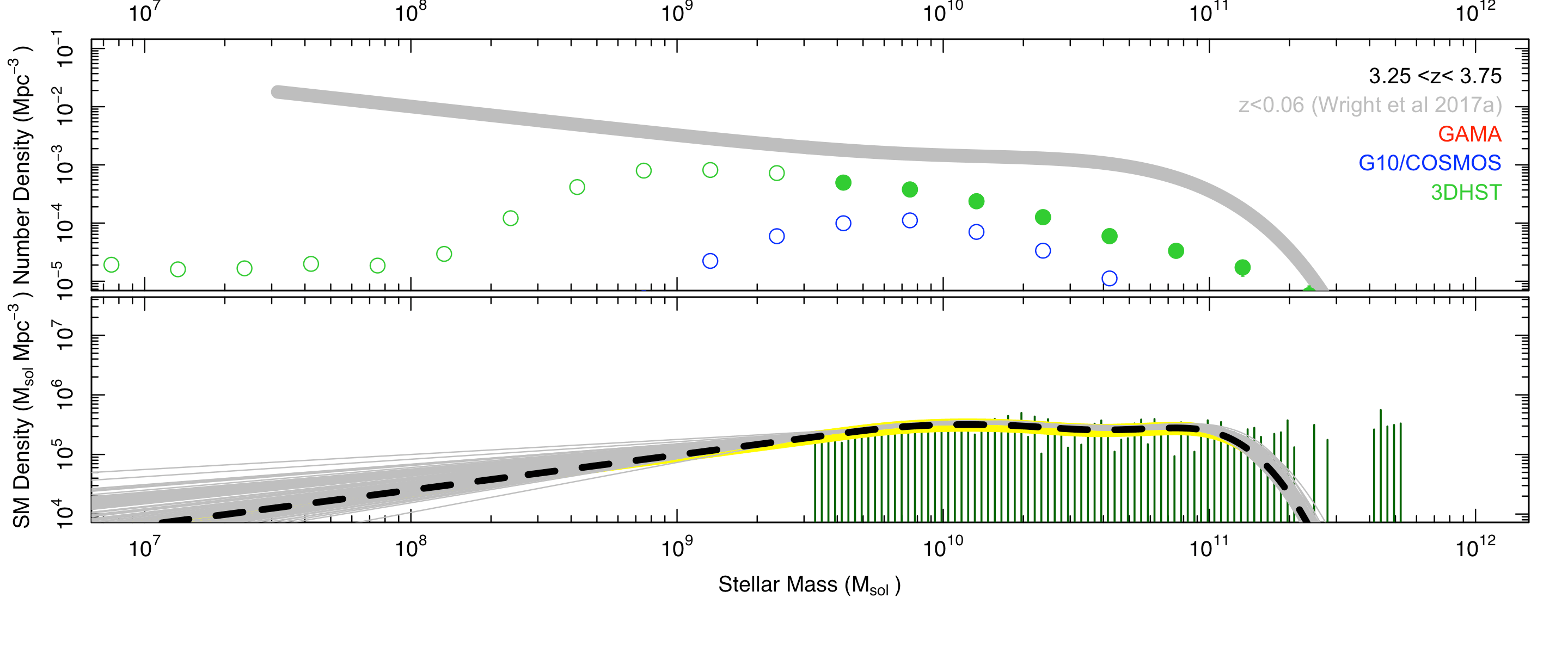}
\vspace{-0.5cm}
\caption{As for Fig.~\ref{fig:methods} except for the redshift range indicated.}
\end{center}
\end{figure}


\begin{figure}
\begin{center}
\includegraphics[width=\columnwidth]{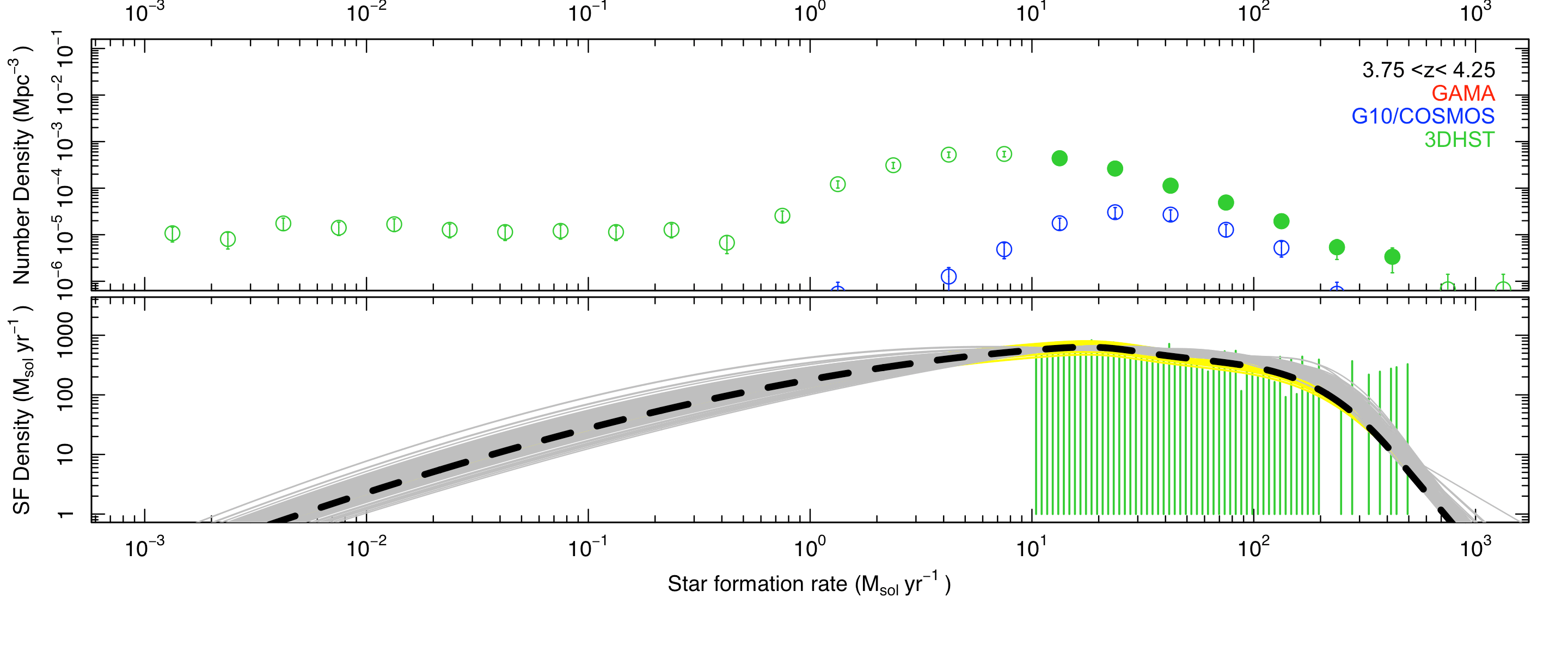}
\vspace{-0.5cm}
\includegraphics[width=\columnwidth]{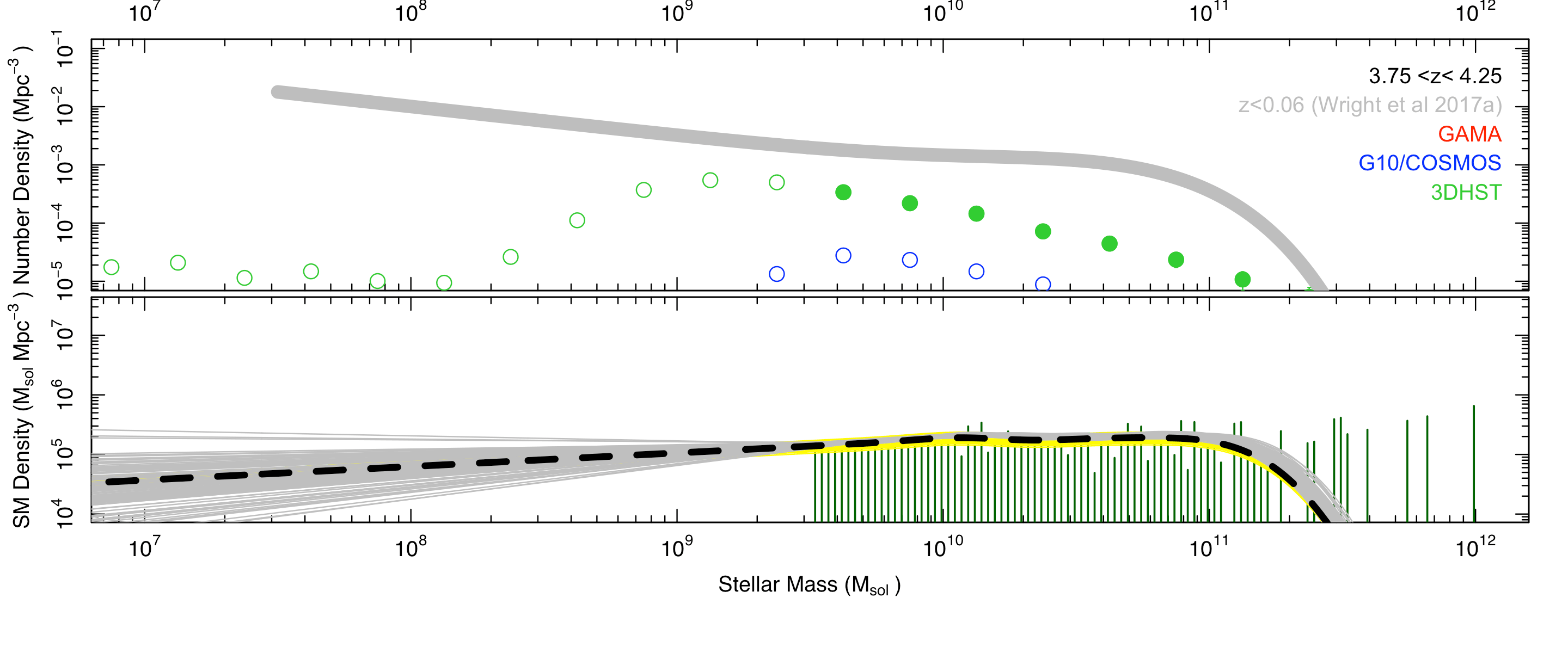}
\vspace{-0.5cm}
\caption{As for Fig.~\ref{fig:methods} except for the redshift range indicated.}
\end{center}
\end{figure}


\begin{figure}
\begin{center}
\includegraphics[width=\columnwidth]{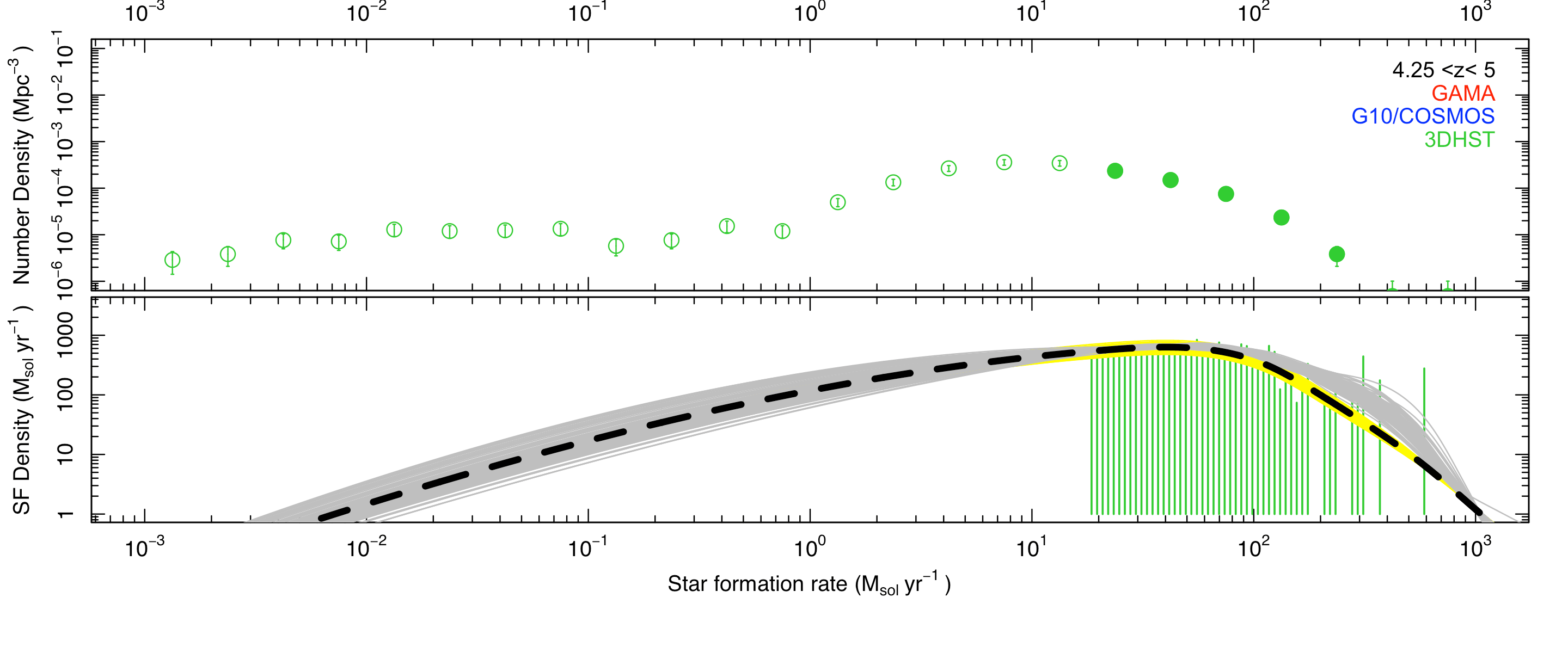}
\vspace{-0.5cm}
\includegraphics[width=\columnwidth]{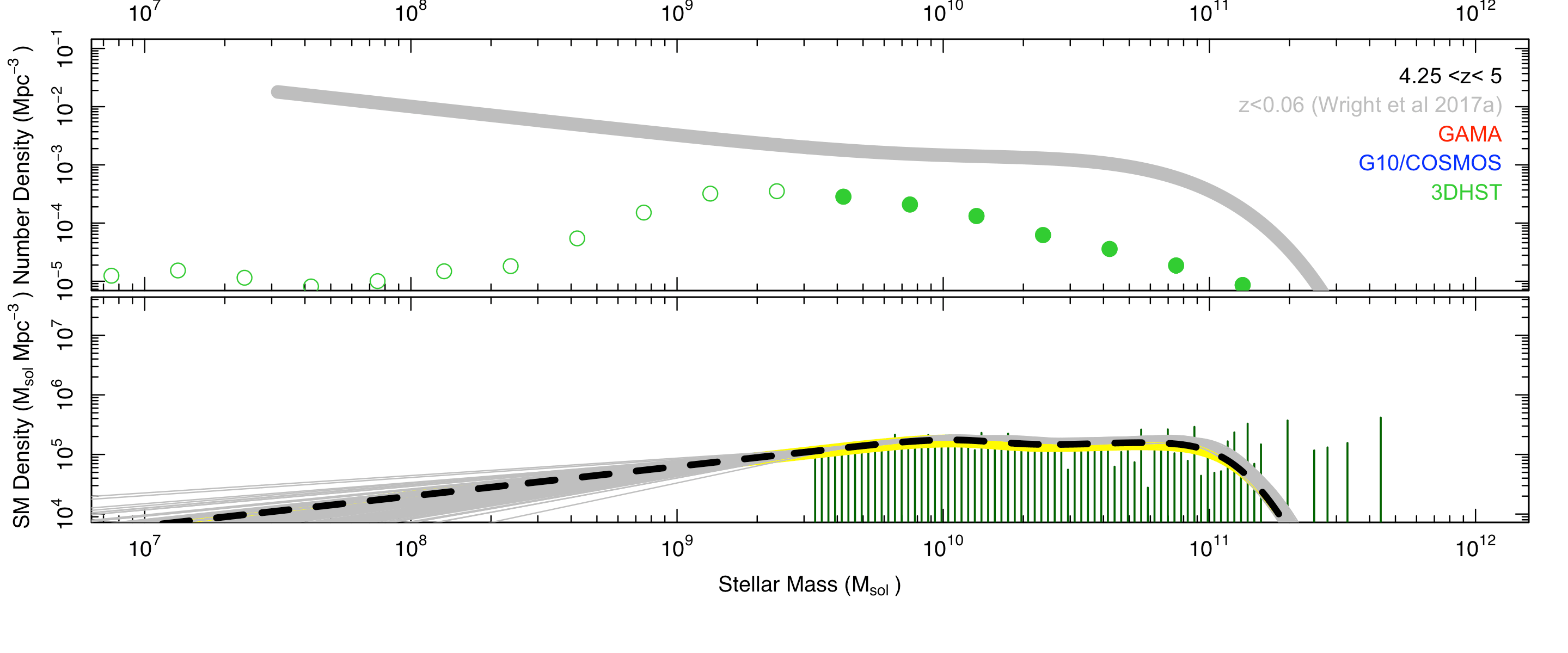}
\vspace{-0.5cm}
\caption{As for Fig.~\ref{fig:methods} except for the redshift range indicated.}
\end{center}
\end{figure}


\section{CSFH, SMD and DMD versus redshift}
\begin{figure*}
\begin{center}
\includegraphics[width=\textwidth]{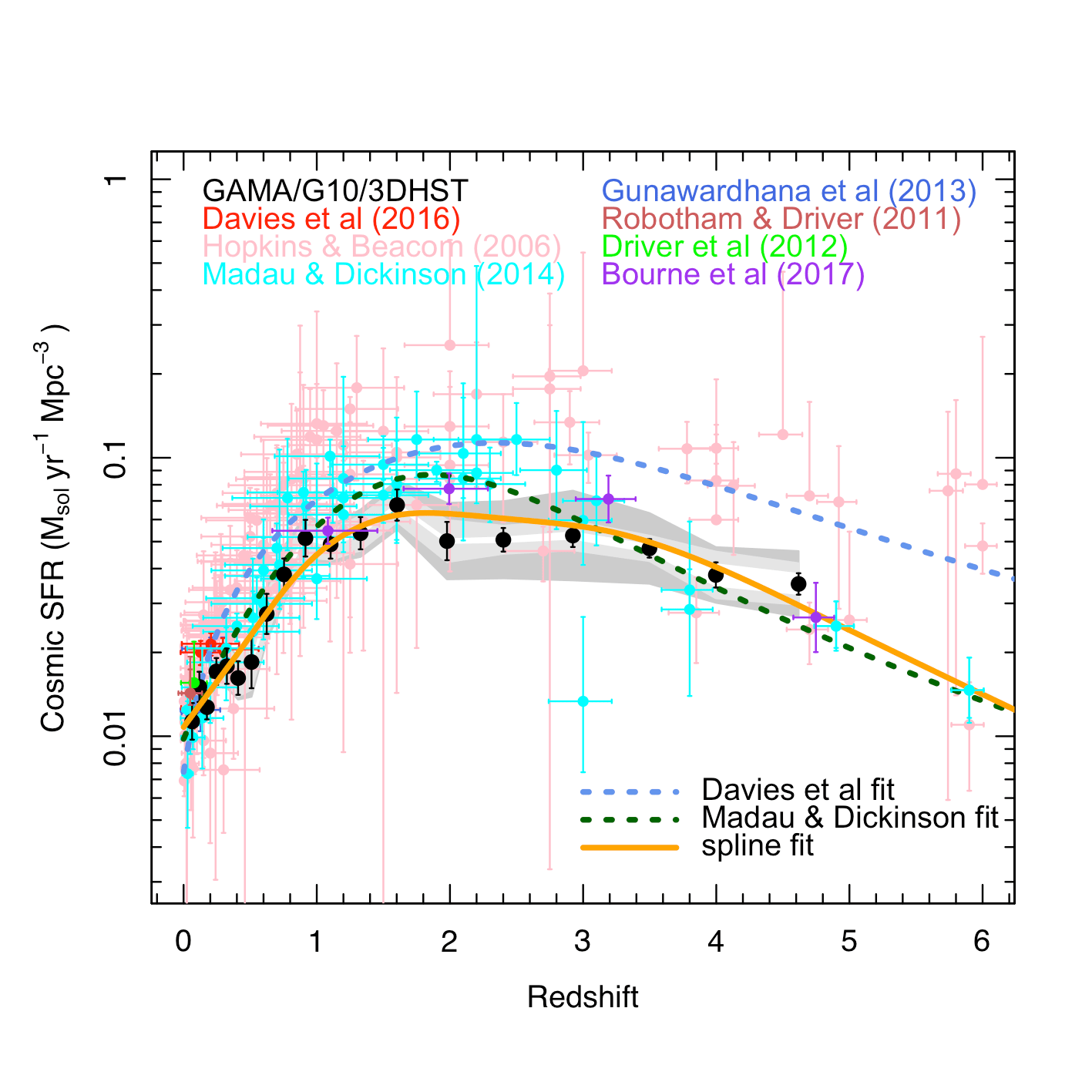}
\caption{An identical copy to Fig.~\ref{fig:data_sfr} except with a linear redshift axis for those who prefer to view in a high-z weighted manner. \label{fig:plotall_z}}
\end{center}
\end{figure*}

For those who prefer to view their data in the highly biased linear
redshift plane we include Fig.~\ref{fig:plotall_z} which replicates
the data shown on Fig.~\ref{fig:data_sfr} but not with a linear redshift
axis, which some folk might find useful.


\end{document}